\documentclass[aps,twocolumn,prc,superscriptaddress,showpacs,nofootinbib,floatfix,amssymb,amsfonts,amsmath]{revtex4-2}
\pdfoutput=1

\usepackage{graphicx}
\usepackage{dcolumn}
\usepackage{bm}
\usepackage{amssymb}
\usepackage{color}
\usepackage{ltxcmds}

\begin{document}

\title{Covariant density functional theory input for r-process simulations in actinides and 
superheavy nuclei: the ground state and fission properties}

\author{A.\ Taninah}
\affiliation{Department of Physics and Astronomy, Mississippi
State University, MS 39762}

\author{S.\  E.\ Agbemava}
\affiliation{Department of Physics and Astronomy, Mississippi
State University, MS 39762}
\affiliation{Ghana Atomic Energy Commission, National
Nuclear Research Institute, P.O.Box LG80, Legon, Ghana.}

\author{A.\ V.\ Afanasjev}
\affiliation{Department of Physics and Astronomy, Mississippi
State University, MS 39762}

\date{\today}

\begin{abstract}

 The systematic investigation of the ground state and fission properties of even-even
actinides and superheavy nuclei with $Z=90-120$ from the two-proton up to two-neutron drip lines with 
proper assessment of systematic  theoretical uncertainties has been performed for the 
first time in the framework of covariant density functional theory (CDFT).  These results 
provide a necessary theoretical input for the r-process modeling in heavy nuclei and, in 
particular, for the study of fission cycling.  Four state-of-the-art globally tested covariant energy 
density functionals (CEDFs), namely,  DD-PC1, DD-ME2, NL3* and PC-PK1, representing 
the major classes of the CDFT models are employed in the present study.  Ground state 
deformations, binding energies, two neutron separation energies, $\alpha$-decay 
$Q_{\alpha}$ values and half-lives and the heights of fission barriers have been 
calculated for all these nuclei. Theoretical uncertainties in these physical observables
and their evolution as a function of proton and neutron numbers have been quantified
and their major sources have been identified.  Spherical shell closures at
 $Z=120$, $N=184$ and $N=258$ and the structure of the single-particle (especially, high-$j$) 
states  in their vicinities as well as nuclear matter properties of employed CEDFs are
two major factors contributing into theoretical uncertainties. However, different 
physical observables are affected in a different way by these two factors. For example, 
theoretical uncertainties in calculated ground state deformations are affected mostly by former 
factor, while theoretical uncertainties in fission barriers depend on both of these
factors.

\end{abstract}

\maketitle

\section{Introduction}
\label{sect-intro}

  The majority of the nuclei found in nature are formed in the astrophysical 
 rapid neutron-capture process (r-process). Indeed, the r-process is responsible for the 
synthesis of approximately half of the nuclei in nature beyond Fe \cite{MP.08} and it is 
the only process which leads to the creation of nuclei heavier than Bi \cite{LMHCF.18}. 
It takes place at extremely high neutron densities ($N_n\geq 10^{20}$ cm$^{-3}$)
which are high enough to make neutron capture faster than $\beta$-decay even for the
nuclei with  neutron excess between 15 to 30
units from the stability line.  The production of 
neutron-rich nuclei located in the vicinity of the neutron dripline is enabled under 
these conditions via neutron capture and $(\gamma, n)$ photodisintegration during 
the r-process.   Once the neutron source ceases, the progenitor nuclei decay either  
via $\beta^-$ decay or $\alpha$ emission  or by fission processes (such as neutron-induced, 
$\beta$-delayed and spontaneous fissions) towards stability and form the stable isotopes 
of elements up to the heaviest species Th, U and Pu.  The typical timescale of the r-process
is in the seconds range \cite{KMBQR.17,LMHCF.18,r-process-review-19}.

   Over the years different possible astrophysical sites have been and
still are considered  as possible candidates for the r-process.
These include core-collapse
supernovae, magneto-rotational core-collapse supernovae,  accretion disk outflows 
from collapsars, neutron star (NS) mergers and neutron star - black hole mergers
etc \cite{KMBQR.17,LMHCF.18,r-process-review-19,TEPW.17}. 
So far only the NS 
merger is experimentally confirmed as a site of the r-process via the observation
of gravitational waves from the GW170817 neutron star merger \cite{NS-grav-wave-exp.17} with 
simultaneous observation of the AT 2017gfo macronova/kilonova afterglow \cite{MM-bin-NS.17a}. 
 In NS mergers, the r-process material originates in the NS crust, and the composition 
of the crust and how it responds to stress caused by the merger dictates the
amount of the r-process material which is ejected. NS merger produces approximately
$10^{-2}$ M$_{\odot}$ of ejected r-process matter in the dynamic ejecta and similar
amount in the accretion disk outflows \cite{JBPAGJ.15,SKKS.15,r-process-review-19}.
Although some uncertainties still exist, at present the NS  merger is considered as the 
major astrophysical site of the r-process providing  the dominant source of heavy nuclei 
\cite{TEPW.17,LMHCF.18,r-process-review-19}. 

\begin{figure*}[htb]
\centering
\includegraphics[angle=-90,width=17.7cm]{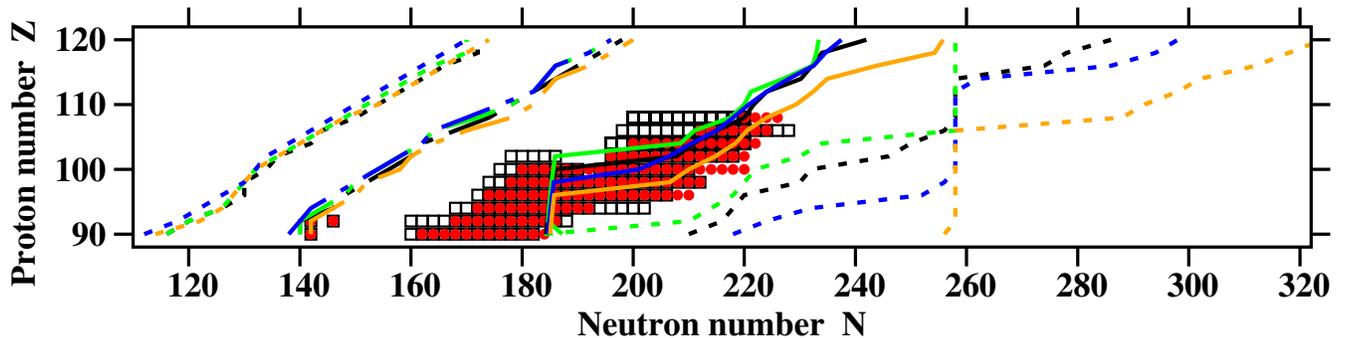}
\caption{The part of nuclear chart under study. Black, green, blue and orange lines are
used for the results obtained with the  DD-PC1, DD-ME2, NL3* and PC-PK1 CEDFs, 
respectively.
Two-proton and two-neutron drip lines predicted by 
four CEDFs are shown by dashed lines.
 Two samples of the distribution of the abundances of heavy 
and superheavy elements in the r-process simulations are shown by open squares (based on 
bottom panel of Fig. 8 of Ref.\ \cite{PLMPRT.12}) and red circles (based on Fig. 3 of Ref.\ \cite{GMRW.17}).
The former results correspond to hot r-process conditions and are based on the ETFS/ETFSI
combination of fission barriers and mass predictions (see Ref.\ \cite{PLMPRT.12}). The latter results have been
obtained in Ref.\ \cite{GMRW.17}  based on fission properties obtained with the BCPM energy density functional.
Note that these r-process calculations are restricted to the $Z\leq 110$ nuclei.
The r-process path is shown here approximately by solid lines corresponding
to two-neutron separation energy $S_{2n}=4.0$ MeV\footnote{The r-process proceeds along the lines 
of constant neutron separation energies towards heavy nuclei that for typical conditions during 
the r-process corresponds to  $S^0_n$ approximately located between 2 and 3 MeV
 \cite{GA.96,MP.08}. However, due to neutron
pairing being stronger at even neutron numbers $N$ the most abundant isotopes always have even 
$N$ values. For this reason, we follow Refs.\ \cite{GA.96,MP.08} and characterize the r-process path
[the path in the $(Z,N)$ plane corresponding to an isotope with highest abundance in each
isotopic chain]  as the path which satisfies the condition that two-neutron separation energy
$S_{2n}$ has the value $S_{2n} = 2S^0_n$.}.
Dot-dashed lines show the beta-stability lines for four functionals.
}
\label{chart-under-study}
\end{figure*}

 The modeling of the r-process in such neutron-rich environments depends  sensitively 
on nuclear masses, $\alpha$- and $\beta$-decay half-lives, neutron capture and fission
properties of the nuclei the majority of which will never be measured in laboratory 
conditions \cite{MWLMB.15,TEPW.17,LMHCF.18,r-process-review-19}. Nuclear masses determine the 
flow path of the r-process, $\beta$-decay rates are responsible for the speed with which 
the r-process moves matter to heavier nuclei, $\alpha$-decays  become important in 
heavy nuclei as a competing decay channel and neutron captures drive the nuclei towards 
neutron-rich side of nuclear landscape. Of special interest in the context of the present 
manuscript are fission properties.   Fission needs to be considered in the r-process simulations
if the neutron-to-seed ratio is large enough to produce fissioning nuclei 
\cite{MMZKL.07,StefG.15,GSLPDHBJ.13,Eichler_2015}.  If the initial neutron-to-seed ratio is large 
($\geq 100$) the r-process can reach the region near and beyond neutron shell closure at 
$N=184$, where fission plays a dominant role (the examples of the distribution of abundances 
of actinides and superheavy elements as obtained in a pair of the r-process simulations are 
shown in Fig. \ref{chart-under-study}). This is exactly the case for the NS mergers \cite{StefG.15}.  In this 
case, all fission channels (neutron induced, beta-delayed, neutrino induced and spontaneous 
fissions)  need to be considered. 
Fission leads to the termination of  hot r-process by means of fission-cycling which returns
matter to lighter nuclei \cite{MWLMB.15}. It also determines the strength of fission cycling, the ratio of 
the actinides to light and medium mass r-process nuclei and thus the shape of final
element abundance pattern.  In addition, it defines the possibility of the formation of 
neutron-rich superheavy nuclei in the r-process \cite{PLMPRT.12}.

  Thus, in the situation when experimental data are not known the outcome of the r-process modeling 
sensitively depends on the quality of employed theoretical frameworks and associated theoretical 
uncertainties and their propagation on going to neutron-rich nuclei. By tradition, the output of different 
theoretical frameworks is used for different physical observables (such as masses, the rates and 
half-lives of different decay channels and reactions etc.) in the
the r-process modeling. 
Existing r-process calculations, which include information on fission properties, are based 
on the fission barrier heights obtained in  non-relativistic models 
\cite{Eichler_2015,MWLMB.15,GPR.18}.  So far fission barrier heights obtained in finite range 
droplet model (FRDM), Thomas-Fermi (TF) model and
extended Thomas-Fermi model with Strutinsky integral approach 
(ETFSI-Q), Hartree-Fock-Bogoliubov (HFB) model with Skyrme HFB-14 energy density
functional (EDF) and BCPM EDF have been used in these calculations.  Moreover the sets 
of fission barriers relevant for the r-process simulations have been generated in the FRDM in 
Ref.\ \cite{GPR.18}, in the ETFSI-Q approach with the SkSC4 functional  in Ref.\ \cite{MPRT.98}, 
in the HFB models with Skyrme HFB-14, SV-min, SLy6, SkI3, SV-bas EDFs 
\cite{HFB14-BSk14,ELLMR.12,Reinh.18},  Gogny D1M* \cite{RHR.20} and BCPM \cite{GPR.18} 
functionals. Note that all these calculations assume axial symmetry of the nuclei.

  Covariant density functional theory (CDFT)~\cite{VALR.05} is an approach alternative to 
above mentioned non-relativistic methods and so far it has not been applied for a 
systematic study of fission properties of the nuclei relevant for the r-process modeling.
However, this theory has a number of advantages over non-relativistic methods
which are discussed below. 
Covariant energy density functionals (CEDF) exploit basic properties of QCD at 
low energies, in particular symmetries and the separation of scales~\cite{LNP.641}. 
They provide a consistent treatment of the spin degrees of freedom and spin-orbit 
splittings (\cite{BRRMG.99,LA.11}); the latter has an essential influence on the underlying 
shell structure. In addition, these
functionals include {\it nuclear magnetism} \cite{KR.89}, i.e. a consistent
description of currents and time-odd mean fields important for odd-mass nuclei \cite{AA.10},
the excitations with unsaturated spins, magnetic moments \cite{HR.88} and nuclear
rotations \cite{AR.00,TO-rot}. Because of Lorentz invariance no new adjustable parameters
are required for the time-odd parts of the mean fields \cite{AA.10}. This is contrary to the 
case of non-relativistic Skyrme DFTs in which several prescriptions for fixing time-odd 
mean fields exist
\cite{DD.95,SDMMNSS.10}\footnote{ Unfortunately,
the role of time-odd mean fields in Gogny DFTs has not been studied so far and it is 
unknown whether they are uniquely defined.}. 
This fact could be extremely important  in the applications to fission processes including 
dynamical correlations since time-odd mean fields have a significant impact on collective 
masses \cite{HLNNV.12,GR-plb.18}.

   The goal of this study is to close this gap in our knowledge and  to perform first systematic 
investigation within the CDFT framework of the ground state and fission properties of the nuclei 
with proton numbers $Z=90-120$ located between two-proton and two-neutron drip lines
(see Fig.\ \ref{chart-under-study}). This study will not only provide an input for the r-process 
modeling but also evaluate the extension of nuclear landscape up to two-neutron drip
line as well as estimate relevant theoretical uncertainties and their sources
 in the description of physical
observables of interest.  In addition, it will allow for the first time to compare the predictions
for fission barriers in the nuclei relevant for r-process modeling obtained in relativistic 
and non-relativistic frameworks.

   Considering the region of the nuclear chart in which the r-process is expected to take 
place  and the fact  that there are no experimental data to benchmark theoretical results, it is 
important to estimate theoretical uncertainties in the predictions of physical observables 
of interest \cite{RN.10,DNR.14,AARR.14}. Theoretical uncertainties emerge from the underlying 
theoretical approximations. In the DFT framework, there are two major sources of these approximations, namely,
the range of interaction and the form of the density dependence of the effective interaction
\cite{BHP.03,BB.77}. In the non-relativistic case one has zero range Skyrme and finite range 
Gogny forces and different density dependencies \cite{BHP.03}. A similar situation exists 
also in the relativistic case: point coupling and meson exchange models have
an interaction of zero and of finite range, respectively \cite{VALR.05,DD-ME2,NL3*,DD-PC1}. 
The density dependence is introduced either through an explicit dependence of the coupling 
constants \cite{TW.99,DD-ME2,DD-PC1} or via non-linear meson couplings \cite{BB.77,NL3*}. 
This ambiguity in the definition of the range of the interaction and its density dependence 
leads to several major classes of the covariant energy density functionals (CEDF) which 
were discussed in Ref.\ \cite{AARR.14}.

   Since statistical uncertainties in the physical observables are smaller than systematic ones 
(see Ref.\ \cite{AAT.19}), we focus here on the latter ones.  They are related to the choice of 
EDF.  We follow our previous publications on this topic \cite{AARR.14,AARR.17,AAR.16,AA.16}
and define systematic theoretical uncertainty for a given physical observable (which we call 
in the following ``spreads'')  via the spread of theoretical  predictions as 
\cite{AARR.14}
\begin{equation}
\Delta O(Z,N) = |O_{max}(Z,N) - O_{min}(Z,N)|,
\label{spread} 
\end{equation}
where $O_{max}(Z,N)$ and $O_{min}(Z,N)$ are the largest and smallest
values of the physical observable $O(Z,N)$ obtained within the set
of CEDFs under investigation
for the $(Z,N)$ nucleus. Note that these spreads are only a 
crude approximation to the {\it systematic} theoretical errors discussed 
in Ref.\ \cite{DNR.14} since they are obtained with a very small number of 
functionals which do not form an independent statistical ensemble.
Note also that these {\it systematic} errors are not well defined in unknown 
regions of nuclear chart or deformation since systematic biases of 
theoretical models could not be established in these regions in the 
absence of experimental data and/or an exact theory. 

   In order to consider several possible scenarios in the evolution of  physical 
observables as a function of proton and neutron numbers and to evaluate 
systematic theoretical uncertainties, the CEDFs NL3* \cite{NL3*}, DD-ME2 
\cite{DD-ME2}, DD-PC1 \cite{DD-PC1} and PC-PK1 \cite{PC-PK1} are used 
here\footnote{The compilation of Ref.\ \cite{RMF-nm} published in 2014 
indicates the existence of 263 CEDFs ranging from simplest ansatz 
non-linear meson exchange functionals such as NL1 \cite{NL1} and NL3* \cite{NL3*} 
to more microscopically motivated CEDFs such as G1, G2 \cite{FST.97} and DD-ME$\delta$
\cite{DD-MEdelta}. In addition, a number of new functionals were fitted
in the time period between 2014 and 2020 (see, for example, Refs.\ \cite{NIV.17,G3,DD-PCX,AAT.19,TAAR.20}) 
so at  present the total number of available CEDFs is likely to be in the vicinity of 300.
Because of extremely time-consuming nature of numerical calculations in this project, 
we use only the indicated last generation functionals.  They outperform previous 
generation functionals in terms of the accuracy of global description of ground state observables 
such as binding energies and charge radii \cite{AARR.14,AA.16,AANR.15,LLLYM.15}, 
properly describe the regions of octupole deformation \cite{AAR.16,AA.17-oct}  and 
are able to reproduce experimentally known fission barriers in actinides 
\cite{AAR.10,PNLV.12,LZZ.14}.}
%
 for all $Z=90-120$ even-even nuclei located between two-proton and 
two-neutron drip lines\footnote{Present
study partially builds on previous results obtained by us. These are ground 
state properties of the $Z=90-104$ nuclei located between two-proton and
two-neutron drip lines obtained in reflection
symmetric RHB calculations with the NL3*, DD-ME2 
and DD-PC1 CEDFs in Ref.\ \cite{AARR.14}, ground state properties of
octupole deformed nuclei with $N<210$ obtained in reflection asymmetric
RHB calculations with NL3*, DD-ME2,
DD-PC1 and PC-PK1 functionals in Ref.\ \cite{AAR.16,AA.17-oct}  and 
ground state properties  and the heights of inner fission barriers of 
superheavy nuclei with $Z=100-120, N\leq196$ obtained in reflection symmetric
RHB calculations with NL3*, DD-ME2,
DD-PC1 and PC-PK1 functionals in Refs.\ \cite{AANR.15,AARR.17}. 
Additional information on the extension
of nuclear landscape to  $Z>120$ obtained with DD-PC1 CEDF can 
be found in Refs.\ \cite{AAG.18,AATG.19,AAT.20-acta}.}.
These are state-of-the-art functionals representing the 
major classes of CDFTs (for more details see the discussion in Sect. II of 
Ref.\ \cite{AARR.14} and the introduction to Ref.\ \cite{AANR.15}).
Their performance and related theoretical uncertainties have recently 
been analyzed globally in Refs.\ \cite{AARR.14,ZNLYM.14,AA.16,AAR.16} 
and in particular in superheavy nuclei in Refs.\ \cite{AANR.15,AARR.17}.
They  are characterized by an improved accuracy of the  description of 
experimental data as compared with the previous generation of CEDFs. 
The fact that the NL3*, DD-PC1 and PC-PK1 functionals reproduce empirical
data on fission barrier heights in actinides \cite{AAR.10,PNLV.12,LZZ.12,LZZ.14} 
is especially important in the context of the present study.

  The manuscript is organized as follows. Theoretical framework and the 
details related to the calculations of the ground states and fission barriers are discussed 
in Sec.\ \ref{sect-theory}.  Sec.\ \ref{sect-ground} is devoted to the analysis of the results of 
the calculations for ground state properties.  Theoretical results for $\alpha$-decay
properties and related theoretical uncertainties are presented in Sec.\ \ref{sect-alpha}. The 
heights of primary fission barriers, their distribution in the $(Z,N)$ plane, related 
theoretical uncertainties and the comparison with non-relativistic results are
considered in Sec.\ \ref{sect-fission}. Finally, Sec.\ \ref{sect-concl} summarizes the results of our
work.

\begin{figure*}[htb]
\centering
\includegraphics[angle=0,width=5.53cm]{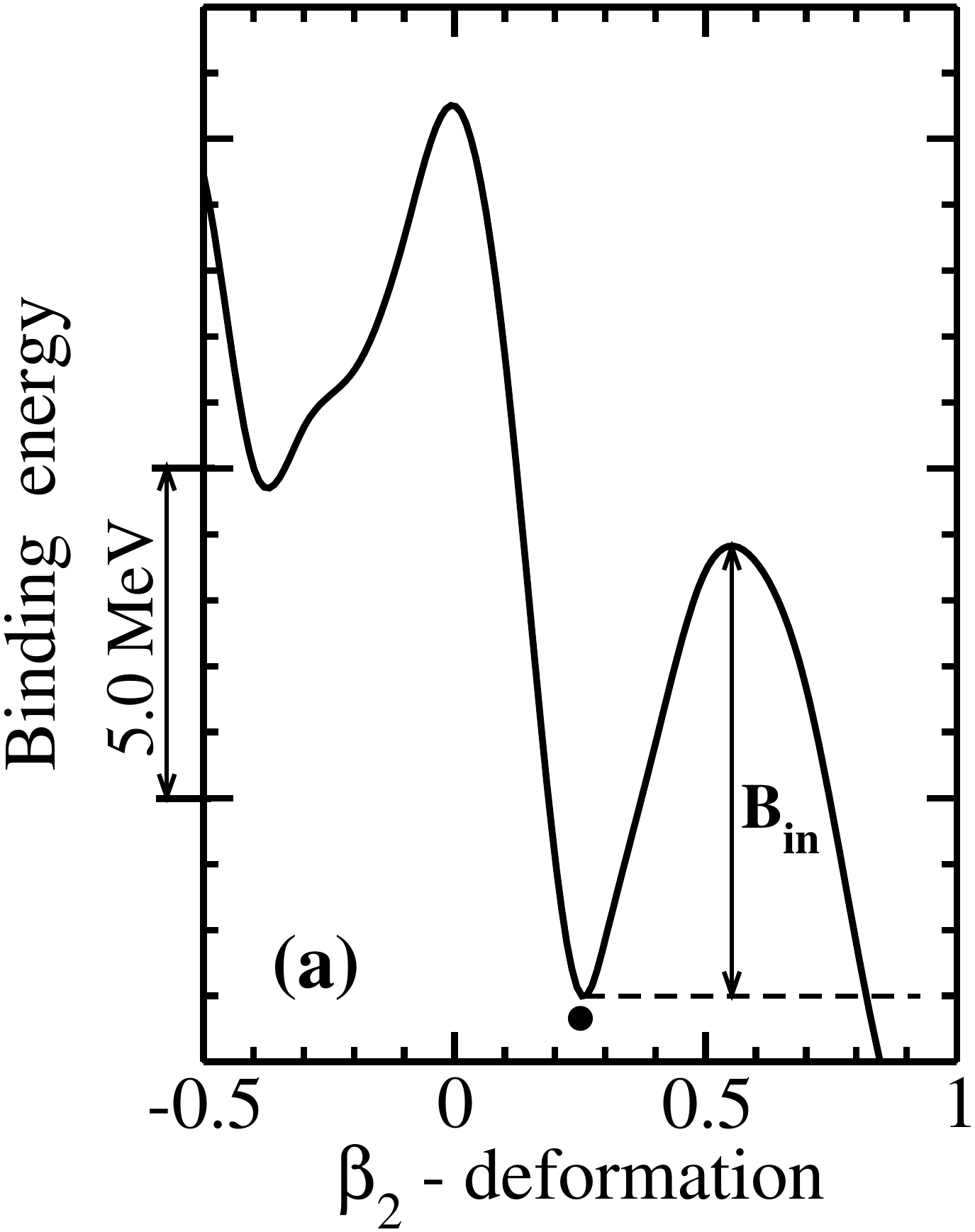}
\includegraphics[angle=0,width=5.0cm]{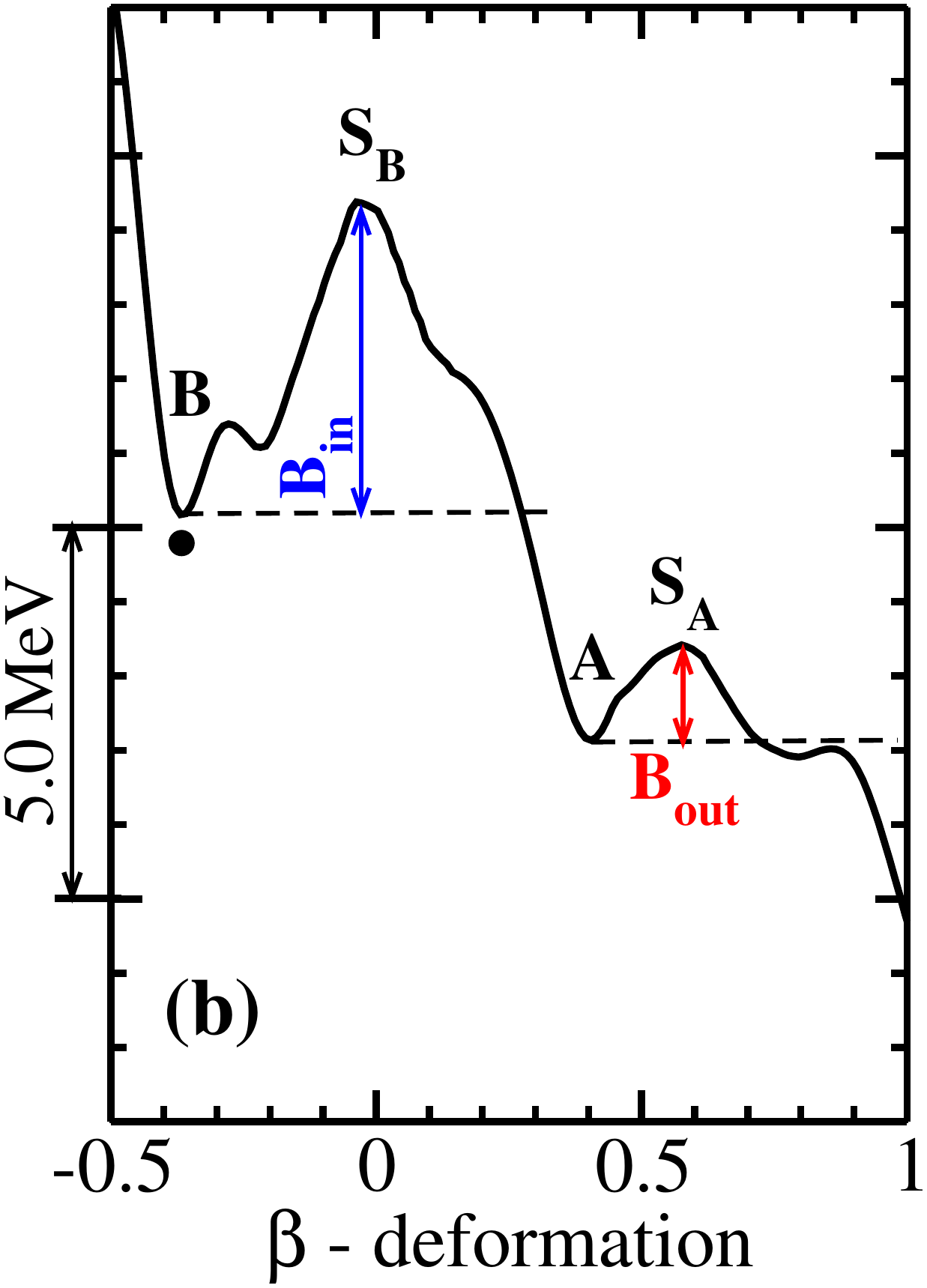}
\includegraphics[angle=0,width=5.0cm]{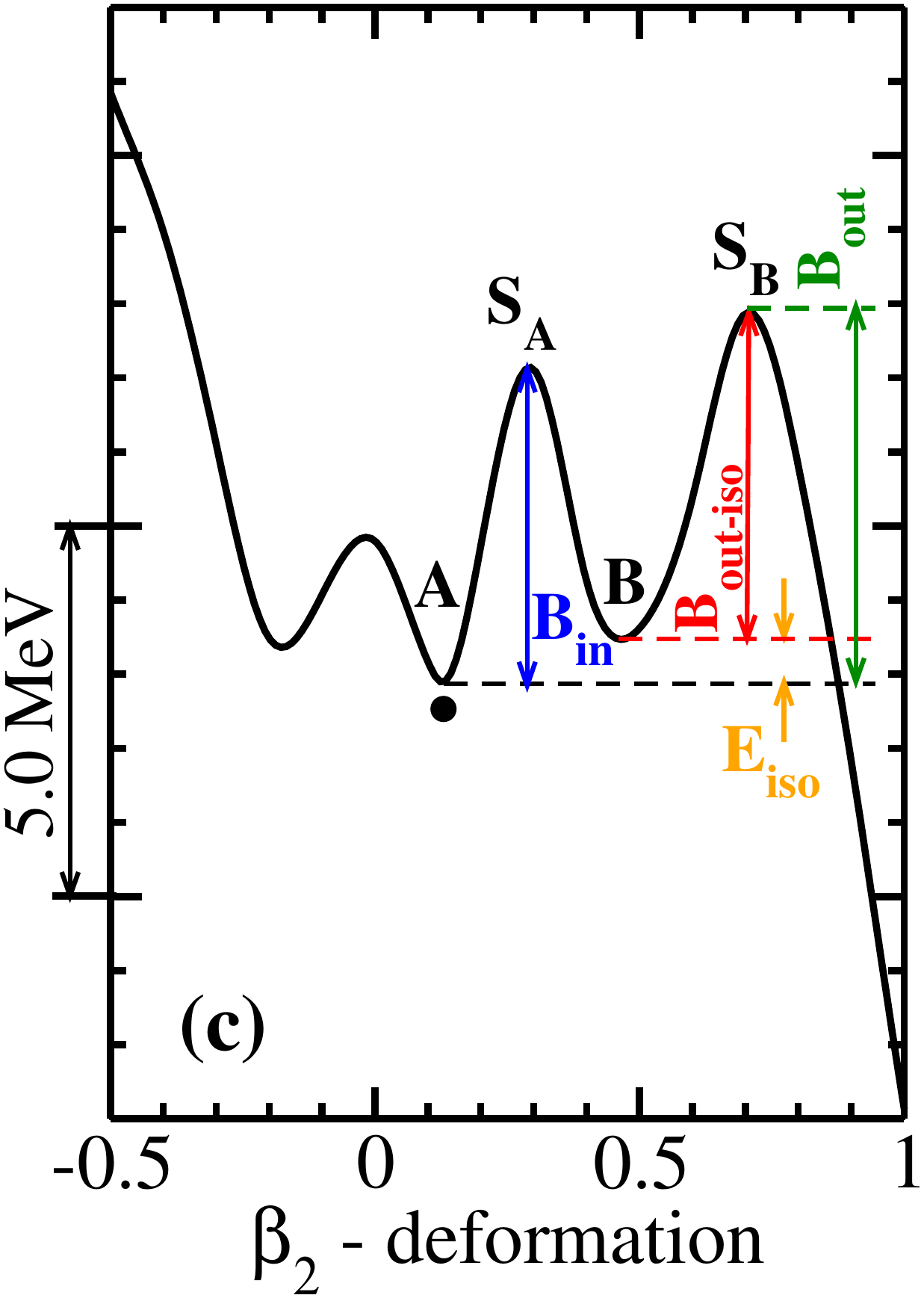}
\caption{Schematic illustration of different types of deformation energy curves and
the selection of respective ground states (see text for details).  Local minima are labelled 
by the letters A and B and the saddle points of respective fission barriers by $\rm S_A$ 
and $\rm S_B$.  Solid circles indicate the minima selected as
the ground states. The heights of inner and outer 
fission barriers with respect of corresponding minima (shown by dashed lines) are 
indicated by $\rm B_{in}$ and $\rm B_{out}$. $\rm B_{out-iso}$ is  the height of outer 
fission barrier with respect of fission isomer. 
}
\label{def-curve-sel}
\end{figure*}

\section{Theoretical framework}
\label{sect-theory}

\begin{figure*}[htb]
\centering
\includegraphics[angle=0,width=5.9cm]{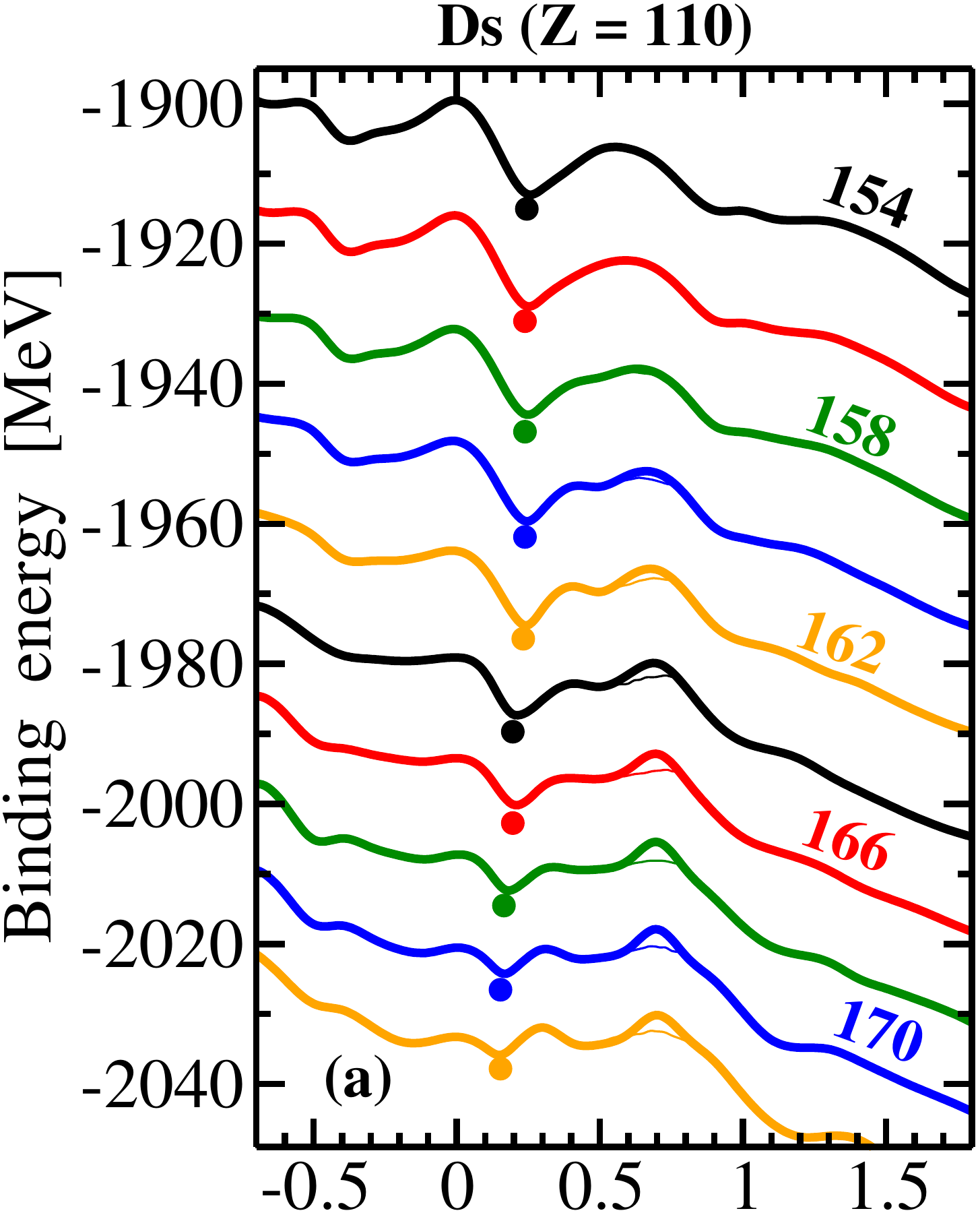}
\includegraphics[angle=0,width=5.9cm]{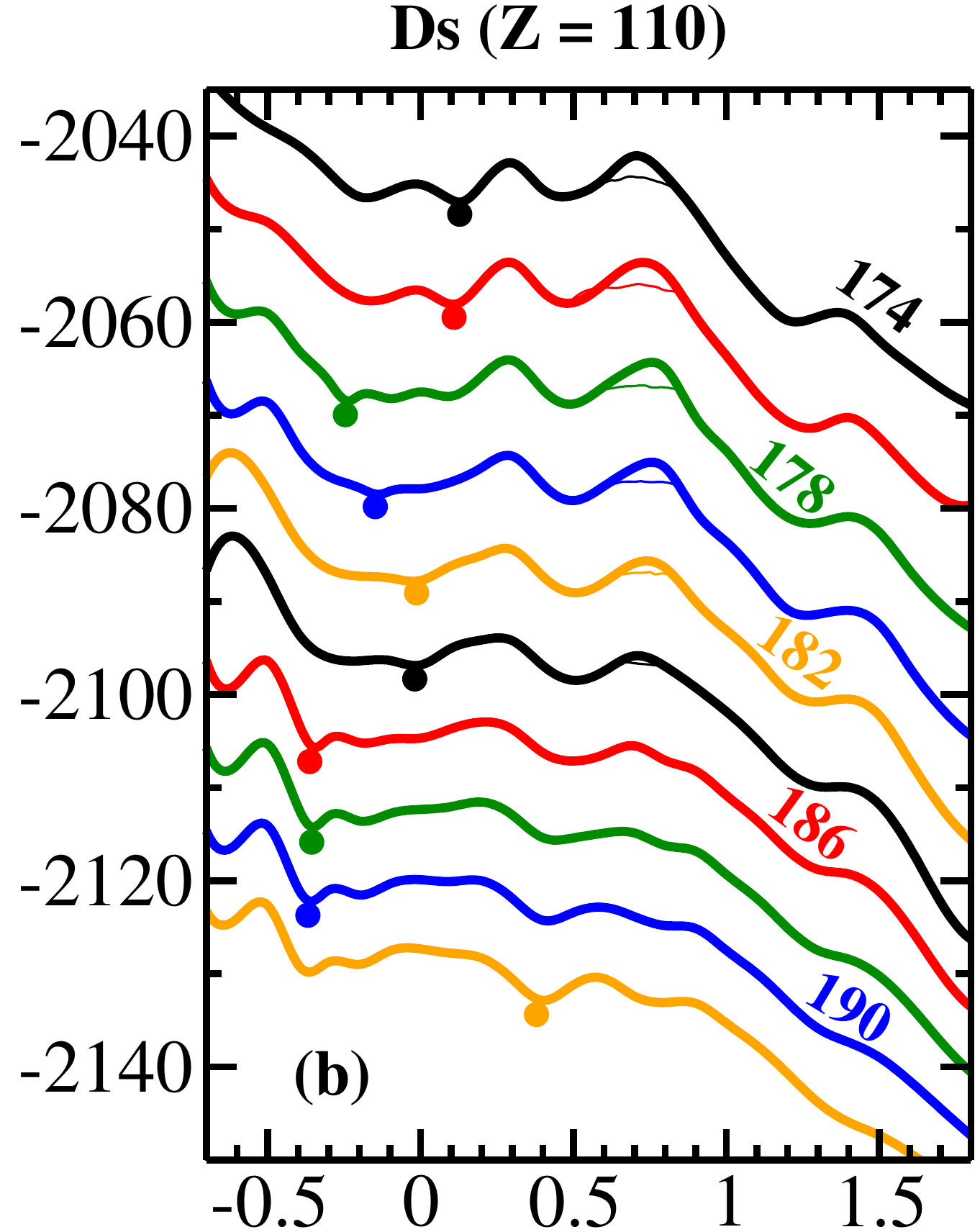}
\includegraphics[angle=0,width=5.9cm]{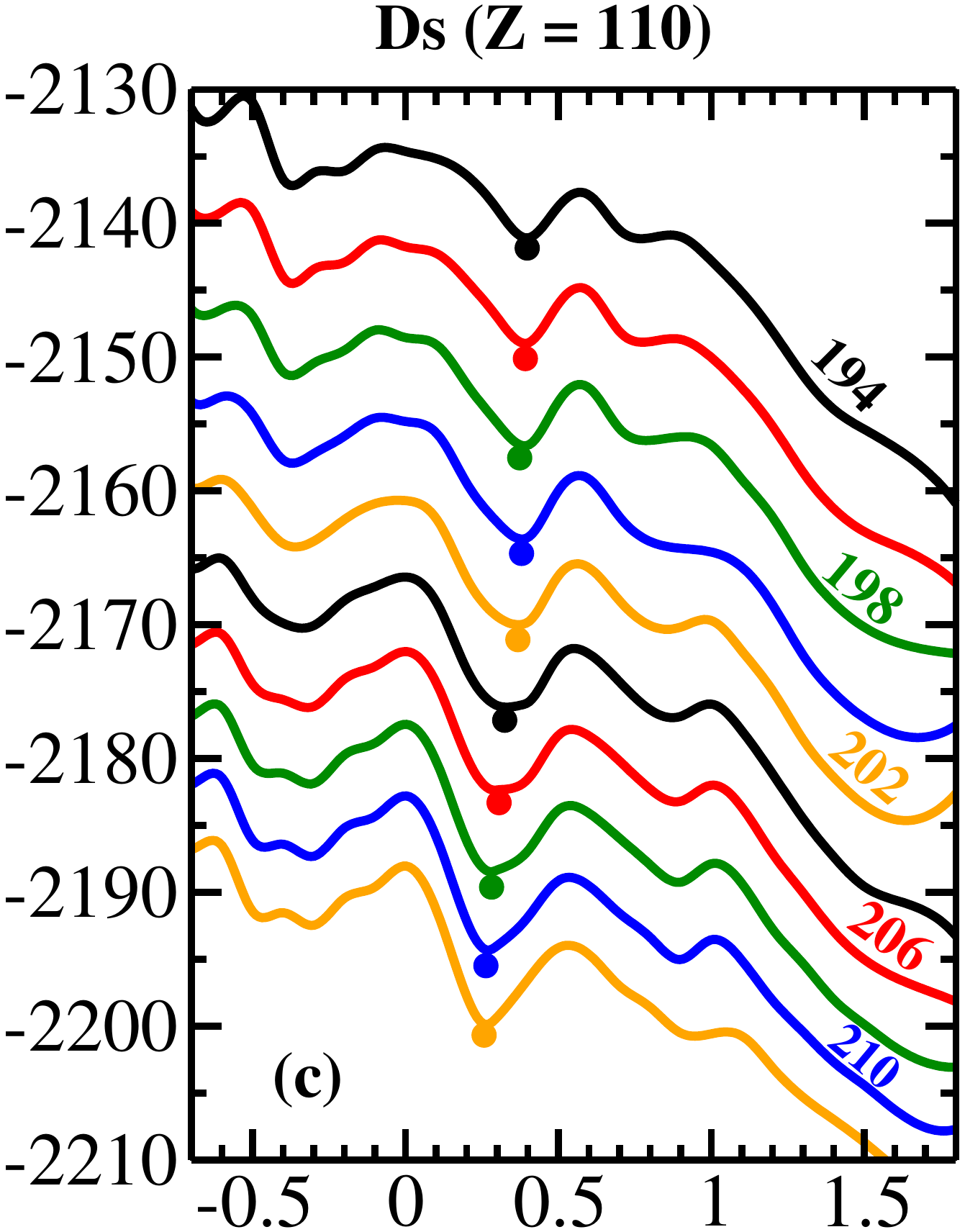}
\includegraphics[angle=0,width=5.9cm]{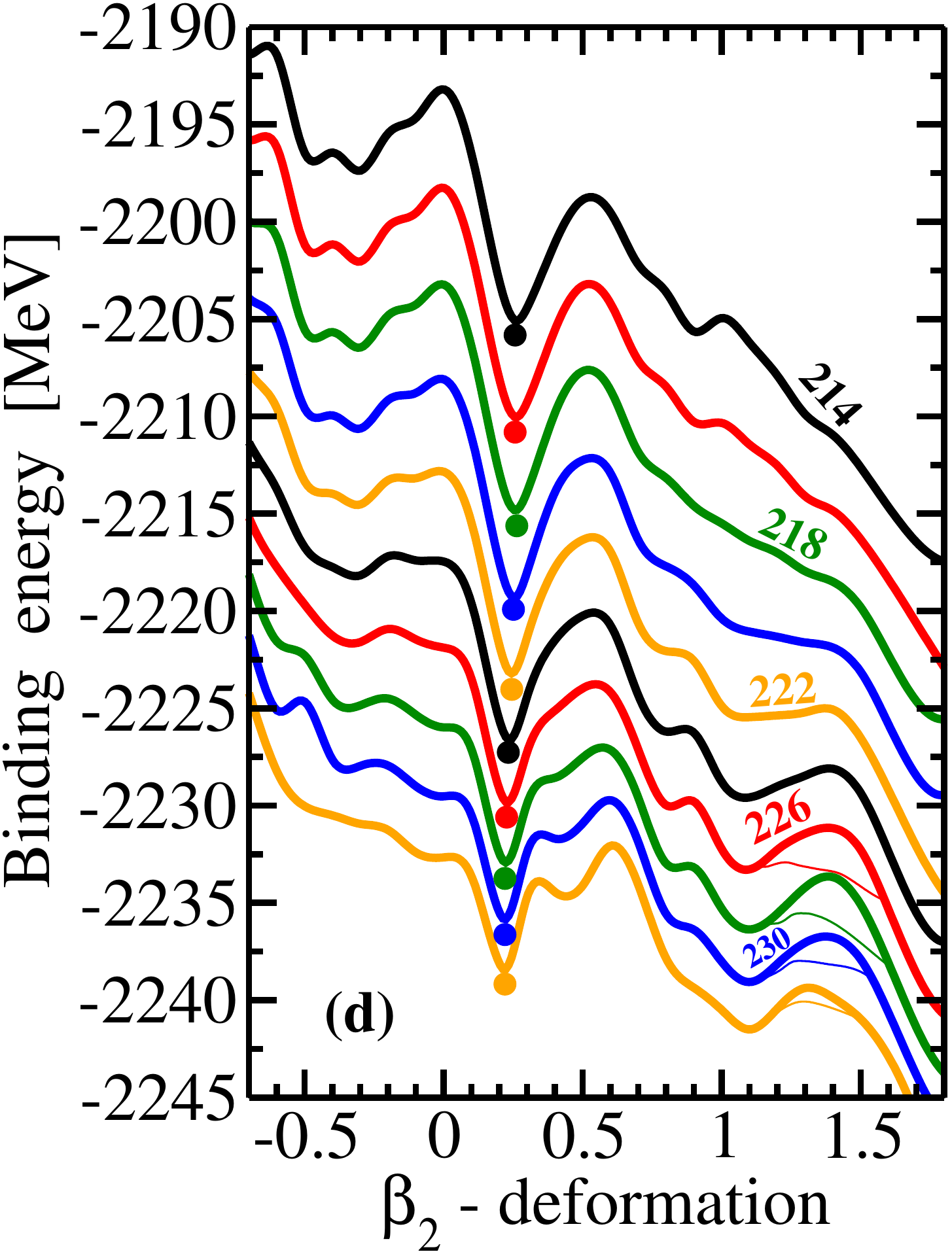}
\includegraphics[angle=0,width=5.9cm]{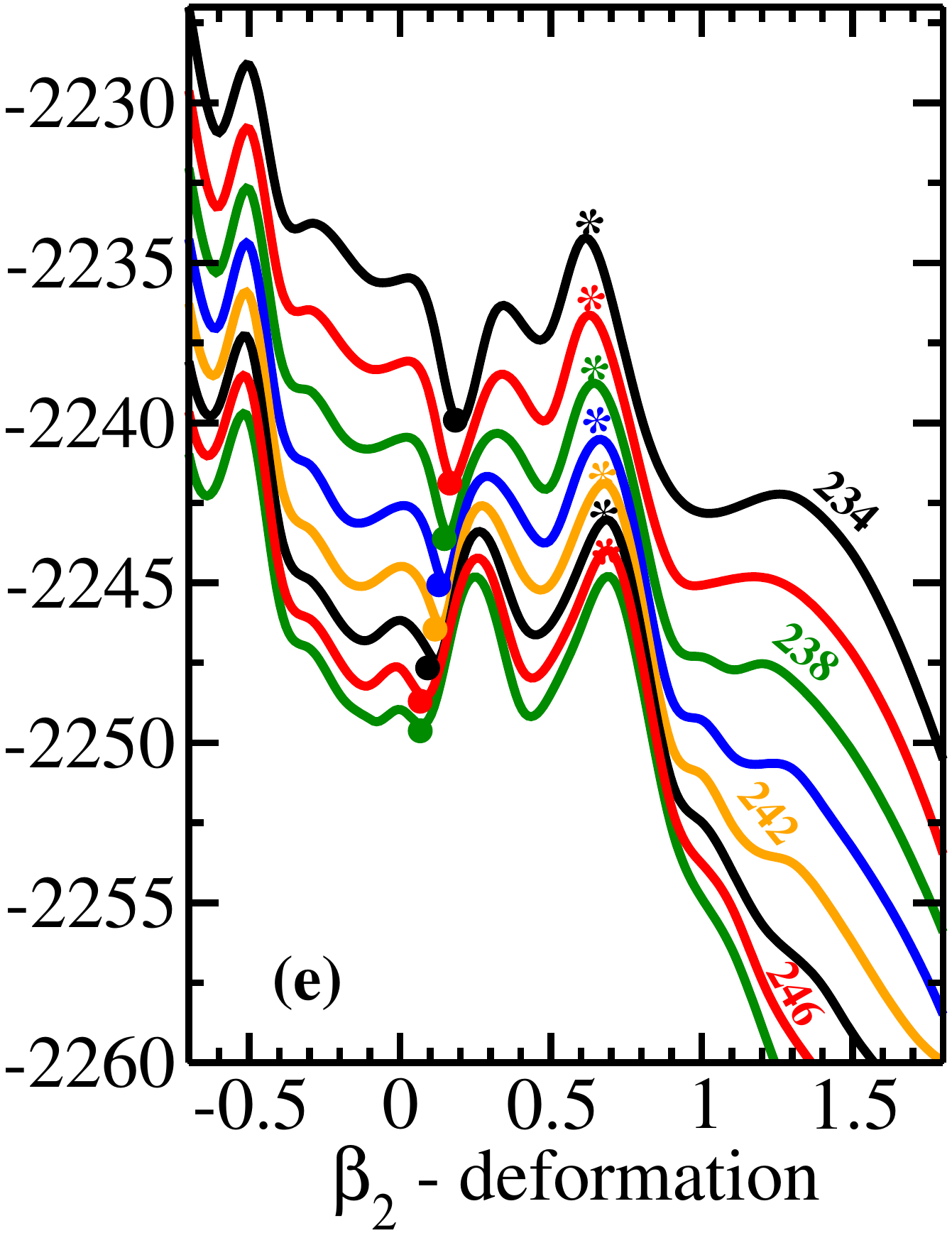}
\includegraphics[angle=0,width=5.9cm]{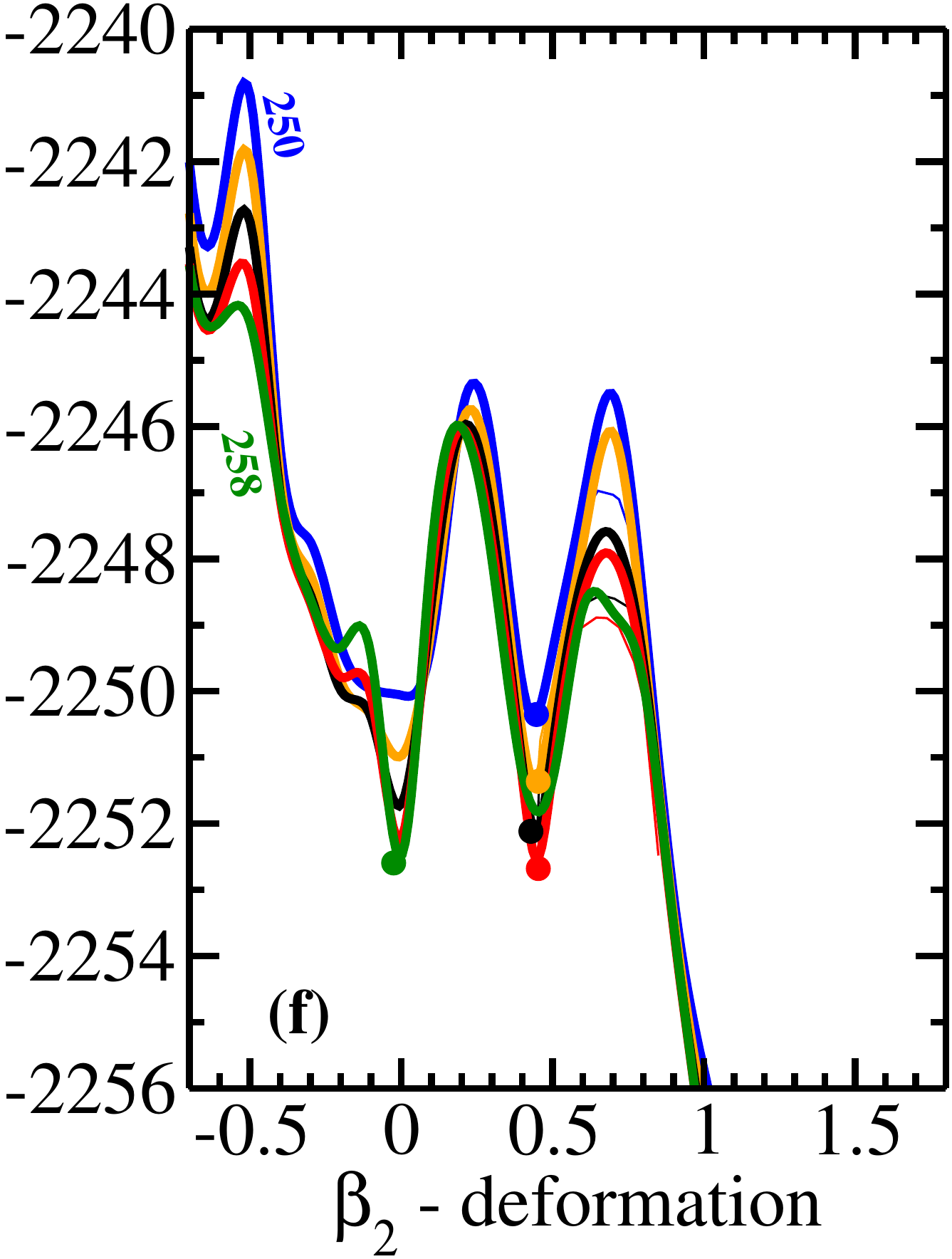}
\caption{Deformation energy curves obtained in axial RHB calculations with DDPC1 
functional for the Ds isotopes. The isotopes are indicated by respective neutron 
numbers. Thick and thin solid lines are used for the RS-RHB and RA-RHB 
results, respectively. The results of the RA-RHB calculations are shown only in 
the deformation range in which they are lower in energy than the RS-RHB ones.
Solid circles indicate the ground states and the asterisks  denote the saddles of outer  
fission barriers which are not affected by octupole deformation.  Blue, orange, black, 
red and green lines are used to indicate neutron numbers the last digits of which are 
0, 2, 4, 6, and 8, respectively. Note that the energy on the vertical axis spans 
different ranges in different panels.
\label{deformation-curve-Ds}
}
\end{figure*}

\begin{figure*}[htb]
\centering
\includegraphics[angle=0,width=5.9cm]{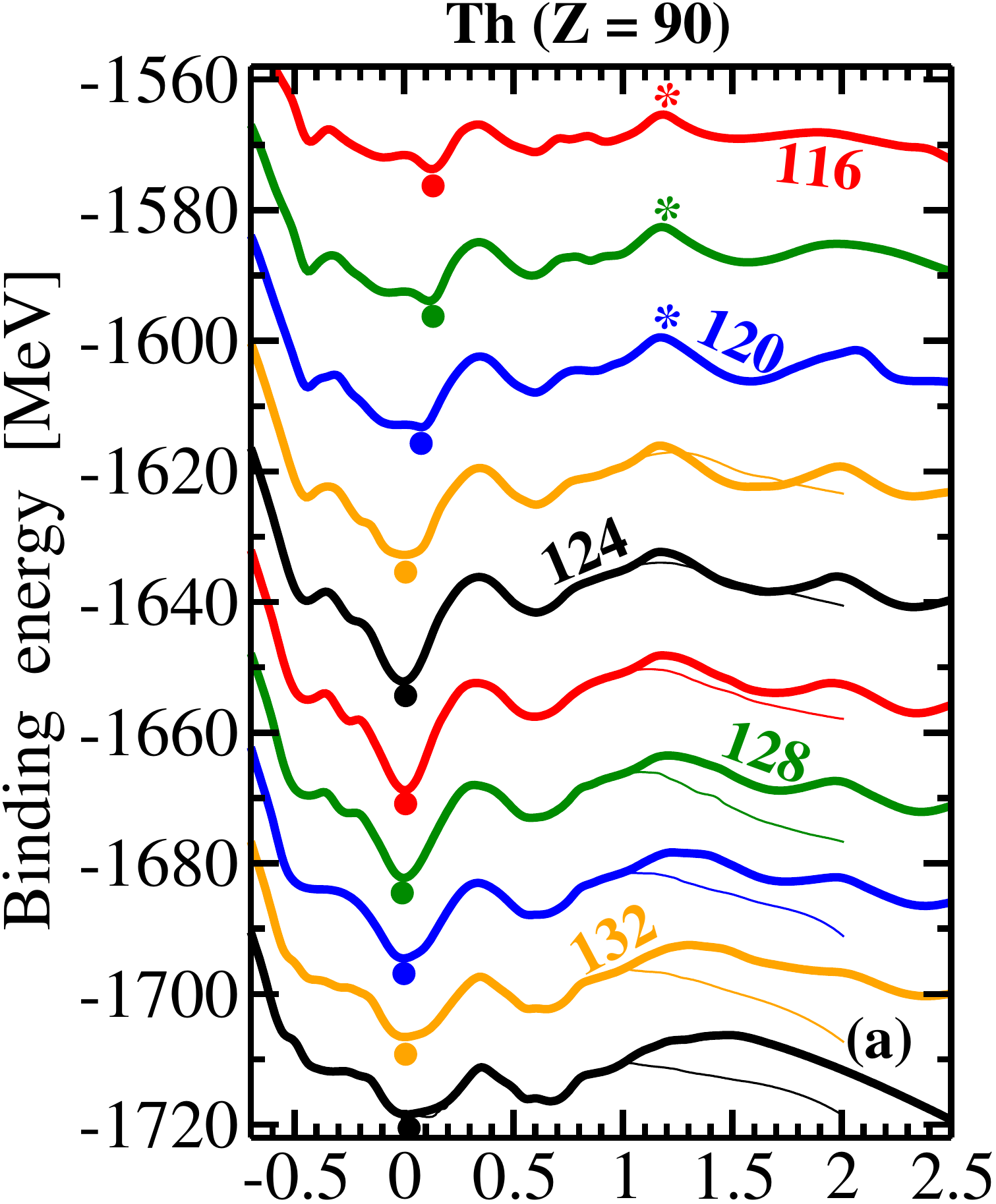}
\includegraphics[angle=0,width=5.9cm]{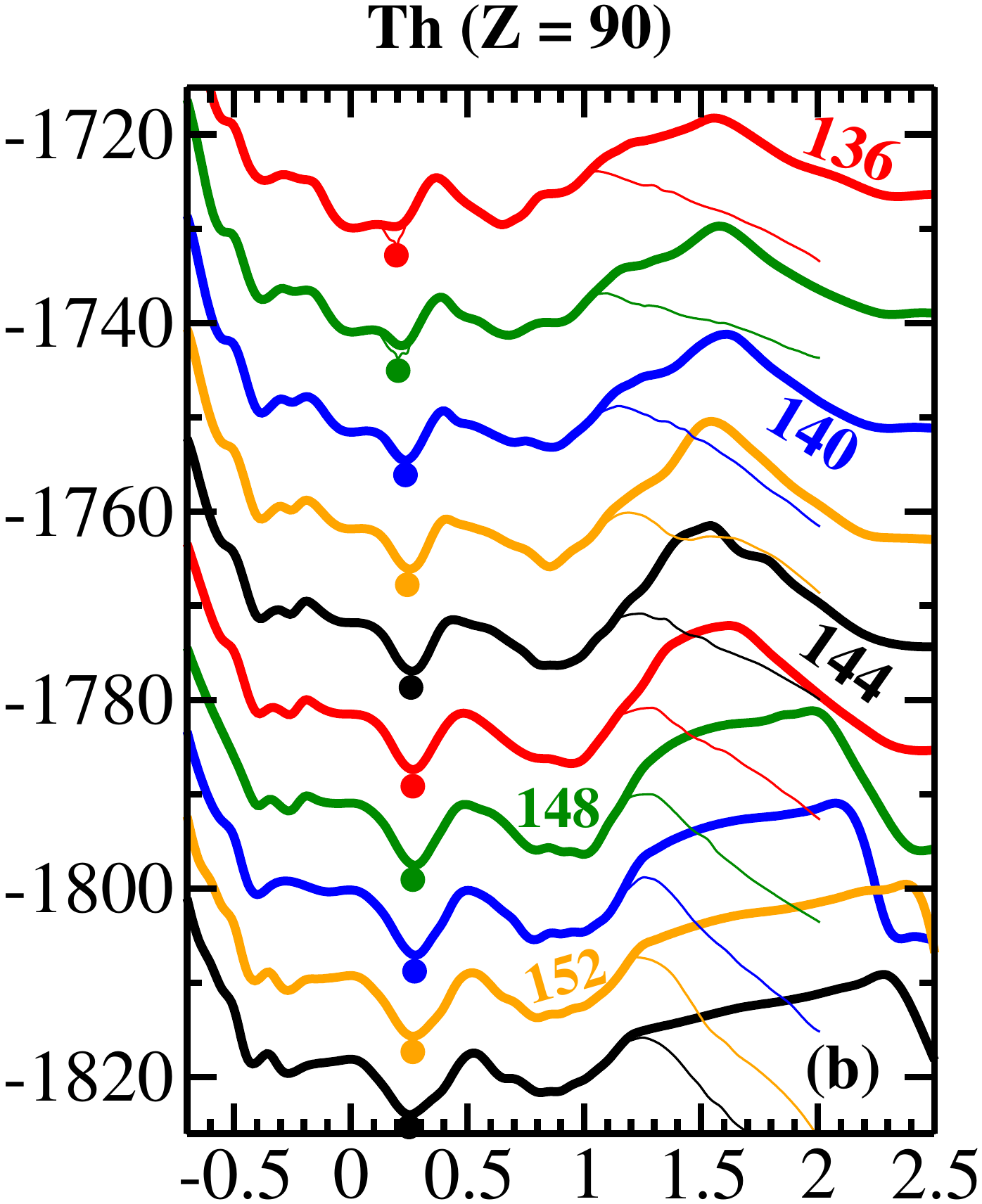}
\includegraphics[angle=0,width=5.9cm]{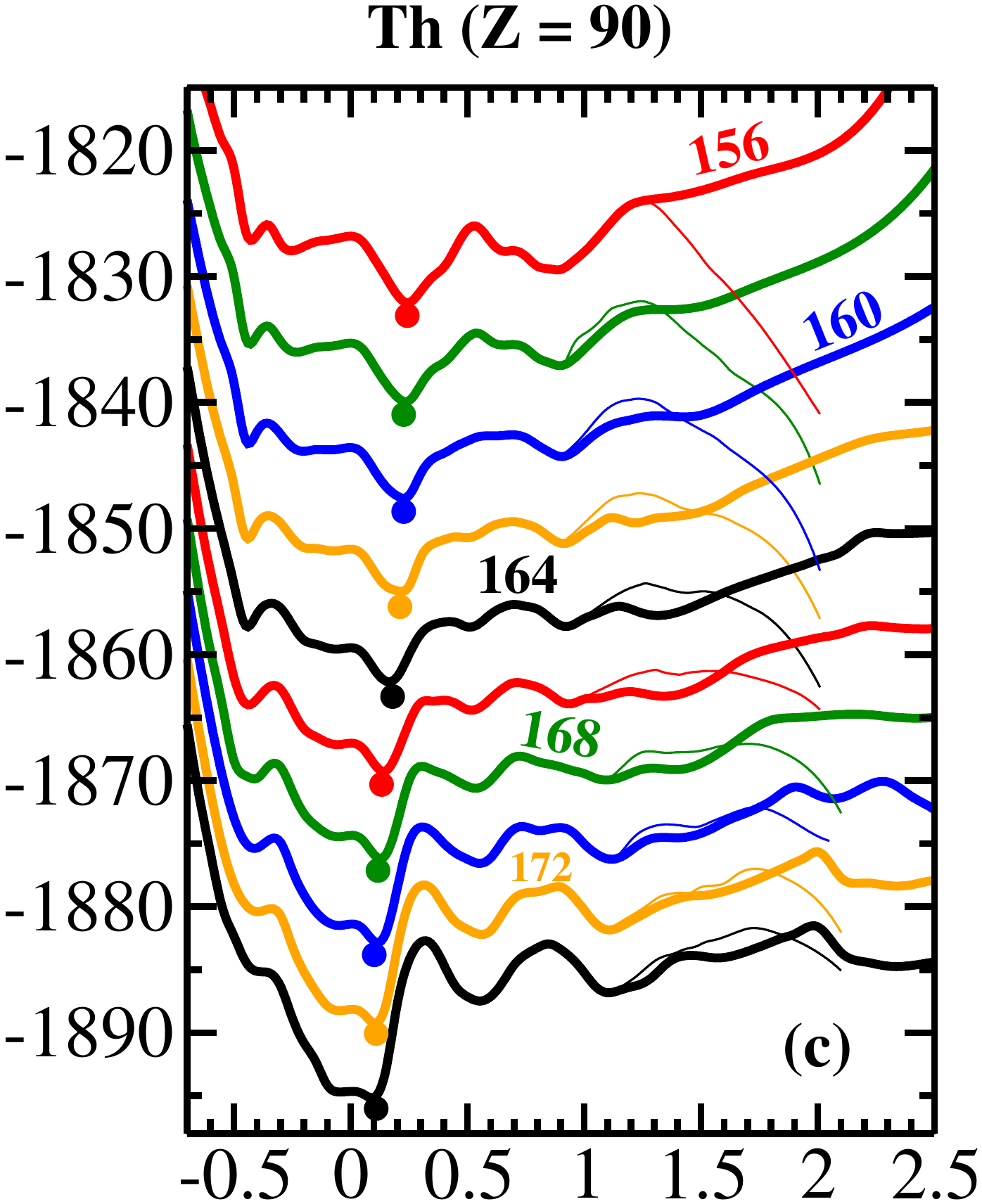}
\includegraphics[angle=0,width=5.9cm]{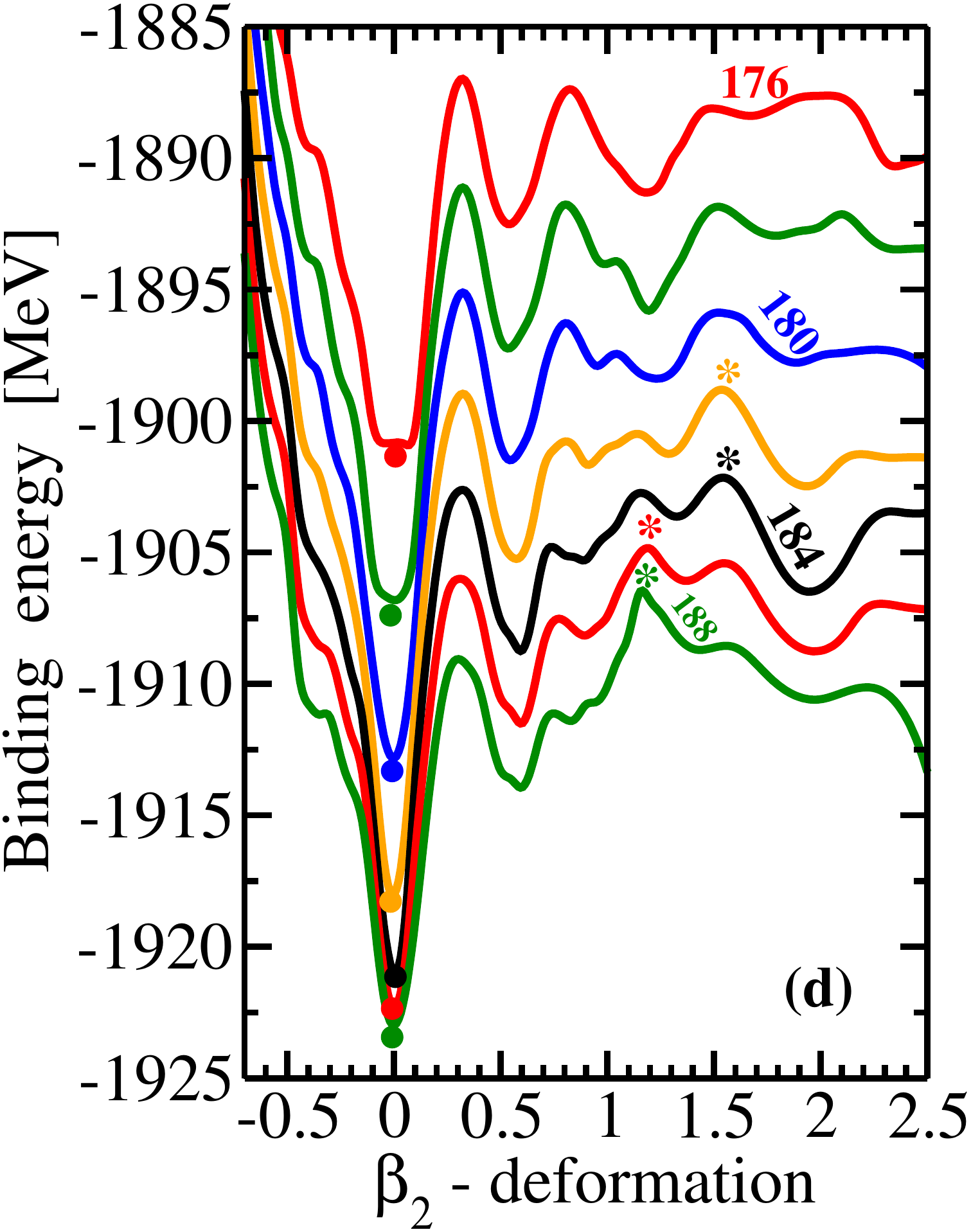}
\includegraphics[angle=0,width=5.9cm]{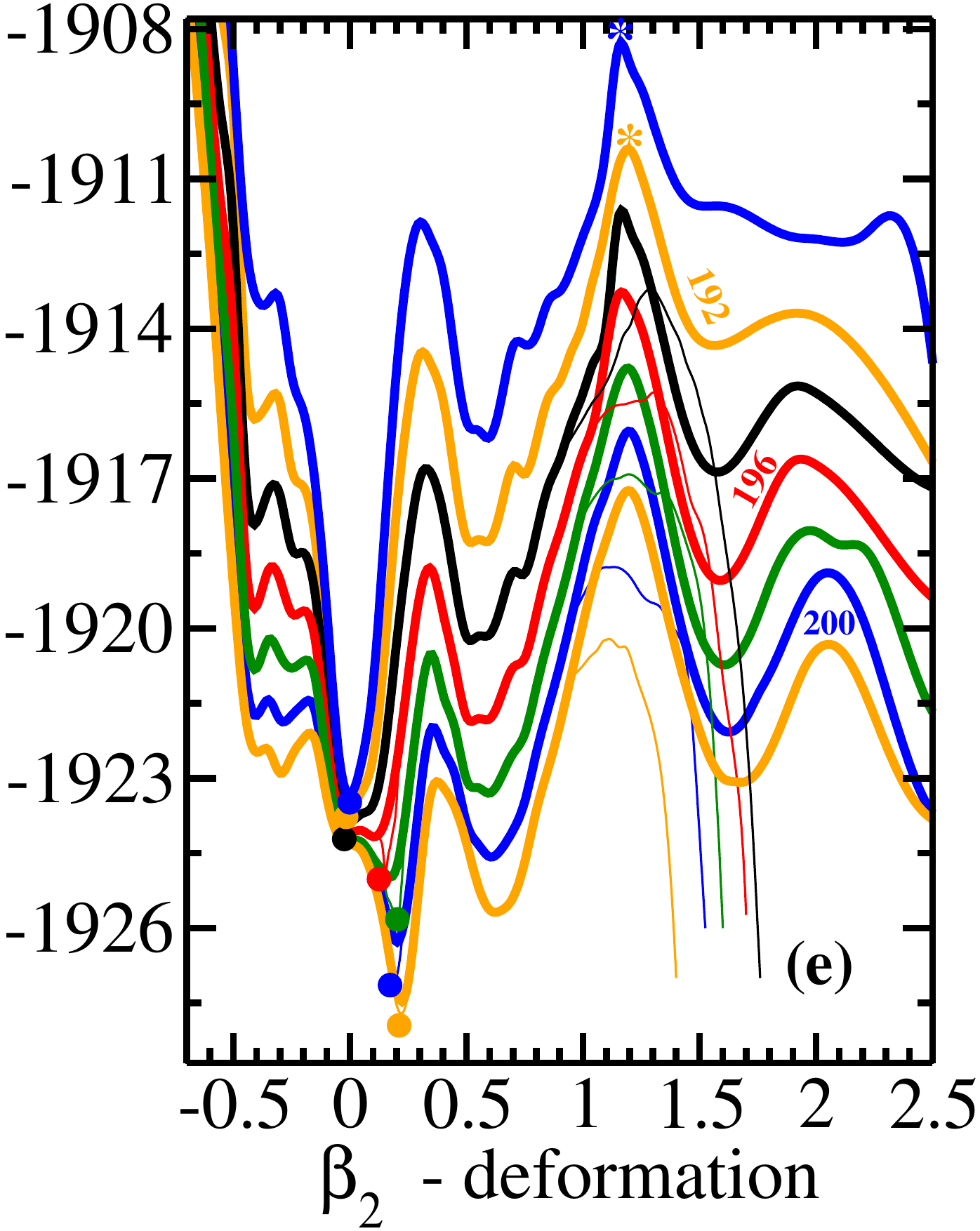}
\includegraphics[angle=0,width=5.9cm]{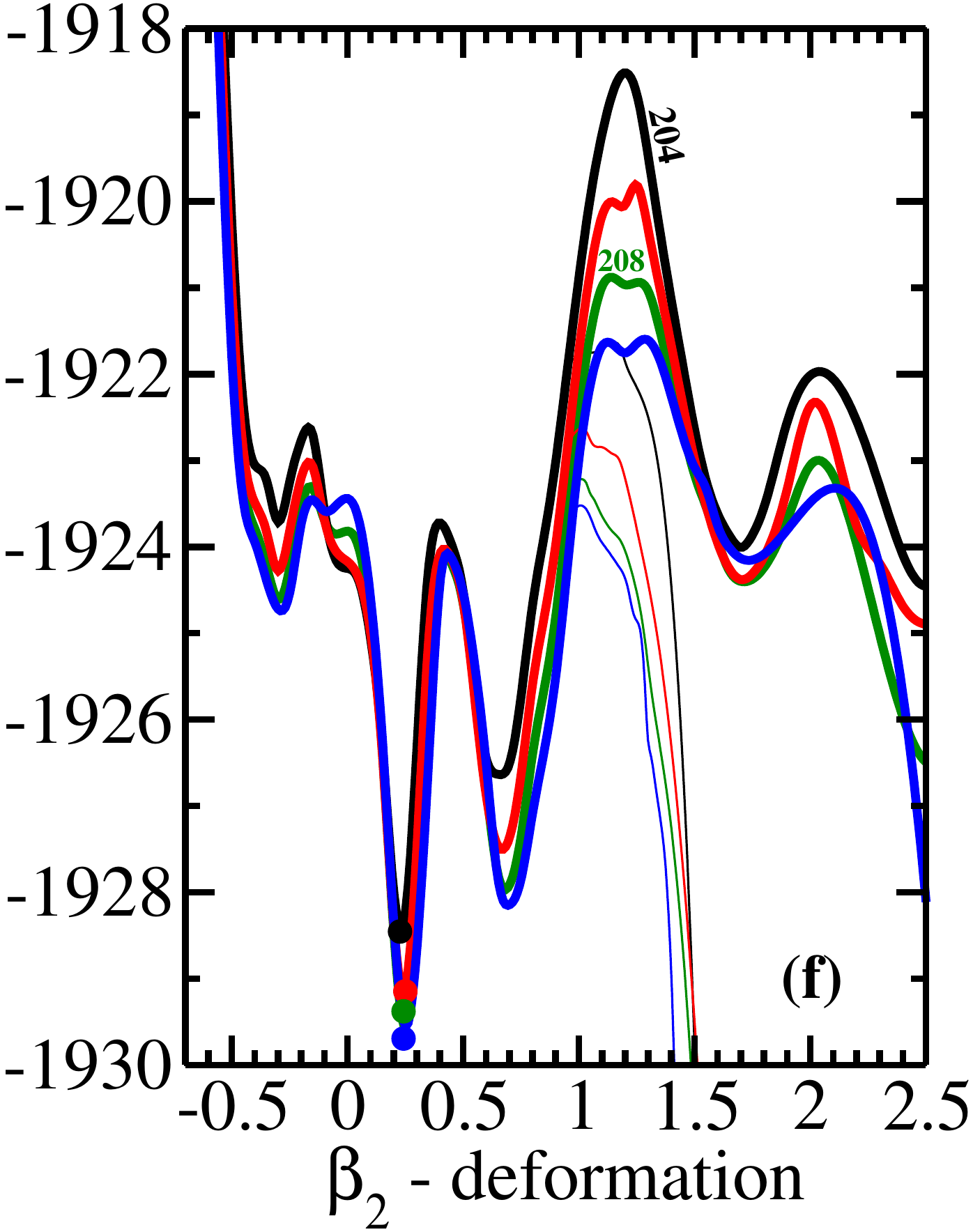}
\caption{The same as Fig.\ \ref{deformation-curve-Ds} but for the Th isotopes.
Note that the deformation range has been extended on horizontal 
axis as compared with Fig.\ \ref{deformation-curve-Ds}.  In order to
save computational time the RA-RHB calculations have been 
carried out only up to $\beta_2=2.0$.
\label{deformation-curve-Th}
}
\end{figure*}

 In the present manuscript, the RHB framework with finite range pairing and its separable limit 
are used for a systematic study of ground state properties of all even-even actinides ($Z=90-102$) 
and superheavy ($Z=104-120$) nuclei from the proton to neutron drip line. It has the proper coupling 
to the continuum at the neutron drip line and, therefore, it allows a correct description of weakly bound 
nuclei close to the neutron drip line.

\begin{figure*}[htb]
\centering
\includegraphics[angle=-90,width=18.0cm]{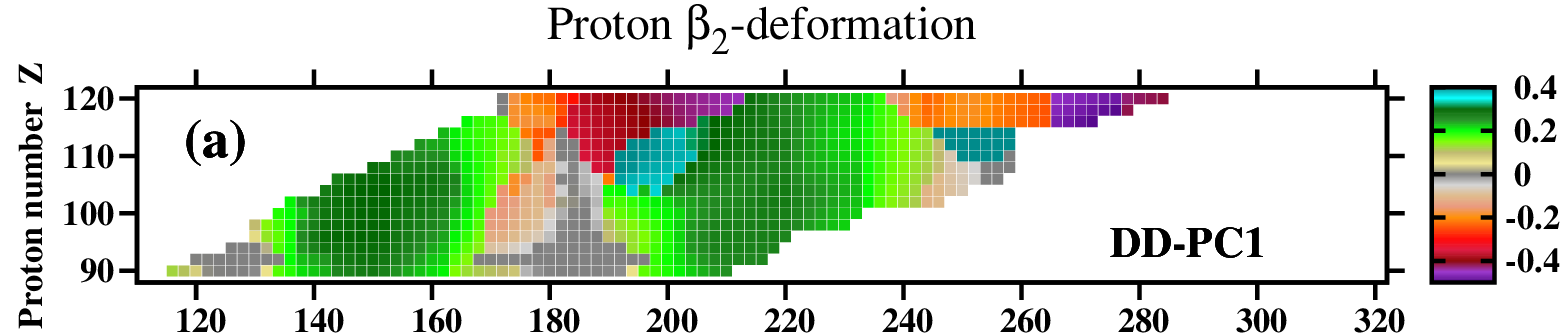}\\
\includegraphics[angle=-90,width=18.0cm]{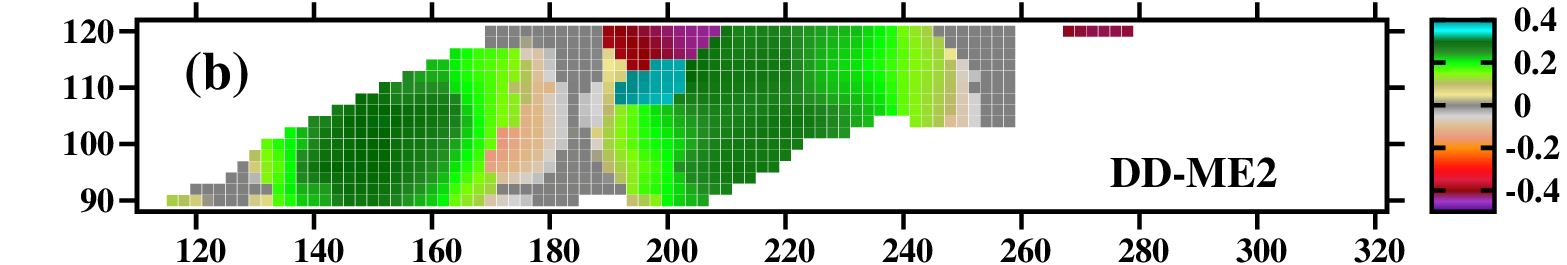}\\
\includegraphics[angle=-90,width=18.0cm]{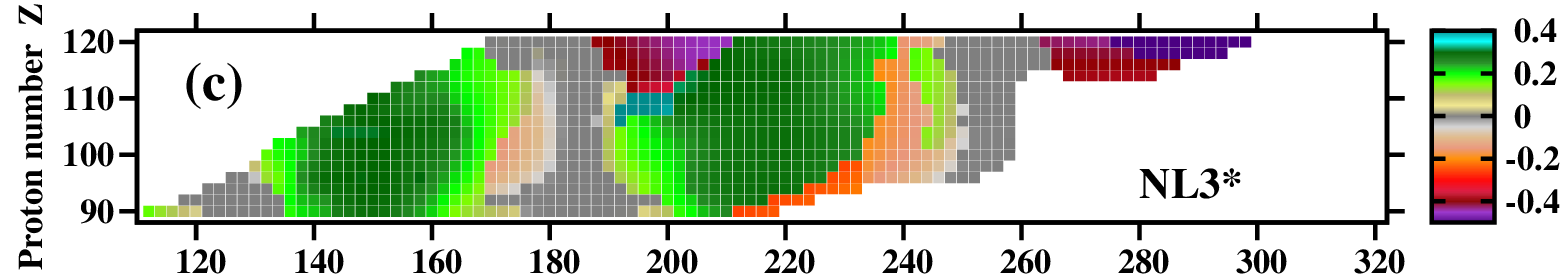}\\
\includegraphics[angle=-90,width=18.0cm]{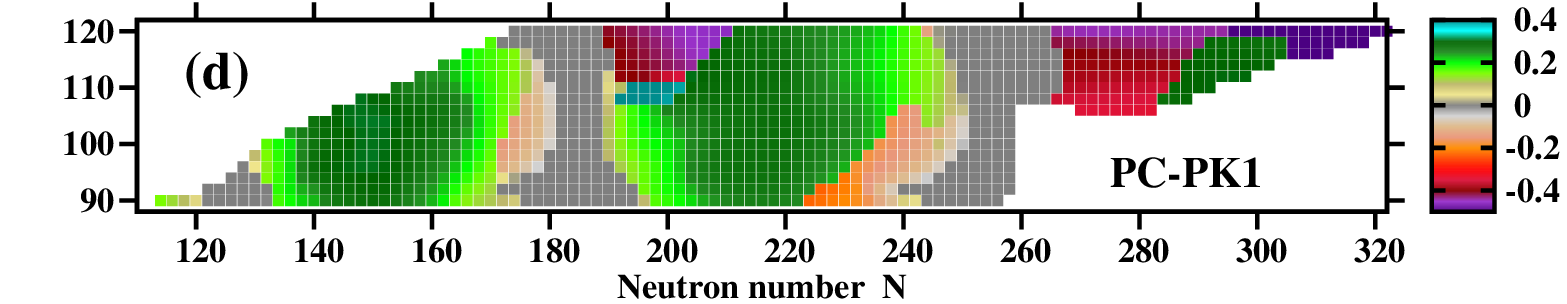}\\
\caption{Proton quadrupole deformations $\beta_2$ obtained in the RS-RHB and
RA-RHB calculations with indicated CEDFs (see Refs.\ \cite{AAR.16,AA.17-oct} for the details 
of the regions of octupole deformation). Note that last bound $Z=120$ nucleus appears
at $N=324$ in the calculations with the CEDF PC-PK1 (see Table \ref{Table-drip-lines} below); 
for simplicity it is not shown in panel (d).
\label{deformations}
}
\end{figure*}

\begin{table}[htb]
\centering
\caption{The nuclei in which extremely superdeformed minimum is the lowest in 
energy in the calculations with CEDF DD-PC1.
The columns 1 and 2 show the proton and neutron numbers of the nuclei.
Third column displays  the energy $E_{N-S}$ [in MeV] by which the ESD 
minimum is lower  than the normal-deformed minimum.  The 
deformations of the ESD minimum $\beta_2^{min}$ and the saddle 
of second fission barrier $\beta_2^{saddle}$,  $\beta_3^{saddle}$ 
are presented in the columns 4 and 5, respectively.  The energies [in MeV] 
of the second fission barrier with respect of  the  ESD minimum, obtained in 
the RS-RHB and RA-RHB  calculations, are shown in the columns 6 and  7.
Note that the values presented in the columns 5-7 are obtained in the 
calculations with $N_F=26$;  this is done in order to have a comparable 
numerical accuracy with the one obtained at normal deformed minimum. 
}
\begin{tabular}{| c| c| c|  c| c| c| c|} 
\hline
       Z      &     N      &    $E_{N-S}$    &   $\beta_2^{min}$ & $\beta_2^{saddle}, \beta_3^{saddle}$ & $E_B^{II}$[MeV] &  $E_B^{II}$ [MeV]  \\  
               &             &      [MeV]                &                           &                                                          &  (RS-RHB)  &  (RA-RHB)  \\  \hline
  1 &  2   &    3          &    4      &   5       &   6       &    7         \\ \hline
 98& 142&    0.603   &  0.88 & 1.21, 0.33  & 7.207  &  2.625    \\ \hline
 98& 144&    0.306   &  0.90 & 1.34, 0.46  & 7.373  &  3.503    \\ \hline
 98& 228&    2.341   &  1.01 & 1.29, 0.34  & 6.622  &  2.906    \\ \hline
 98& 230&    2.083   &  1.01 & 1.30, 0.35  & 5.072  &  3.203    \\ \hline 
                        
100& 146&    0.876   &  0.97 & 1.32, 0.37  & 5.890  &  2.985    \\ \hline
100& 230&    2.431   &  1.01 & 1.30, 0.35  & 5.750  &  3.270    \\ \hline
100& 232&    2.336   &  1.01 & 1.30, 0.37  & 4.060  &  2.739    \\ \hline    
                                              
102& 146&    2.038   &  0.99 & 1.32, 0.34  & 4.813  &  2.749     \\ \hline
102& 148&    0.629   &  0.98 & 1.28, 0.31  & 4.093  &  2.250   \\ \hline
102& 232&    2.591   &  1.02 & 1.29, 0.36  & 3.476  &  2.840    \\ \hline
102& 234&    3.567   &  1.03 & 1.31, 0.31  & 2.674  &  2.304    \\ \hline
                                              
104& 146&    3.435   &  0.99 & 1.33, 0.28  & 6.042  &  2.410    \\ \hline
104& 148&    1.921   &  1.00 & 1.31, 0.27  & 5.579  &  2.204    \\ \hline
104& 150&    0.924   &  1.00 & 1.29, 0.27  & 5.370  &  2.505    \\ \hline         

106& 148&    3.638   &  1.08 & 1.32, 0.26  & 4.112  &  2.340   \\ \hline
106& 150&    2.374   &  1.10 & 1.30, 0.23  & 3.753  &  2.345    \\ \hline
 
\end{tabular}
\label{Table-HD}
\end{table}

\begin{figure*}[htb]
\centering
\includegraphics[width=8.0cm]{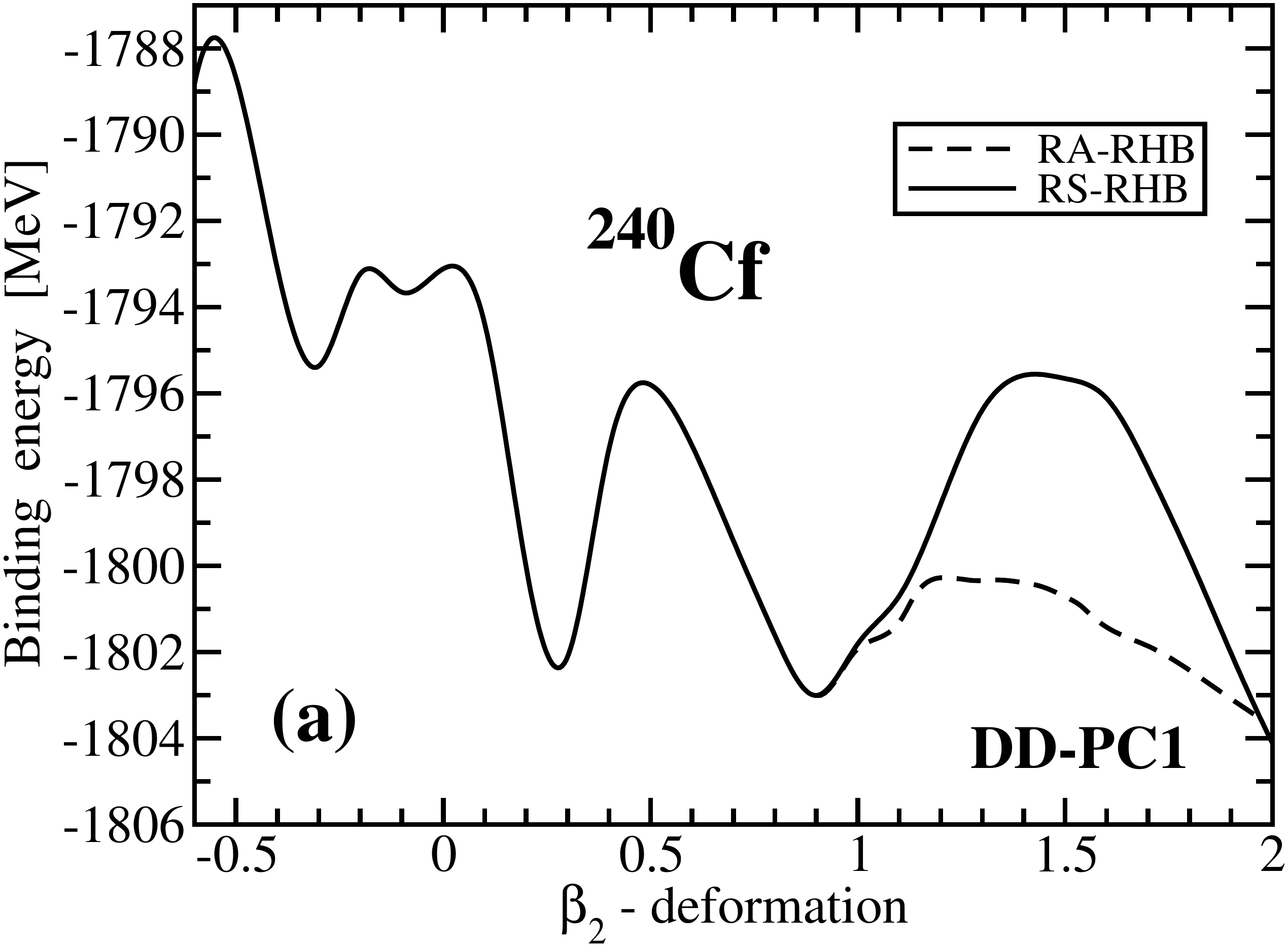} 
\includegraphics[width=8.0cm]{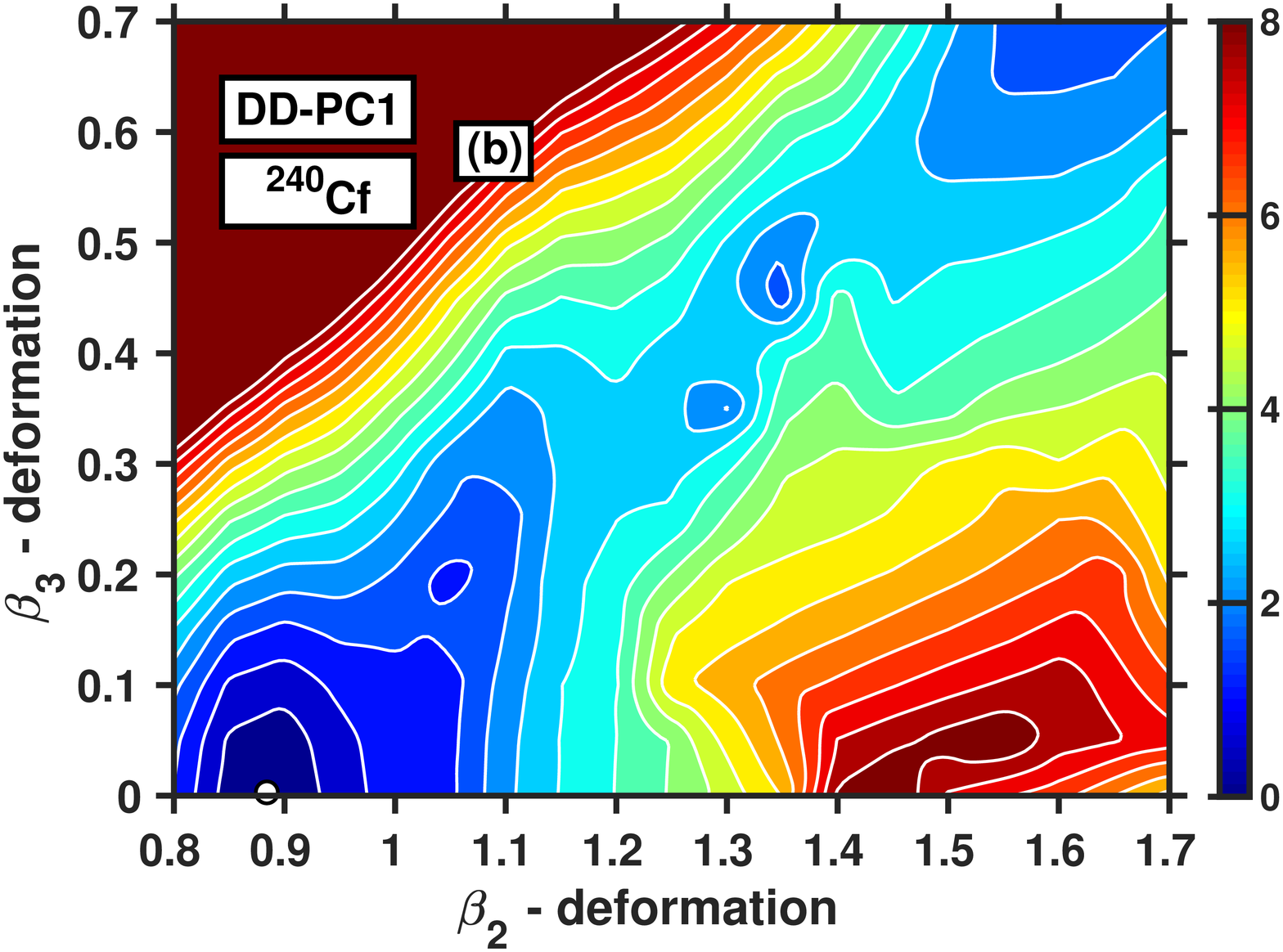}
\caption{(a) Deformation energy curves obtained in  axial RS-RHB calculations 
with the CEDF DD-PC1 for the $^{240}$Cf  nucleus. (b) Potential energy 
surface in the $(\beta_2, \beta_3$) plane obtained in the RA-RHB calculations. 
Extremely superdeformed minimum is indicated by open white circle.
}
\label{HD-minima}
\end{figure*}

\begin{figure*}[htb]
\centering
\includegraphics[angle=-90,width=18.0cm]{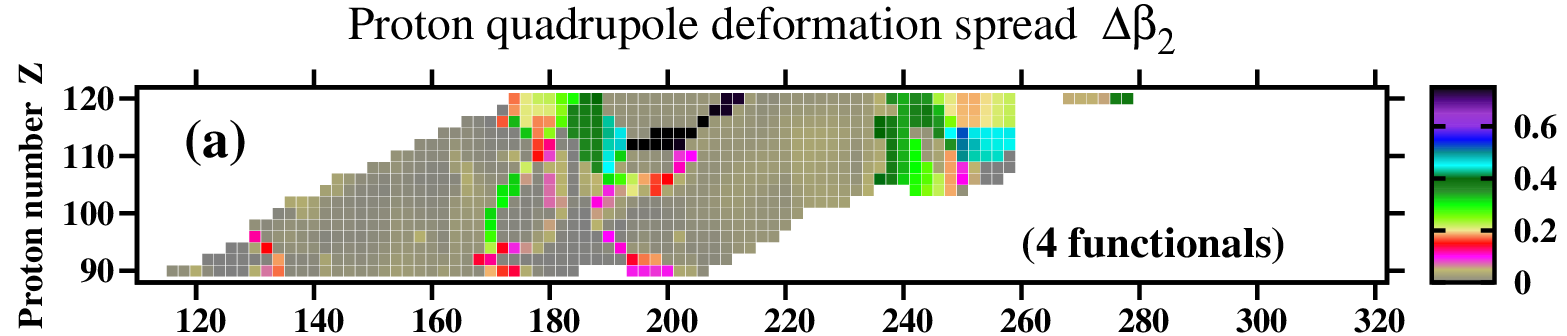}\\
\includegraphics[angle=-90,width=18.0cm]{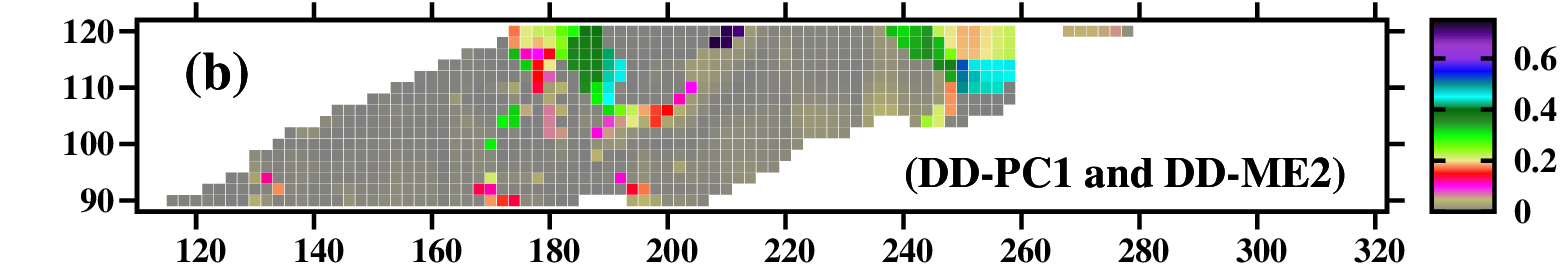}\\
\includegraphics[angle=-90,width=18.0cm]{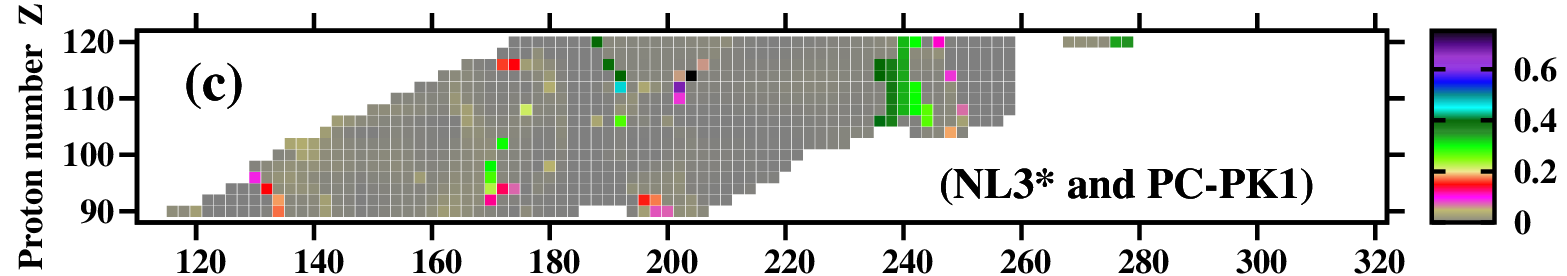}\\
\includegraphics[angle=-90,width=18.0cm]{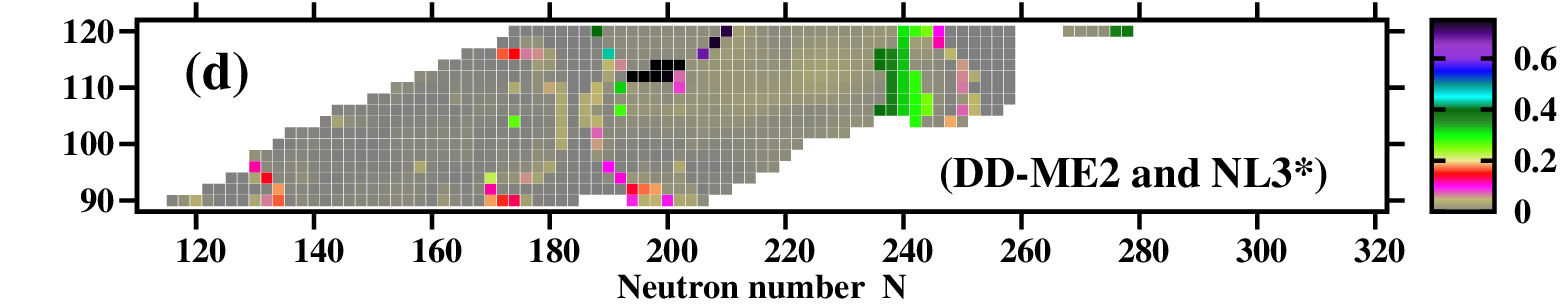}\\
\caption{Proton quadrupole deformation spreads
$\Delta \beta_2(Z,N)$ as a function of proton and neutron number.
$\Delta \beta_2(Z,N)=|\beta_2^{\rm max}(Z,N)-\beta_2^{\rm min}(Z,N)|$,
where $\beta_2^{\rm max}(Z,N)$ and $\beta_2^{\rm min}(Z,N)$ are the
largest and smallest proton quadrupole deformations obtained
with four employed CEDF for  the $(Z,N)$
nucleus.
\label{deformation-spreads}
}
\end{figure*}

The RHB equations for the fermions are given by \cite{CRHB}
\begin{eqnarray}
\begin{pmatrix}
  \hat{h}_D-\lambda  & \hat{\Delta} \\
 -\hat{\Delta}^*& -\hat{h}_D^{\,*} +\lambda
\end{pmatrix}
\begin{pmatrix}
U({\bm r}) \\ V({\bm r})
\end{pmatrix}_k
= E_k
\begin{pmatrix}
U({\bm r}) \\ V({\bm r})
\end{pmatrix}_k,
\end{eqnarray}
Here, $\hat{h}_D$ is the Dirac Hamiltonian for the nucleons with mass
$m$; $\lambda$ is the chemical potential defined by the constraints on
the average particle number for protons and neutrons;
$U_k ({\bm r})$ and $V_k ({\bm r})$ are quasiparticle Dirac
spinors~\cite{Kucharek1991_ZPA339-23,Ring1996_PPNP37-193,CRHB} and
$E_k$ denotes the quasiparticle energies. The Dirac Hamiltonian
\begin{equation}
\label{Eq:Dirac0}
\hat{h}_D = \boldsymbol{\alpha}(\boldsymbol{p}-\boldsymbol{V}) + V_0 + \beta (m+S).
\end{equation}
contains an attractive scalar potential
\begin{eqnarray}
S(\bm r)=g_\sigma\sigma(\bm r),
\label{Spot}
\end{eqnarray}
a repulsive vector potential
\begin{eqnarray}
V_0(\bm r)~=~g_\omega\omega_0(\bm r)+g_\rho\tau_3\rho_0(\bm r)+e A_0(\bm r),
\label{Vpot}
\end{eqnarray}
and a magnetic potential
\begin{eqnarray}
\bm V(\bm r)~=~g_\omega\bm\omega(\bm r)
+g_\rho\tau_3\bm\rho(\bm r)+e\bm A(\bm r).
\label{Vmag}
\end{eqnarray}
The last term breaks time-reversal symmetry and induces currents.
For example, time-reversal symmetry is broken when the time-reversed
orbitals are not occupied pairwise; this takes place in odd-mass
nuclei \cite{AA.10}. 
However,  nuclear magnetism \cite{KR.89}, i.e. currents and time-odd mean fields,
plays no role in the studies of ground states and fission barriers in
even-even nuclei. Thus, magnetic potential is neglected in the present
RHB calculations.

  In order to avoid the uncertainties connected with the definition of
the size of the pairing window\ \cite{KALR.10}, we use the separable form 
of the finite range Gogny pairing interaction introduced by Tian et al 
\cite{TMR.09}. Its matrix elements in $r$-space have the form
\begin{eqnarray}
\label{Eq:TMR}
V({\bm r}_1,{\bm r}_2,{\bm r}_1',{\bm r}_2') &=& \nonumber \\
= - G \delta({\bm R}-&\bm{R'}&)P(r) P(r') \frac{1}{2}(1-P^{\sigma})
\label{TMR}
\end{eqnarray}
with ${\bm R}=({\bm r}_1+{\bm r}_2)/2$ and ${\bm r}={\bm r}_1-{\bm r}_2$
being the center of mass and relative coordinates. The form factor
$P(r)$ is of Gaussian shape 
\begin{eqnarray}
P(r)=\frac{1}{(4 \pi a^2)^{3/2}}e^{-r^2/4a^2}
\end{eqnarray}
The two parameters $G=728$ MeV$\,$fm$^3$ and $a=0.644$ 
fm of this interaction are the same for protons and neutrons and 
have been derived in Ref.\ \cite{TMR.09} by a mapping of the 
$^1$S$_0$ pairing gap of infinite nuclear matter to that of the 
Gogny force D1S~\cite{D1S}.  This pairing provides a reasonable 
description of pairing properties in the actinides (see Refs.\ 
\cite{AO.13,AARR.14,DABRS.15}) and has been used in our 
previous studies of different phenomena in actinides, super- and 
hyperheavy nuclei in Refs.\  
\cite{AARR.14,AANR.15,AAR.16,AA.17-oct,AAG.18,AATG.19}.

 The truncation of the basis is performed in such a way that all states
belonging to the major shells up to $N_F = 20$ fermionic shells for
the Dirac spinors and up to $N_B = 20$ bosonic shells for the meson
fields are taken into account. Note that the latter applies only to the 
NL3* and DD-ME2 functionals which contain meson exchange. 
As follows from investigation of Refs.\ 
\cite{AARR.14,AAR.12} this truncation of basis provides sufficient
numerical accuracy.

The calculations are performed in the following way:
\begin{itemize}

\item
 Reflection-symmetric constrained axial RHB calculations (further RS-RHB)
are performed  for each nucleus and the potential energy curve is defined in a large 
deformation range from $\beta_2=-1.0$ up to $\beta_2=1.6$ in step
of $\Delta \beta_2=0.05$ by means of the constraint on the quadrupole 
moment $q_{20}$.  The calculations are performed by successive 
diagonalizations using the method
of quadratic constraints \cite{RS.80}. The parallel version of computer
code allows simultaneous
calculations for a significant number of nuclei and deformation points in each
nucleus. For each nucleus, we minimize
\begin{equation}
E_{quad} = E_{RHB} + C_{20} (\langle\hat{Q}_{20}\rangle-q_{20})^2
\label{constr-quad}
\end{equation}
where $E_{RHB}$ 
is the total energy  and
$\langle\hat{Q}_{20}\rangle$ denotes the expectation value of
the mass quadrupole operator,
\begin{equation}
\hat{Q}_{20}=2z^2-x^2-y^2
\end{equation}
$q_{20}$ is the constrained value of the multipole moment, and
$C_{20}$ the corresponding stiffness constant~\cite{RS.80}.
In order to provide the convergence to the exact value
of the desired multipole moment we use the method suggested in
Ref.~\cite{BFH.05}. Here the quantity $q_{20}$ is replaced by the
parameter $q_{20}^{eff}$, which is automatically modified during
the iteration in such a way that we obtain
$\langle\hat{Q}_{20}\rangle = q_{20}$ for the converged solution.
This method works well in our constrained calculations.

\item
 In addition, reflection-asymmetric (octupole deformed) constrained axial RHB
calculations (further RA-RHB)
are performed  in discussed below cases using parallel version of
the code developed in Ref.\ \cite{AAR.16}.  In these calculations, the constraint 
\begin{equation}
E_{quad} + C_{30} (\langle\hat{Q}_{30}\rangle-q_{30})^2
\end{equation}
is employed in addition to the constraint on quadrupole moment (see Eq.\ 
(\ref{constr-quad})).  Here $\langle\hat{Q}_{30}\rangle$ denotes the expectation 
value of the mass octupole operator
\begin{equation}
\hat{Q}_{30}=z(2z^2-3x^2-3y^2).
\end{equation}
Note that we also fix the (average) center of mass of the nucleus at the origin
with the constraint 
\begin{equation} 
\langle \hat{Q}_{10}\rangle =0
\end{equation}
on the center-of-mass operator ${\hat Q}_{10}$ in order to avoid a spurious motion
of the center of mass. In the present paper, reflection asymmetric RHB calculations have been performed for the 
ground states of the nuclei not covered in previous systematic studies of octupole deformation 
in CDFT (see Refs.\ \cite{AAR.16,AA.17-oct}). We have not found any additional (as compared with 
those given in Refs.\ \cite{AAR.16,AA.17-oct}) nuclei which possess octupole deformation in the 
ground state. So full information on the octupole deformation of the ground states can be found in 
these references. The information (which follows from Refs.\ \cite{AAR.16,AA.17-oct})
about the gain in binding energy due to octupole deformation at  the ground state and its impact on 
ground state quadrupole deformation and fission barrier  heights  is fully taken into account in the 
present  paper.  In addition, RA-RHB calculations have been performed in some nuclei in order to 
define the heights of outer fission barriers (see the discussion below for more details).

\end{itemize}

\begin{figure*}[htb]
\centering
\includegraphics[angle=-90,width=18.0cm]{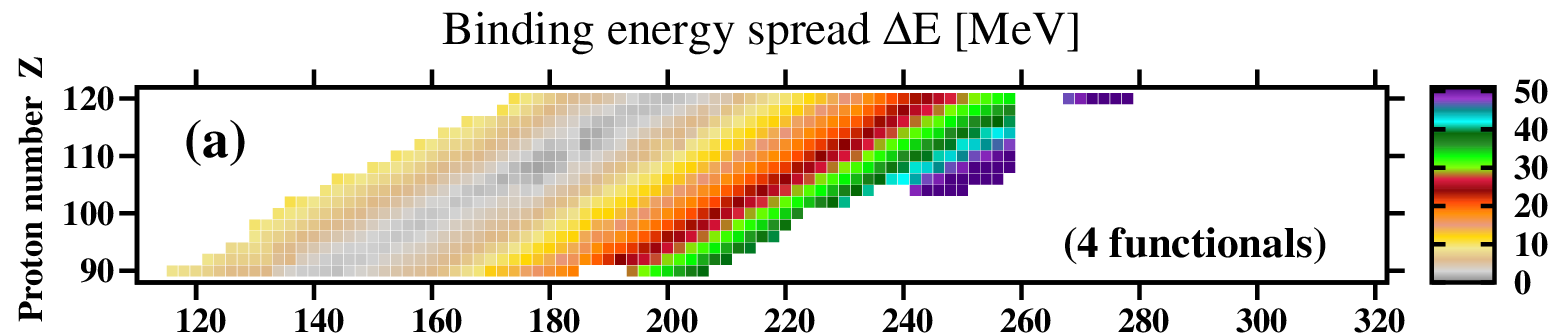}
\includegraphics[angle=-90,width=18.0cm]{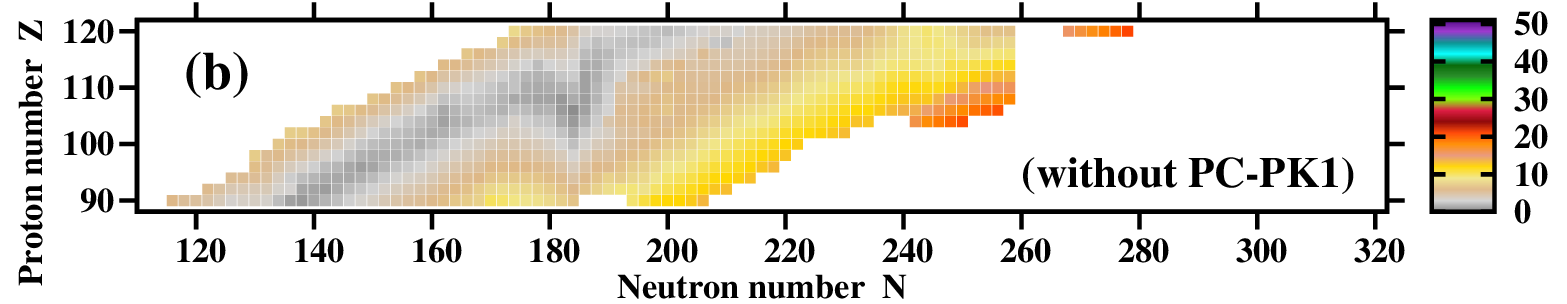}
\caption{The binding energy spreads $\Delta E(Z,N)$ as a function of proton and neutron number.
$\Delta E(Z,N)=|E_{\rm max}(Z,N)-E_{\rm min}(Z,N)|$,
where $E_{\rm max}(Z,N)$ and $E_{\rm min}(Z,N)$ are the
largest and  smallest  binding energies obtained
with employed set of CEDFs for  the $(Z,N)$ nucleus. Upper panel shows the binding energy
spreads for four employed functionals, while the bottom one the spreads for a set of
functionals in which PC-PK1 is excluded.
\label{energy-spreads}
}
\end{figure*}

  The charge quadrupole and octupole moments are defined as
\begin{eqnarray}
Q_{20} &=& \int d^3r \rho({\bm r})\,(2z^2-r^2_\perp), \\
Q_{30} &=& \int d^3r \rho({\bm r})\,z(2z^2-3r^2_\perp)
\end{eqnarray}
with $r^2_\perp=x^2+y^2$. In principle these values can be directly
compared with experimental data. However, it is more convenient to
transform these quantities into dimensionless deformation
parameters $\beta_2$ and $\beta_3$ using the relations
\begin{eqnarray}
Q_{20}&=&\sqrt{\frac{16\pi}{5}} \frac{3}{4\pi} Z R_0^2 \beta_2,
\label{beta2_def} \\
Q_{30}&=&\sqrt{\frac{16\pi}{7}}\frac{3}{4\pi} Z R_0^3 \beta_3
\label{beta4_def}
\end{eqnarray}
where $R_0=1.2 A^{1/3}$. These deformation parameters are more
frequently used in experimental works than quadrupole and octupole
moments. In addition, the potential energy surfaces (PES) are plotted in
this manuscript in the ($\beta_2,\beta_3$) deformation plane.

   Because of different patterns of deformation energy curves (see Figs.\ \ref{deformation-curve-Ds}
and  \ref{deformation-curve-Th}), a special care  is used when assigning a specific minimum 
to the ground state. A basic rule in this process is the assumption that local minimum surrounded
by the barrier, the height of which
is less than 2 MeV, is considered as extremely unstable\footnote{This low fission barrier of
2 MeV or less would translate into a high penetration probability for spontaneous fission 
so that such minima are expected to be  extremely unstable. In addition, the inclusion of octupole 
deformation (or triaxial deformation in some nuclei \cite{AAR.12})
in the case of superdeformed minima surrounded by such low fission barriers
could either completely eliminate or substantially reduce them (see Refs.\ \cite{BBM.04,AAR.12,AATG.19}).}. 
The procedure of the selection of the ground state is discussed below.  The 
situation shown in Fig.\ \ref{def-curve-sel}a is the simplest one: single-humped (inner) fission 
barrier acts against the fission into two fragments and when $\rm B_{in} > 2$ MeV the assignment 
of the normal-deformed prolate minimum to the ground state is straightforward.  It
changes  if the height of this fission barrier decreases and becomes smaller than 2 MeV
(see Fig.\ \ref{def-curve-sel}b). Then highly-deformed oblate minimum B becomes a ground 
state; it has larger and broader fission barrier as compared with minimum A. Thus, it is expected
that this ground state will live significantly longer than the state associated with minimum A.

\begin{figure*}[htb]
\centering
\includegraphics[angle=-90,width=18.0cm]{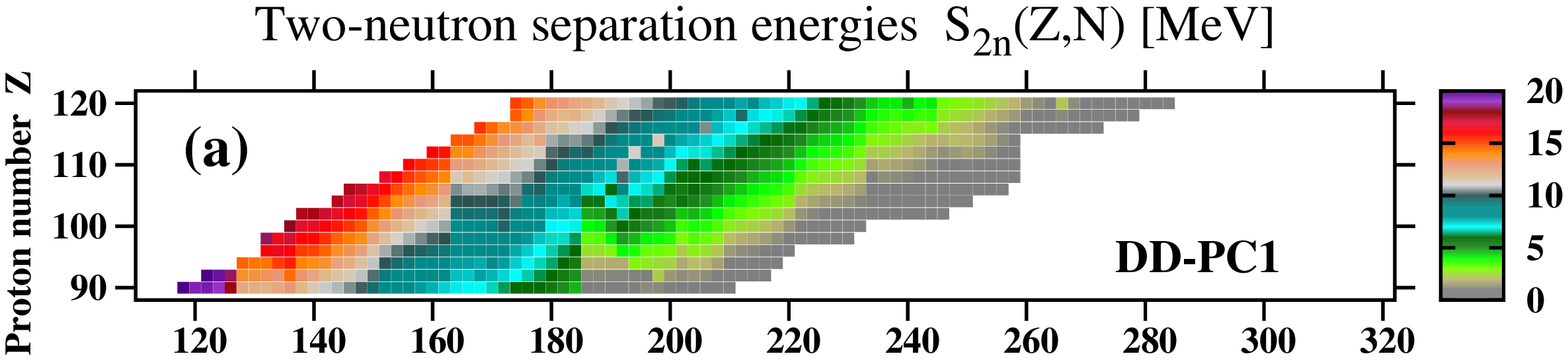}\\
\includegraphics[angle=-90,width=18.0cm]{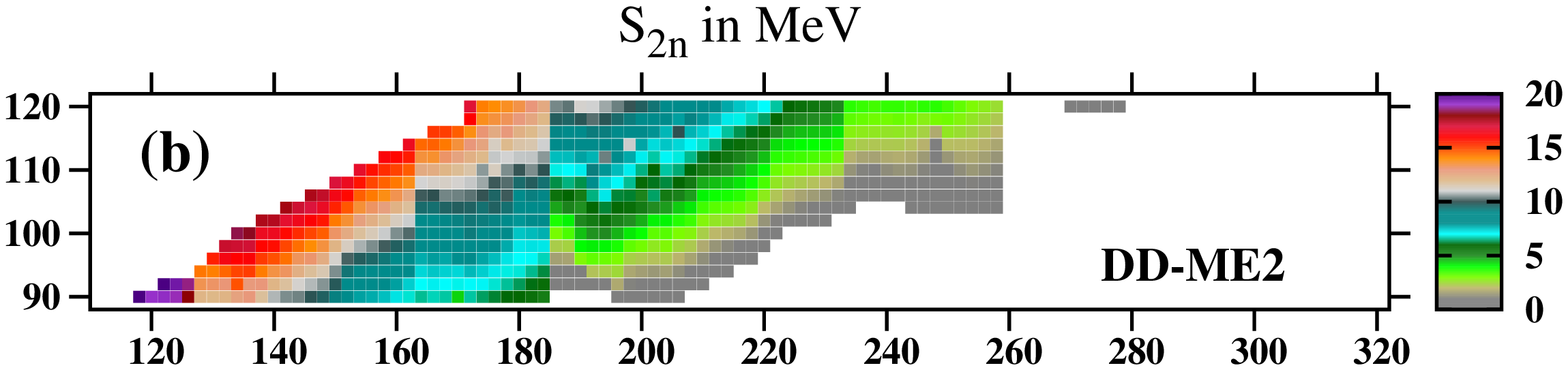}\\
\includegraphics[angle=-90,width=18.0cm]{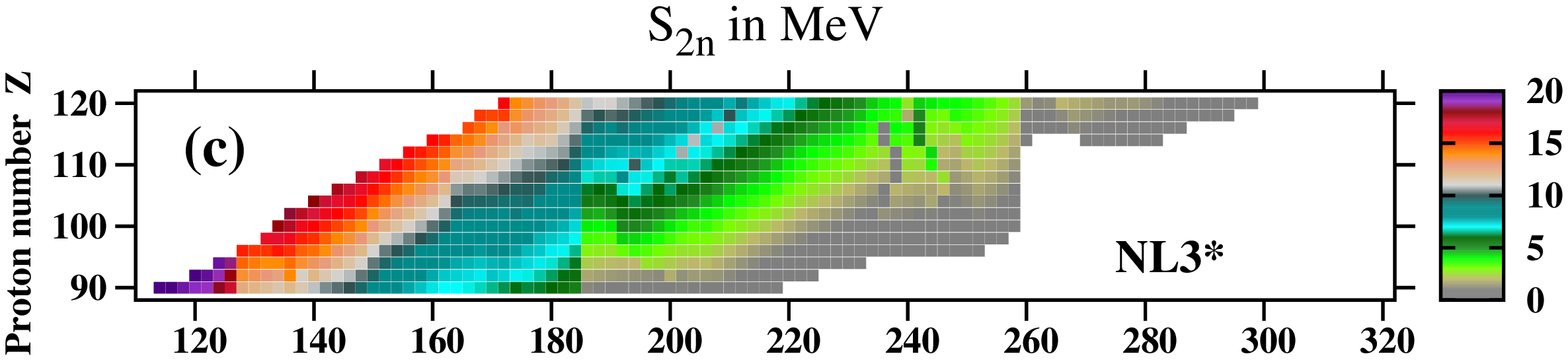}\\
\includegraphics[angle=-90,width=18.0cm]{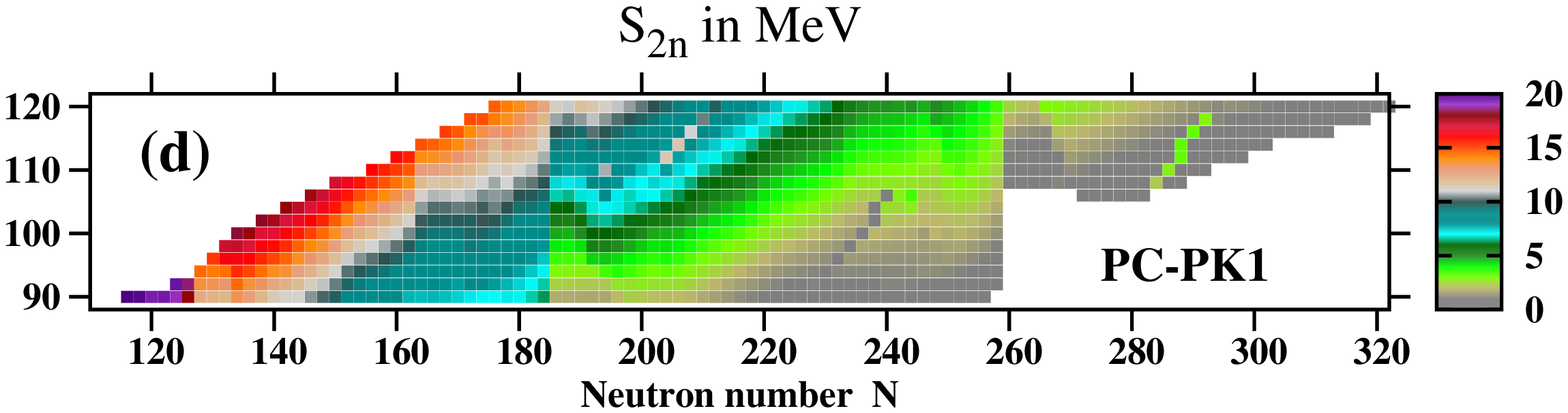}\\
\caption{Two-neutron separation energies $S_{2n}(Z,N)$ obtained in the RHB calculations 
with indicated CEDFs.
\label{2n-separation-energies}
}
\end{figure*}

 More complex situation involving two humped  fission barrier is shown in Fig.\ \ref{def-curve-sel}(c). 
If $B_{out-iso} < 2$ MeV, then the minimum B corresponding to fission isomer is considered extremely  
unstable and the minimum A is associated with the ground state. Note that in some cases the 
minimum B can be lower in energy than minimum A. If that is the case and if $B_{out-iso} > 2$ MeV 
then the minimum B is considered as the ground state.  Note that the heights of outer fission barriers 
are frequently lowered when octupole deformation is included in the calculations
\cite{BBM.04,SBDN.09,MSI.09,LZZ.12,AAR.12,PNLV.12}.  Thus, if $B_{out-iso} > 2$ MeV 
in RS-RHB calculations, we perform RA-RHB calculations in the 
region of the $(\beta_2, \beta_3)$ plane  covering the minimum B and the saddle of outer fission barrier 
on the grid with the steps of  $\Delta \beta_2 = \Delta \beta_3 =0.05$. This allows to establish whether 
minimum B could be considered as relatively stable or unstable. Similar calculations are also performed 
in the cases when $\rm B_{in} < B_{out}$ in the RS-RHB calculations. This is because we consider only 
the height of the primary (highest) fission barrier  (PFB) in Sec.\ \ref{sect-fission} in the case of 
doubly-humped structure of the barrier to minimize the computational cost. Note that the calculations 
leading to the definition of the fission path and the saddle point in the RA-RHB code are by roughly two 
orders of magnitude more time consuming than those in the RS-RHB code.

 The procedure outlined above takes into account potential stability of the nuclei in respective energy
minima with respect of fission  and it is especially important in superheavy nuclei some local minima of 
which are characterized by small fission barriers (see Fig.\ \ref{deformation-curve-Ds}). Note that after 
defining the minimum corresponding to the ground state, the RS-RHB and RA-RHB calculations without 
constraint  are performed  in it and precise binding  energy and equilibrium of the ground  state is 
determined.  In addition, the height(s) of the fission barrier(s) is(are) defined.
 
 In the calculations with the PC-PK1 and NL3* functionals  there are two small islands of the nuclei 
located in the $Z\approx 114-118, N \approx 238-240$ and $Z\approx 106-110, N\approx 190-194$ regions 
in which calculated deformation energy curves reveal several local minima (somewhat similar to the
deformation energy curves shown at the bottom  of Fig.\ \ref{deformation-curve-Ds}b). However, all these 
minima are surrounded by very low fission barriers with the heights smaller than 2.0 MeV. Moreover, 
many of these minima have even lower heights of respective fission barriers (on the level of 
1.0 MeV or smaller).  These nuclei are expected to be unstable against fission and in principle it does 
not matter which  of the calculated fission barriers is used.  For these nuclei, we select the ground state 
guided by the flow of the $\beta$-decays in the r-process: 
the selected local minimum (and thus the corresponding ground state deformation and 
fission barrier height) in the $(Z,N)$ nucleus has the deformation closest to the one of 
the well established ground state in the $(Z-2, N+2)$ nucleus.



\begin{figure*}[htb]
\centering
\includegraphics[angle=-90,width=18.0cm]{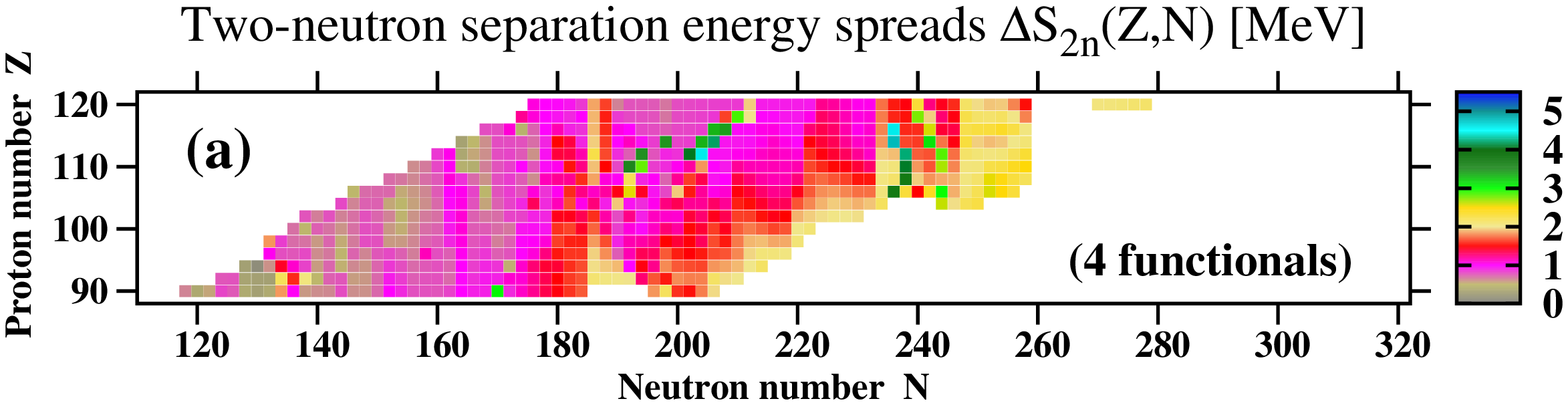}
\includegraphics[angle=-90,width=18.0cm]{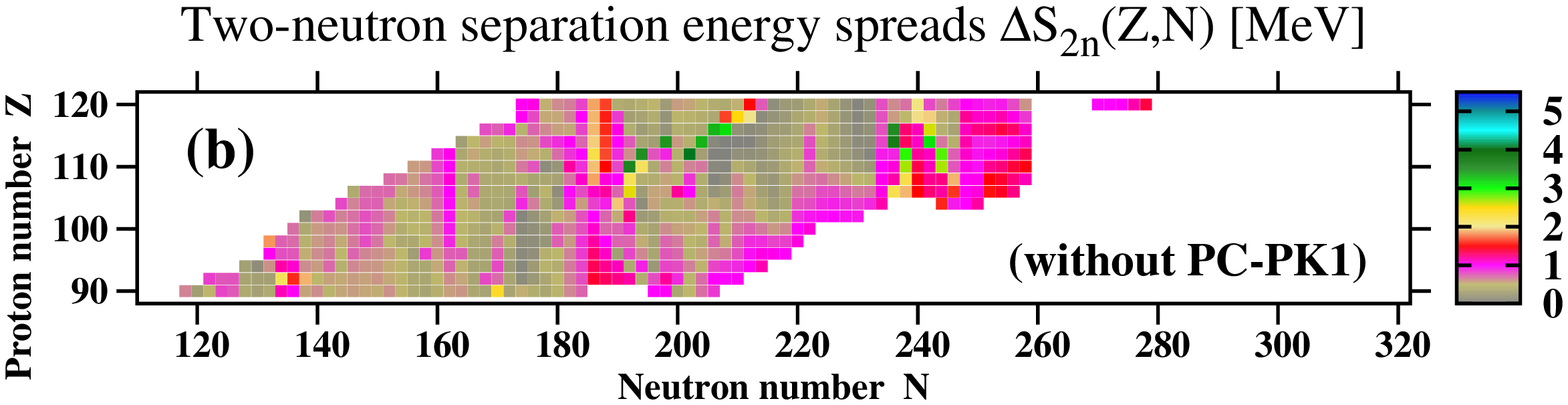}
\caption{The $S_{2n}$ spreads $\Delta S_{2n}(Z,N)$ as a function of proton and neutron 
number. $\Delta S_{2n}(Z,N)=|S_{2n}^{\rm max}(Z,N)-S_{2n}^{\rm min}(Z,N)|$,
where $S_{2n}^{\rm max}(Z,N)$ and $S_{2n}^{\rm min}(Z,N)$ are the
largest and smallest  $S_{2n}$ values obtained with four employed  CEDFs for  
the $(Z,N)$ nucleus.
}
\label{S_2n_spread}
\end{figure*}

   So far all existing global calculations of the fission barriers for the r-process simulations have been 
performed in non-relativistic models. These include the calculations within the FRDM \cite{MS.99,GPR.18},
the TF and ETFSI-Q approaches \cite{MPRT.98,PLMPRT.12}, the HFB approaches with different Skyrme 
functionals \cite{ELLMR.12,Reinh.18,AG.20}, Gogny  D1M* \cite{RHR.20} and BCPM \cite{GPR.18} EDFs.
Because of their global character,  all these calculations are restricted to axial symmetry.  We also 
assume axial symmetry in our calculations because triaxial RMF+BCS and RHB calculations (see Refs.\ 
\cite{AAR.10,AAR.12,AARR.17,AATG.19,SALM.19}) are too time consuming to be performed on a global 
scale.  Note also that dynamical correlations are not taken into account in our calculations of
fission barriers because of the reasons discussed in Appendix A.

\section{Ground state properties}
\label{sect-ground}

    The distributions of calculated proton deformations $\beta_2$ in the
$(Z,N)$ plane obtained with four employed CEDFs are shown in Fig.\  
\ref{deformations}. The width of the gray region  (the gray color corresponds to spherical 
and near-spherical shapes) along a specific magic number corresponding to a 
shell closure indicates the impact of this shell closure on the structure of 
neighboring nuclei. Note that proton and neutron shell gaps act simultaneously
in the vicinity of doubly magic spherical nuclei. Thus, the effect of a single gap 
is more quantifiable away from these nuclei.  One can see that neutron $N=184$
and $N=258$\footnote{ Note that appreciable $N=258$ spherical shell gap 
appears also in the calculations of some superheavy nuclei
with other CEDFs such as G1, G2 
\cite{SPSCV.04}, NL3 \cite{BNR.01,SPSCV.04,ZMZGT.05}, 
NL-Z2 \cite{BNR.01},  NL1, NLSH, TM1, TW99, DD-ME1, PK1, and 
PK1R \cite{ZMZGT.05}. However, these calculations are restricted to spherical 
shapes and thus it is not clear how large is the impact of this gap on ground state
deformations in the region near the $N=258$ line. 
} 
spherical shell gaps have a pronounced impact on calculated deformations
while the impact of the proton $Z=120$ spherical shell closure is limited  to the $N\sim 170-184$ 
nuclei (see Refs.\ \cite{AARR.15,AANR.15} for examples of  their size dependence 
on the functional). In addition, 
as illustrated in Ref.\ \cite{SALM.19} on the example of the PC-PK1 functional,  the inclusion
of the correlations beyond mean field washes out the impact of the $Z=120$
shell closure and leads to oblate deformed ground states in the majority of the
$Z=120$ nuclei with $N=172-186$. The predictions of the DD-PC1 functional differs substantially
from other CEDFs: the impact of above mentioned shell closures are substantially
reduced in it and as a consequence the regions with
$Z\sim 120, N\sim 184$ and $Z\sim 120, N\sim 258$ are dominated by oblate
ground states contrary to spherical states in other functionals. Note that this 
functional provides the best global description of experimental binding energies 
(see Ref.\  \cite{AARR.14}). This, however, does not guarantee that it will outperform
other functionals  in the description of
physical observables of interest in the region of superheavy nuclei
(see Table I in Ref.\ \cite{AANR.15}).

   The calculations reveal a number of nuclei scattered across the part of nuclear  chart 
under study which have extremely superdeformed (ESD) minimum with $\beta_2 \sim 1.0$ located 
at lower energy than normal-deformed minimum (see Fig.\  \ref{HD-minima}a and 
Table\ \ref{Table-HD}).  In these nuclei the ESD minimum is surrounded  by outer fission barrier the 
height of which exceeds 2.0 MeV  (being typically in the range of $2.0-3.0$ MeV) in the RA-RHB 
calculations (see Fig.\ \ref{HD-minima}b and Table\ \ref{Table-HD}).  Although the fission barrier is 
low, some of these ESD minima could be potentially stabilized against fission  for physically sufficient 
time because of  broad fission barrier in the $(\beta_2, \beta_3)$  plane. If that would be a case, they 
would represent the ground states. However, they are not included into Fig.\ \ref{deformations} because 
of the following reasons. First, there are significant theoretical uncertainties in the predictions of fission 
barriers (see present paper and Refs.\  \cite{AAR.12,AARR.17})  as  well as in relative energies of 
the minima with different deformations \cite{KALR.10}.  Second, only few nuclei in the 
$Z\approx 100, N\approx 230$ region  could be potentially extremely supedeformed 
in the ground states (see Table  \ref{Table-HD}). However, the flow of matter in the 
r-process between two nuclei with drastically different deformations of the ground states will 
be most likely significantly suppressed because of  considerable differences in the  wave 
functions of these ground states.  Thus, it will proceed mostly along the dominant deformation
of the ground states in the region, namely, normal deformation, even if such states are excited
in energy in a few nuclei. Third,  the majority of the nuclei in Table \ref{Table-HD} are neutron 
poor $Z\approx 102$ nuclei which do not play a role in the r-process.  There are  experimental 
data on the $^{240,242}$Cf, $^{246}$Fm and $^{254}$Rf nuclei  but only for their ground
states \cite{Eval-data}. At present, these data do not allow to define the deformations of the ground 
states. However, since it has been obtained in the reactions (such as 
$\alpha$-decay,  $\beta$-decay, electron  capture and the reactions on spherical Pb
isotopes) which do not favor significant shape changes, these ground states are most
likely normal-deformed.  More detailed and focused experimental studies are needed in order to see 
whether ESD states exist in such nuclei.

\begin{table}[ht]
\caption{Selected properties of symmetric nuclear matter at saturation: 
the incompressibility $K_0$, the symmetry energy $J$ and its slope $L_0$.
Top four lines show the values for indicated CEDFs, while bottom two lines show two sets 
(SET2a and SET2b) of the constraints on the experimental/empirical ranges for the 
quantities of interest defined in Ref.\ \cite{RMF-nm}. The CEDF values which are located 
outside  the limits of the SET2b constraint set are shown in bold. }
\label{tab-nuclear-matter}
\begin{center}
\begin{tabular}{|c|c|c|c|}\hline
CEDF                               &   $K_0$ [MeV]    & $J$ [MeV]     & $L_0$ [MeV]      \\ \hline
      1                                 &       2                   &       3             &      4                   \\ \hline
NL3* \cite{NL3*}               &    258                  & {\bf 38.68}     & {\bf 122.6}          \\
DD-ME2 \cite{DD-ME2}    &    251                 & 32.40            &  49.4                   \\
DD-PC1 \cite{DD-PC1}     &    230                 & 33.00            &  68.4                   \\
PC-PK1 \cite{PC-PK1}      &    238                 & {\bf 35.6}      &  {\bf 113}             \\ 
 SET2a                              &  190-270            &  25-35          &   25-115              \\ 
 SET2b                              &  190-270            &  30-35          &   30-80                \\ \hline
\end{tabular}
\end{center}
\label{Table-mass-SNMP}
\end{table}

   The spreads of theoretical predictions in quadrupole deformations $\beta_2$ obtained with four employed 
functionals are summarized in Fig.\ \ref{deformation-spreads}(a). The largest spread of $\Delta \beta_2 \approx 0.7$ 
is visible along the line of $N/Z \approx 1.81$ which starts at $Z=104$. This corresponds to the boundary of the 
transition from oblate to prolate shapes the exact position of which in the $(Z,N)$ plane is functional dependent
(see Fig.\ \ref{deformations}).
It is defined by the underlying single-particle structure at prolate and oblate shapes as well as to a degree by the 
heights of outer fission barriers (see the rules for the definition of the ground states described in Sec.\ \ref{sect-theory}).
Second region of the largest spreads in $\Delta \beta_2$
is located along the $N\approx 184$ line starting from $Z\approx 100$ and 
extending up to $Z=120$. Third region is located along the $Z=120$ line from proton-drip line up to  $N\approx 188$. 
These two regions of large spreads in calculated quadrupole deformation emerge  from the differences in the predictions 
of ground state deformations (see Fig.\ \ref{deformations}) which in turn can be traced back to the sizes of the $Z=120$ and 
$N=184$  spherical shell closures and the densities of the single-particle states in their vicinities (see Ref.\ \cite{AANR.15}).  
The  last region of large theoretical uncertainties is located between $N \approx 236$ and $N=258$. In the region around 
$N\approx 236$ these theoretical uncertainties are mostly due to the uncertainties in the predictions of the boundary
of the transition from prolate to oblate shapes.  For higher $N$ 
values, large $\Delta \beta_2$ values emerge from the transition from prolate or oblate shapes to spherical ones and 
to a large degree are defined by the uncertainties in the prediction of the size of the $N=258$ spherical shell closure 
(see Fig. 6d in Ref.\ \cite{AARR.15}) and single-particle densities in its vicinity.  With few exceptions theoretical 
uncertainties in the predictions of ground state deformations in the part of nuclear chart outside of above discussed 
regions are very small (see Fig.\ \ref{deformation-spreads}(a)).

  It is important to understand to what extent the predictions of the ground state 
deformations and related theoretical uncertainties in these predictions are dependent 
on nuclear matter properties of employed CEDFs. All employed CEDFs have the 
density $\rho_0$ and the energy per  particle $E/A$ at the saturation of symmetric 
nuclear matter (SNM) which are very close to each other and to empirical estimates 
(see Table III in Ref.\ \cite{AA.16}).  Thus,  the impact of only selected SNM properties 
listed in Table \ref{Table-mass-SNMP} on the ground state deformations are discussed 
below. Let start from the consideration of  the predictions by the pair of the functionals 
DD-PC1 and DD-ME2. Their SNM properties such as incompressibility $K_0$, the 
symmetry energy $J$ and its slope $L_0$ are close to each other and are located 
within the SET2b constraints on  experimental/empirical ranges for physical 
observables  of interest (see Table  \ref{Table-mass-SNMP}).  Despite that this
pair of the functionals gives the largest contribution into the spreads $\Delta \beta_2$
(compare panels (b) and (a) of Fig.\ \ref{deformation-spreads}). On the contrary,
the pairs of the functionals PC-PK1 and NL3* (which have $J$ and $L_0$ values 
located outside the SET2b constraint range [see Table \ref{Table-mass-SNMP}])
as well as NL3* and DD-ME2 (which have drastically different values of the $J$ 
and $L_0$ parameters [see Table \ref{Table-mass-SNMP}]) have (with very few
exceptions) very similar predictions for the ground state deformations across the
part of nuclear chart under study. These
exceptions are related to some differences in the predictions of the boundaries
between oblate and prolate shapes as well as between prolate and spherical
shapes. 

  These results for ground state deformations together with the analysis of the results 
for binding energies and charge radii of the $Z\leq104$ nuclei presented in Ref.\ 
\cite{AA.16}  strongly indicate that 
\begin{itemize}
\item
the major source of the uncertainties in the predictions of ground state deformations is related 
to local differences in underlying single-particle structure and, in particular, to the size 
of  spherical $Z=120$ and $N=184$ and 258 shell closures and the densities of the 
single-particle states in their vicinities,

\item
strict enforcement of the limits on the nuclear matter properties defined in Ref.\ \cite{RMF-nm}
will not necessary lead to the functionals with good description of ground and excited
state properties and will not reduce theoretical uncertainties in the description of physical
observables of interest in high-$Z$ and/or neutron-rich nuclei.

\end{itemize}

\begin{table}
\centering
\caption{Two-proton and two-neutron drip lines predicted by the NL3* and 
PC-PK1\footnote{The analysis of Ref.\ \cite{XLZLQCLZZKM.18} performed within 
the relativistic continuum Hartree-Bogoliubov theory and PC-PK1 functional leads 
to somewhat different predictions for the position of the two-neutron drip line for
some isotopic chains as compared with our results. This is a consequence of 
the neglect of deformation effects in Ref.\ \cite{XLZLQCLZZKM.18}.} functionals
(see Fig.\ \ref{chart-under-study} for graphical representation of drip lines).
Neutron numbers (columns 2-5) corresponding to these drip lines are given for each 
even proton number $Z$ (column 1). An asterisk at a neutron number at the 
two-neutron drip line indicates isotope chains with additional two-neutron binding 
at higher $N$ values (peninsulas).}
\begin{tabular}{| c| c| c|  c| c|} 
\hline 
\multicolumn{1}{|c|}{ Proton} &\multicolumn{2}{c|}{ Two-proton drip line} &\multicolumn{2}{c|}{ Two-neutron drip line} \\
       number $Z$  &  NL3*  & PCPK1 &   NL3* &PCPK1 \\ 
\hline
1      &  2   &   3    &   4    &     5      \\\hline
90    & 112  & 114 & 218  &  256    \\ 
92    & 118  & 122 & 224  &  258    \\ 
94    & 122  & 126 & 232  &  258    \\ 
96    & 126  & 128 & 252  &  258    \\
98    & 130  & 130 & 256  &  258    \\ 
100   & 132  & 132 & 258  &  258    \\ 
102   & 134  & 136 & 258  &  258    \\ 
104   & 138  & 140 & 258  &  258    \\ 
106   & 142  & 144 & 258  &  258*    \\ 
108   & 146  & 148 & 258  &  288    \\ 
110   & 150  & 154 & 258  &  292    \\ 
112   & 154  & 158 & 258  &  298    \\ 
114   & 158  & 162 & 262*  &  302    \\ 
116   & 162  & 166 & 286  &  312    \\ 
118   & 166  & 170 & 294  &  318    \\ 
120   & 170  & 174 & 298  &  324    \\ \hline
\end{tabular}
\label{Table-drip-lines}
\end{table}

\begin{figure*}[htb]
\centering
\includegraphics[angle=-90,width=18.0cm]{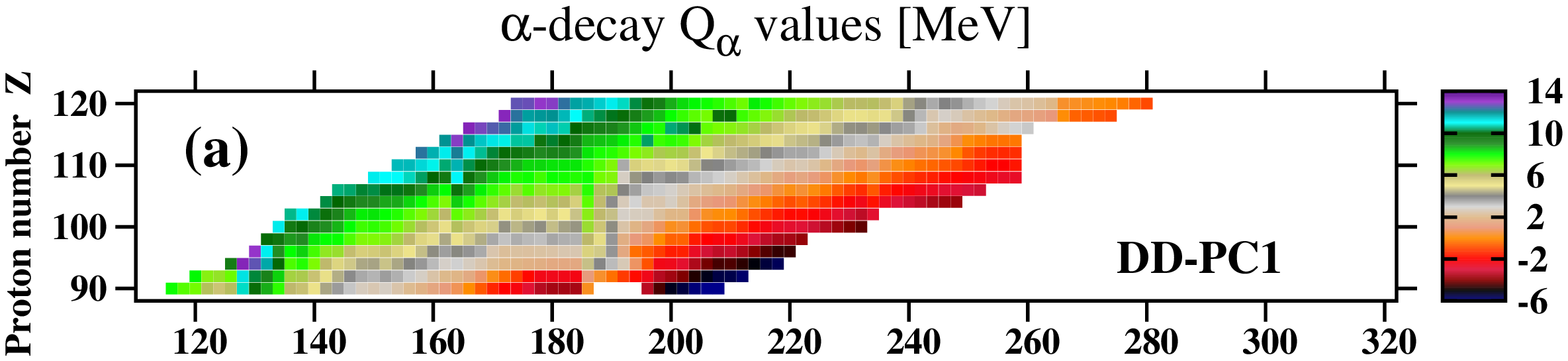}
\includegraphics[angle=-90,width=18.0cm]{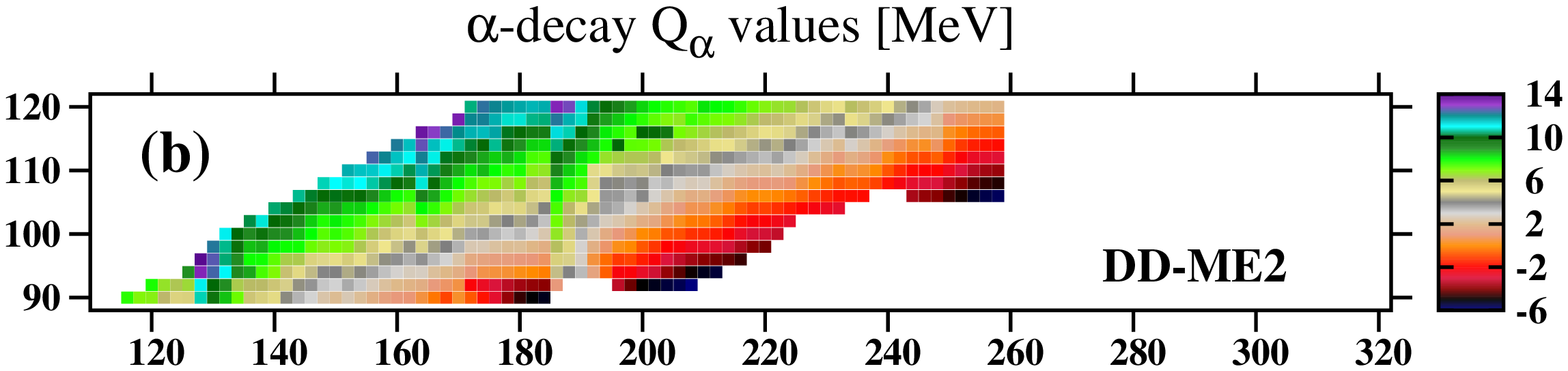}
\includegraphics[angle=-90,width=18.0cm]{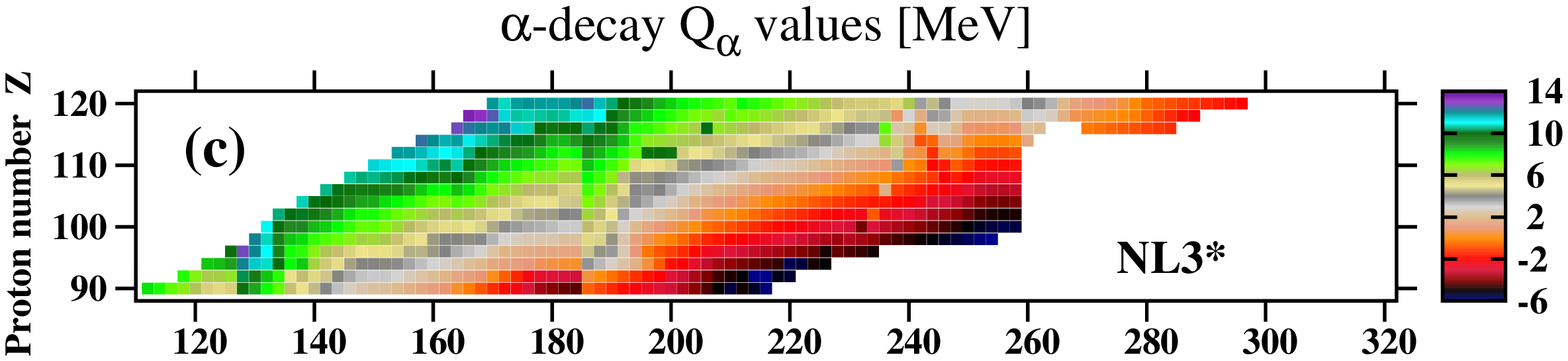}
\includegraphics[angle=-90,width=18.0cm]{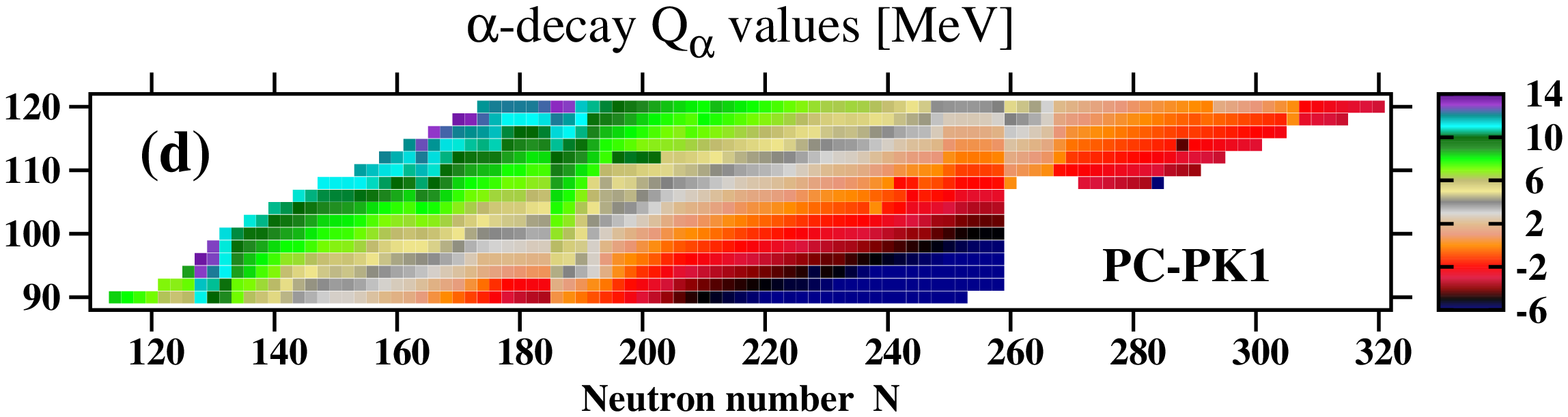}
\caption{The $Q_{\alpha}$ values for even-even actinides and superheavy nuclei
calculated with indicated CEDFs.}
\label{Q_alpha}
\end{figure*}

  Table \ref{Table-drip-lines} compares the positions of the two-proton and two-neutron 
drip  lines obtained  in the RHB calculations with the NL3* and PC-PK1 functionals; the 
results for the DD-PC1 and DD-ME2 functionals could be found in Table IV of Ref.\
\cite{AARR.14}. One can see that the two-proton drip lines are very similar in these two 
functionals; they differ by no more than four neutrons. This is in line with earlier 
observations that theoretical uncertainties in the predictions of the position of two-proton 
drip line are relatively small (see Ref.\ \cite{AARR.13} and Sec. VII in Ref.\ \cite{AARR.14}). 
Note that two-proton drip lines in the isotopic chains of interest obtained with PC-PK1 are 
very close to those obtained with DD-PC1 (compare Table \ref{Table-drip-lines} in the present 
article with Table IV of Ref.\ \cite{AARR.14}).

 Among the considered CEDFs the PC-PK1 functional provides the most neutron-rich
two-neutron drip line and the NL3* provides the second most neutron-rich two-neutron drip 
lines (compare Table \ref{Table-drip-lines} in the present article with Table IV of 
Ref.\ \cite{AARR.14} and see Sec. VIII in Ref.\ \cite{AARR.14}). All employed functionals 
reveal the presence of the shell closure at $N=258$ (see Fig.\ 6d in Ref.\ \cite{AARR.15}).
The size of this gap is almost the same in the NL3* and PC-PK1 functionals, but (i)
it is shifted down in energy by $\approx 400$ keV  for PC-PK1 as compared with NL3*
and (ii) high-$j$ intruder orbitals $1k_{15/2}$ and $2i_{13/2}$, which have a significant
impact on the position of neutron drip line (see discussion in Ref.\ \cite{AARR.15}), are
located at lower energies in the PC-PK1 functional as compared with the NL3* one. 
These features lead to the shift of the two-neutron drip line to substantially higher neutron
numbers in the PC-PK1 CEDF as compared with NL3*.  The sizes of the $N=258$ shell
gaps are smaller by $\approx 10$\% and 20\% in the DD-ME2 and DD-PC1 functionals
as compared with the ones in the PC-PK1 and NL3*. In addition, above mentioned
high-$j$ intruder orbitals in the calculations with the DD-ME2 and DD-PC1 functionals
are located at higher energies as compared with the ones in NL3*. As a consequence,
their two-neutron drip lines are located at lower neutron numbers as compared with
the NL3*. These features are clearly seen in Fig.\ \ref{deformations}.

  Fig.\ \ref{energy-spreads} shows the map of theoretical uncertainties $\Delta E(Z,N)$
in binding energies. These uncertainties increase drastically when approaching the 
neutron-drip line and in some nuclei they reach 50 MeV. Poorly defined isovector 
properties of CEDFs is the major reason for that (see Ref.\ \cite{AARR.14}). Note  
that the $\Delta E(Z,N)$ spreads for the NL3*, DD-PC1 and DD-ME2 functionals are 
relatively modest [see Fig.\ \ref{energy-spreads}(b)]\footnote{The addition of the 
DD-ME$\delta$ functional
to this set of three functionals is not expected to modify significantly  $\Delta E(Z,N)$
(see Fig.\ 9 in Ref.\ \cite{AARR.14}).} and the major contribution to $\Delta E(Z,N)$
is coming from the PC-PK1 functional (compare panels (a) and (b) in Fig.\ \ref{energy-spreads}).  
The fact that isovector properties of the PC-PK1 functional are significanly different from 
those of NL3*, DD-PC1, DD-ME2 and DD-ME$\delta$ is also confirmed by the analysis of 
binding energies in the Yb ($Z=70$) isotopic chain (see Fig.\ 3 in Ref.\ \cite{AA.16}). As 
follows from the analysis of parametric correlations in different classes of  CEDFs performed 
in Ref.\ \cite{TAAR.20}, a possible reason for that  could be related to over-parametrization of 
the isoscalar channel in this class of CEDFs\footnote{The analysis of Ref.\ \cite{TAAR.20} 
suggests that the number of parameters in the isoscalar channel of PC-PK1 CEDF can be 
reduced from 4 to 1.}. This, in turn, may lead to a somewhat wrong balance of the isoscalar 
and isovector  channels in known nuclei which reveals itself in a more pronounced way
via different (as compared with other functionals) isovector dependence of  binding 
energies in neutron-rich one.

   Figure \ref{2n-separation-energies} presents the summary of two-neutron
separation energies $S_{2n}(Z,N) $ obtained  with four employed CEDFs. Note that some 
discontinuities in smooth trends of the $S_{2n}(Z,N)$ distributions as a function of neutron 
number are  either due to the presence of substantial spherical shell gaps at $N=184$ or $N=258$ or due 
to the crossing of the boundaries between prolate and oblate shapes.   
For example, the impact of  the $N=184$ spherical shell gap on the $S_{2n}(Z,N)$  distributions 
is clearly visible in Figs.\ \ref{2n-separation-energies}(b), (c) and (d) (see also Figs.\ \ref{deformations} 
(b), (c) and (d) for deformation distributions). On the contrary,  its impact is substantially suppressed
in superheavy nuclei in the calculations with CEDF DD-PC1 [see Fig.\ \ref{2n-separation-energies}(a)] 
because of the reduced role of the $N=184$ spherical shell gap in this functional. 

    Finally, the spreads $\Delta S_{2n}(Z,N)$ in two-neutron separation energies are 
presented in Fig.\ \ref{S_2n_spread}. They are the lowest in known nuclei but in general 
increase with increasing neutron number. The $\Delta S_{2n}(Z,N)$ values are quite 
large ($\Delta S_{2n}(Z,N) \approx 2.2$ MeV) in the vicinity of two-neutron drip lines and the 
$N=184$ and $N=258$ spherical shell gaps.   However, they become  extremely large 
($\Delta S_{2n}(Z,N) \approx 4.0$ MeV) at the boundaries  between prolate and oblate 
shapes.   Similar to the spreads in binding energies (see discussion of Fig.\ \ref{energy-spreads} 
above),  the largest contribution to the  $\Delta S_{2n}(Z,N)$ values comes from the CEDF  PC-PK1.
If the PC-PK1 functional is excluded from consideration these values on average decrease by a 
factor of 2  (compare panels (b) and (a) in Fig.\ \ref{S_2n_spread}).  It is interesting that in 
neutron-rich deformed $N\approx 190-236$ region the $\Delta S_{2n}(Z,N)$ values are  on  
average comparable with those in known nuclei (see Fig.\ \ref{S_2n_spread}(b). However, they 
still show increased magnitudes at above discussed locations of nuclear chart.

\section{$\alpha$-decay properties}
\label{sect-alpha}

In actinides and superheavy nuclei spontaneous fission and $\alpha$ emission
compete and the shortest half-life determines the dominant decay channel and
the total half-life. Only in the cases where the spontaneous fission half-life is longer
than the half-life of $\alpha$ emission can superheavy nuclei be observed in experiment.
In addition, only nuclei with half-lives longer than $\tau=10$ $\mu$s are observed
in experiments. 

   The $\alpha$ decay half-live depends on the $Q_{\alpha}$ values which are
calculated  according to
\begin{equation}
Q_{\alpha}=E(Z,N)-E(Z-2,N-2)-E(2,2)
\end{equation}
with $E(2,2)=-28.295674$ MeV \cite{AME2012} and $Z$ and $N$ representing
the parent nucleus.
%
%

\begin{figure*}[htb]
\centering
\includegraphics[angle=-90,width=18.0cm]{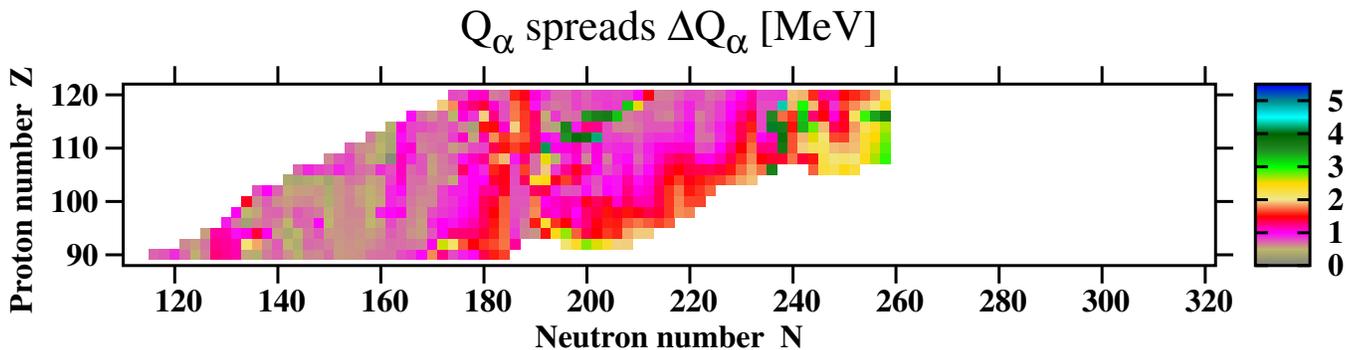}
\caption{The $Q_{\alpha}$ spreads
$\Delta Q_{\alpha}(Z,N)$ as a function of proton and neutron number.
$\Delta Q_{\alpha}(Z,N)=|Q_{\alpha}^{\rm max}(Z,N)-Q_{\alpha}^{\rm min}(Z,N)|$,
where $Q_{\alpha}^{\rm max}(Z,N)$ and $Q_{\alpha}^{\rm min}(Z,N)$ are the
largest and smallest  $Q_{\alpha}$ values obtained with four employed  CEDFs for  
the $(Z,N)$ nucleus. 
}
\label{Q_alpha_spread}
\end{figure*}

  To estimate theoretical uncertainties in the predictions of the $\alpha$-decay half-lives,
they were calculated using three phenomenological expressions, namely,

\begin{itemize} 

\item
the Viola-Seaborg semiempirical formula \cite{VS.66}
\begin{equation}
log_{10}\tau_{\alpha}=\frac{aZ+b}{\sqrt{Q_{\alpha}}}+cZ+d
\label{Viola-Seaborg-eq}
\end{equation}
employing two sets of parametrizations. The first one with the parameters 
$a=1.66175$, $b=-8.5166$, $c=-0.20228$ and $d=-33.9069$ has been 
fitted in Ref.\ \cite{SPC.89}. This set and the results obtained with it
are labelled further as VSS-1989.  Another set has been defined in Ref.\ \cite{DR.05}
and its paramaters are: $a = 1.64062$, $b = -8.54399$, $c = -0.19430$
and $d = -33.9054$. The label VSS-2005 is used for it and its results.

\item
phenomenological first modified Brown fit (mB1) \cite{BBS.16}
\begin{equation}
log_{10}\tau_{\alpha}=\frac{a(Z-2)^b}{\sqrt{Q_{\alpha}}}+c
\label{Brown-eq}
\end{equation}
with the parameters $a = 13.0705$, $b = 0.5182$, $c = -47.8867$.
This set and its results are labeled further as MB-2016.

\item
phenomenological Royer model \cite{Royer.00}
\begin{equation}
log_{10}\tau_{\alpha}=\frac{aZ}{\sqrt{Q_{\alpha}}}+b{A}^\frac{1}{6}{\sqrt{Z}}+c
\label{Royer-eq}
\end{equation}
with the parameters  $a = 1.5864$, $b = -1.1629$ and $c = -25.31$ of Ref. \cite{Royer.00}.
Its results are labeled further as Royer-2000.

\end{itemize}

\begin{figure*}[htb]
\centering
\includegraphics[angle=-90,width=18.0cm]{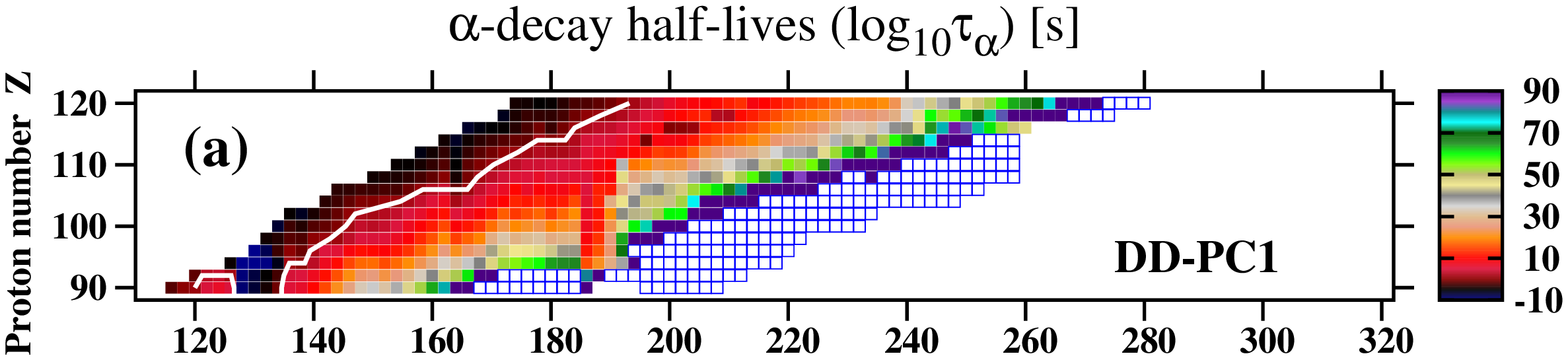}
\includegraphics[angle=-90,width=18.0cm]{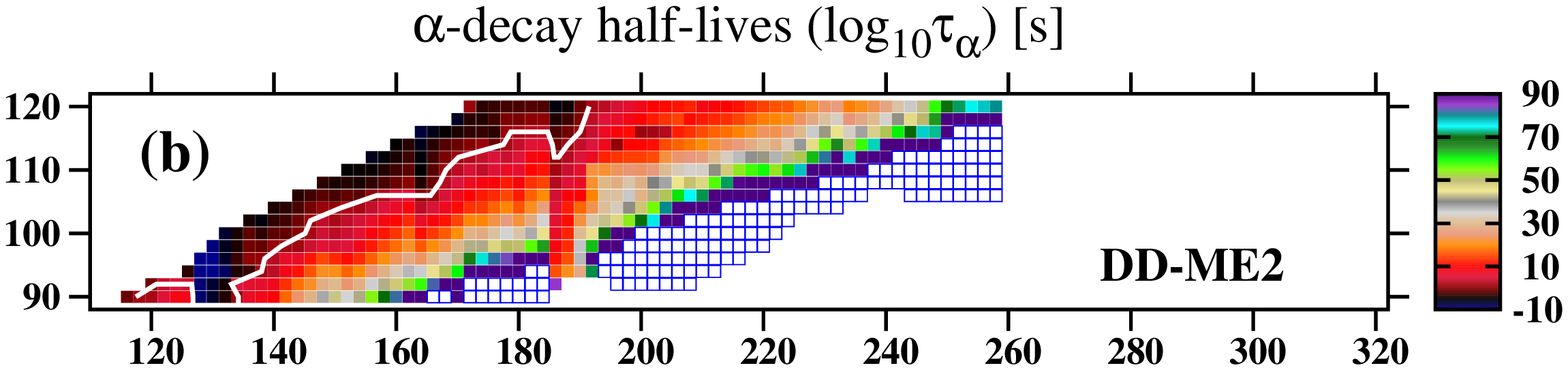}
\includegraphics[angle=-90,width=18.0cm]{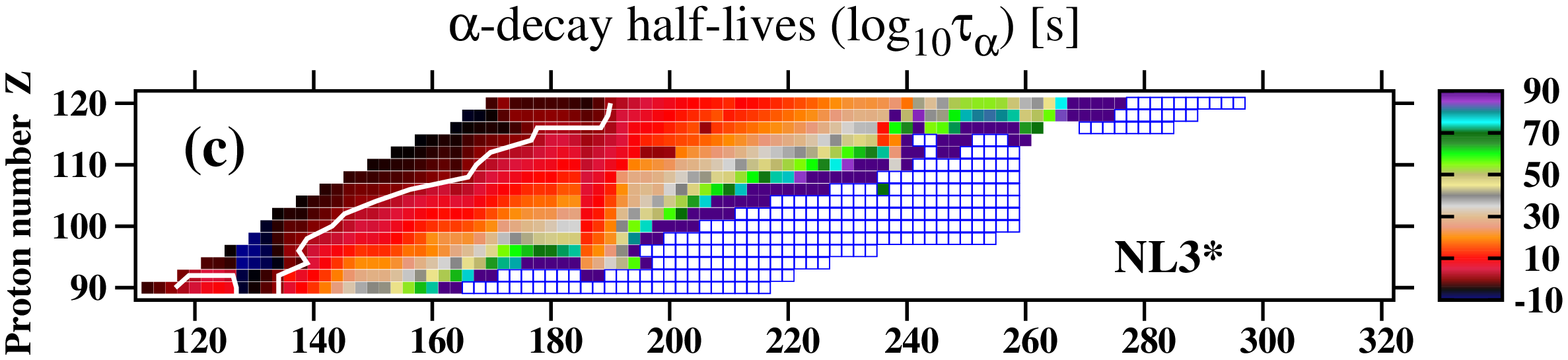}
\includegraphics[angle=-90,width=18.0cm]{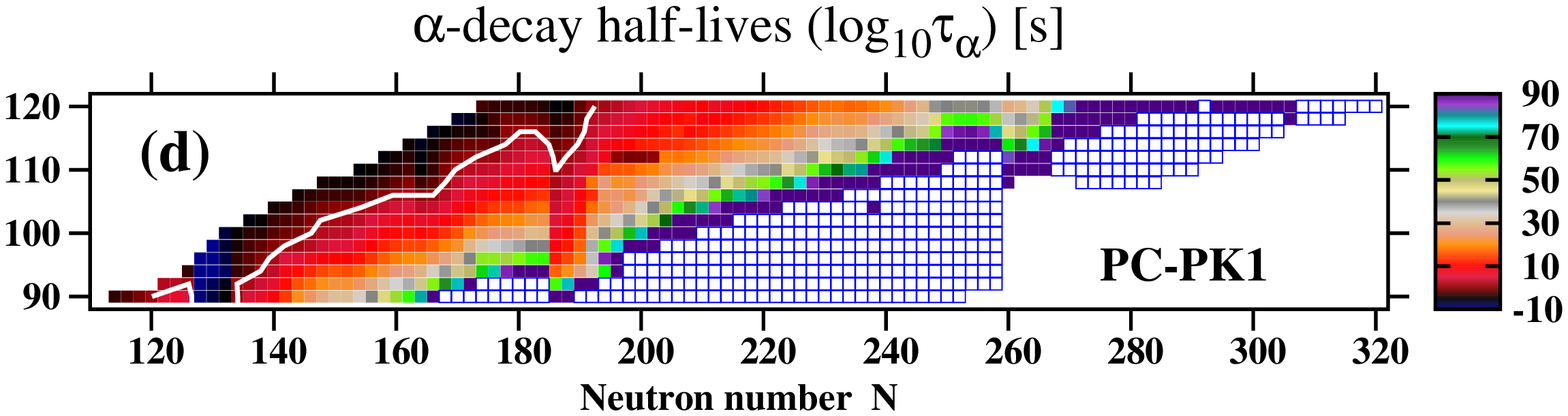}
\caption{
Calculated $log_{10}(\tau_{\alpha})$ values of the $\alpha$-decays for even-even 
superheavy nuclei obtained with the VSS-2005 version of Viola-Seaborg semi-empirical 
formula for four indicated CEDFs. Open squares are used for the nuclei in which $\alpha$-decay 
is energetically forbidden.  The white line corresponding to $log_{10}(\tau_{\alpha})=1.0$ 
outlines the region of nuclei in which the alpha-decay half-live is smaller than 10 s.
}
\label{T_alpha_VSS_2005}
\end{figure*}

 These phenomenological expressions employ different functional dependencies
(in particular, they show different dependencies on proton and mass numbers) and
are fitted to different sets of experimental data. This is expected to lead to different  
predictions  for $\tau_{\alpha}$ in high-$Z$ and neutron-rich nuclei.

\begin{figure*}[htb]
\centering
\includegraphics[angle=-90,width=18.0cm]{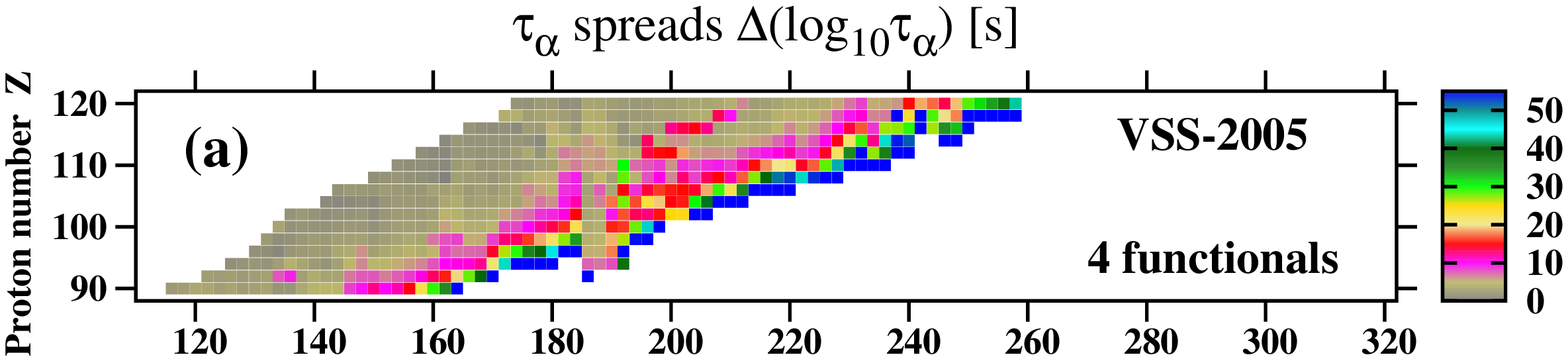}
\includegraphics[angle=-90,width=18.0cm]{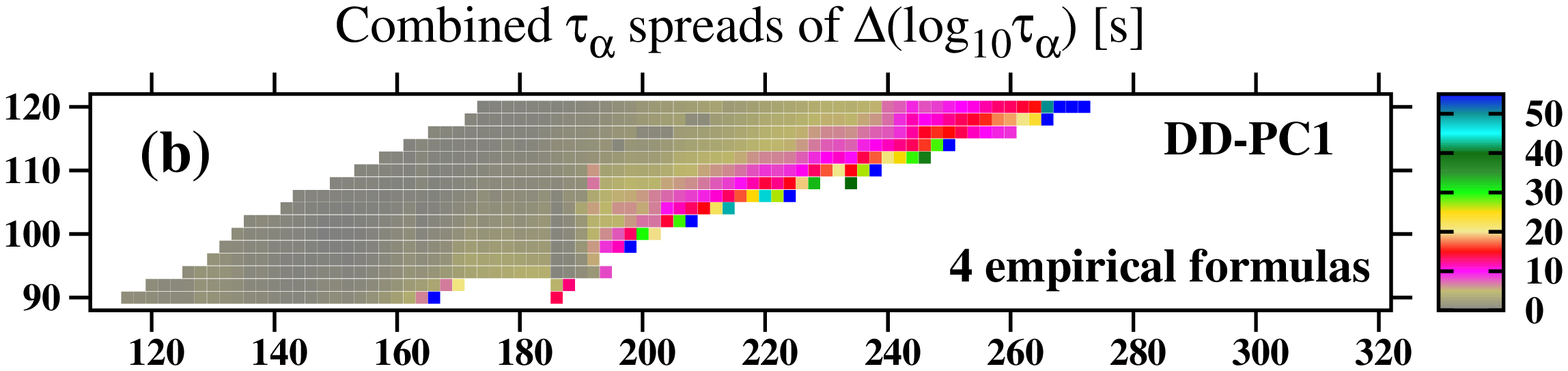}
\includegraphics[angle=-90,width=18.0cm]{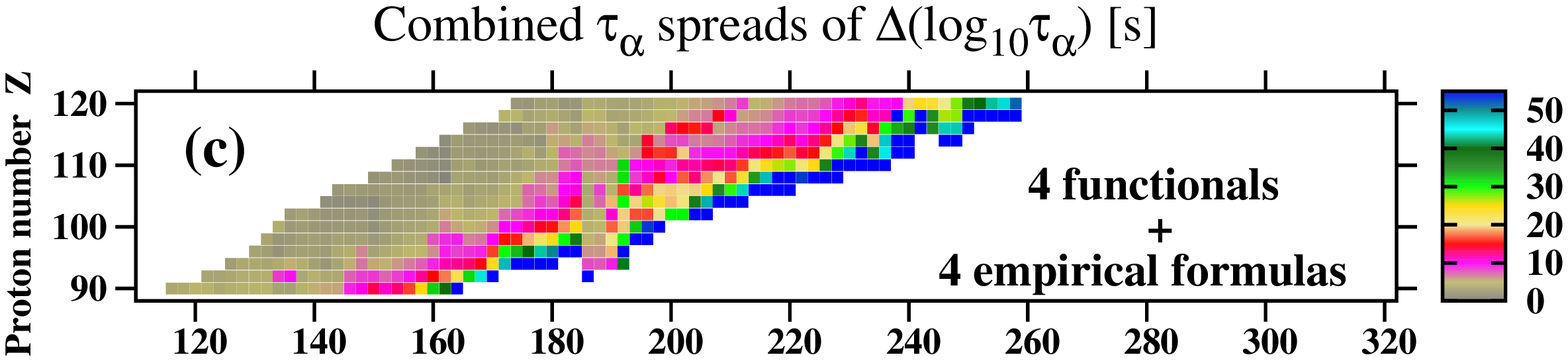}
\includegraphics[angle=-90,width=18.0cm]{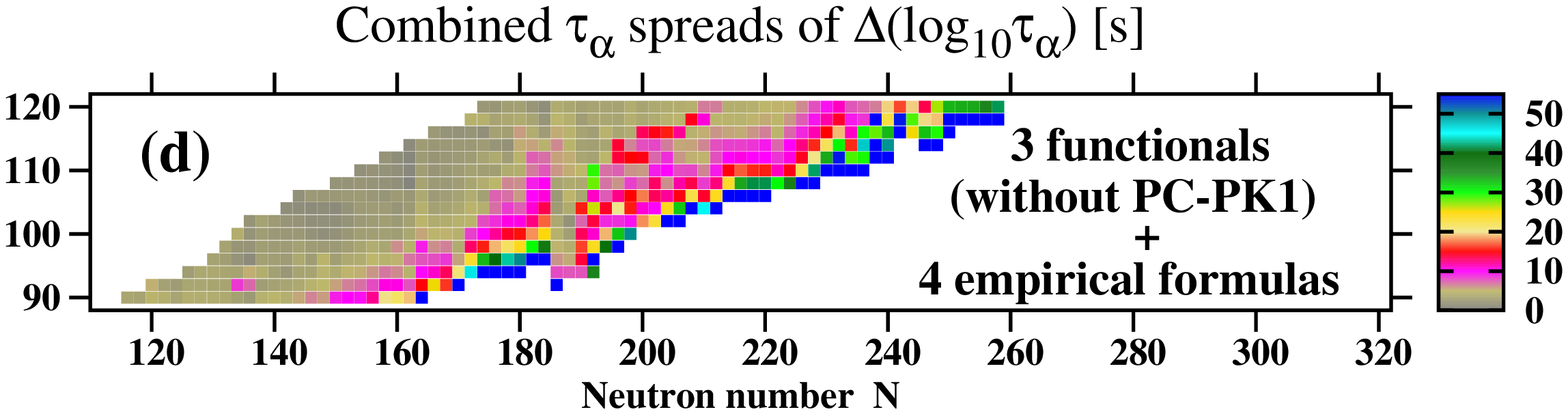}
\caption{The $\tau_{\alpha}$ spreads
$\Delta \tau_{\alpha}(Z,N)$ as a function of proton and neutron number.
$\Delta \tau_{\alpha}(Z,N)=|\tau_{\alpha}^{\rm max}(Z,N)-\tau_{\alpha}^{\rm min}(Z,N)|$,
where $\tau_{\alpha}^{\rm max}(Z,N)$ and $\tau_{\alpha}^{\rm min}(Z,N)$ are the
largest and smallest  $\tau_{\alpha}$ values obtained with selected set of functionals
and empirical formulas. Panel (a) shows these spreads obtained  with four employed  
CEDFs and VSS-2005 empirical formula. DD-PC1 CEDF and four empirical formulas
are used in panel (b).  Panel (c)  presents combined spread of $\tau_{\alpha}$ obtained 
with four CEDFs and four empirical formulas. Panel (d) is the subversion of panel (c) in
which the PC-PK1 functional is excluded.
}
\label{T_alpha_spreads}
\end{figure*}

  The $Q_{\alpha}$ values calculated with the DD-PC1, DD-ME2, NL3* and PC-PK1
functionals are presented in Fig.\ \ref{Q_alpha}. One can see that for a fixed value of $Z$
with increasing neutron number the $Q_{\alpha}$ values in general decrease.  They are
positive in proton-rich nuclei as well as in the nuclei located close to the $\beta$-stability line. 
The $Q_{\alpha}$ values experience a substantial increase at shell closure with $N=184$\footnote{Similar 
increase is also seen in the vicinity of the $N=258$ shell  closure in the calculations with the NL3* and 
PC-PK1 CEDFs [see Fig.\ \ref{Q_alpha}(c) and (d)].}  (see Fig.\ \ref{Q_alpha}
in the present paper as well as Fig.\  14 in Ref.\ \cite{AANR.15}); note that the effect
of this shell closure is washed out in the $Z>110$ nuclei for the DD-PC1 functional.
With subsequent increase of neutron number the $Q_{\alpha}$ values become first smaller,
then they become close (or equal) to zero and with further increase of $N$
they get more and more negative.  Note that  $\alpha$-decay is energetically not possible for $Q_{\alpha}\leq 0$ MeV. 
Thus, very neutron-rich nuclei cannot decay by $\alpha$-emission.

\begin{figure*}[htb]
\centering
\includegraphics[angle=-90,width=18.0cm]{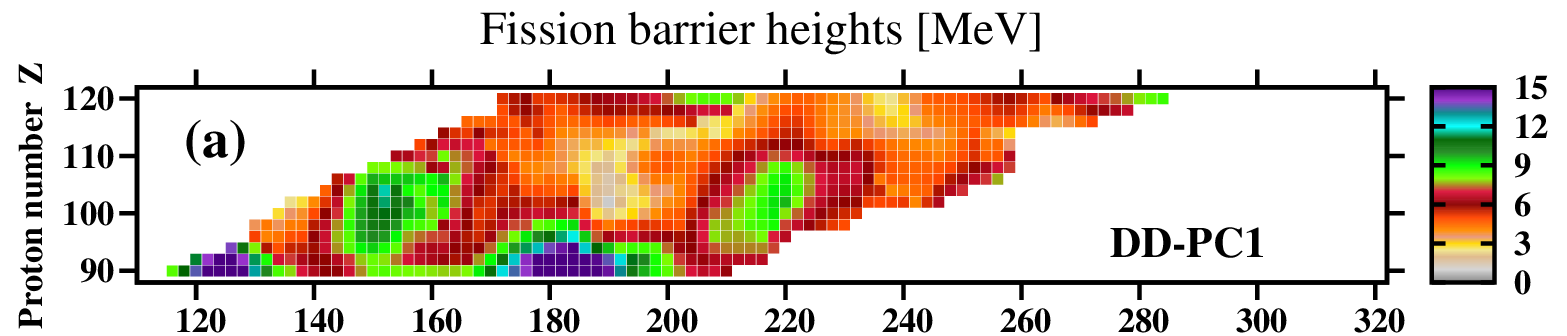}\\
\includegraphics[angle=-90,width=18.0cm]{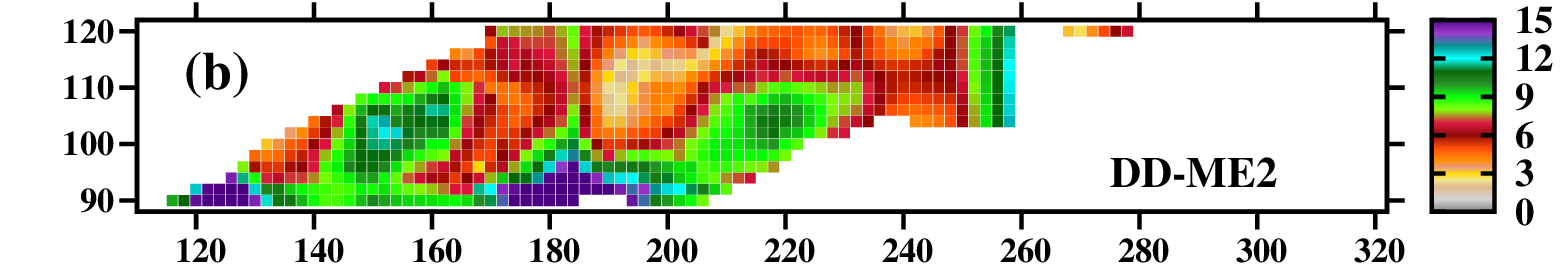}\\
\includegraphics[angle=-90,width=18.0cm]{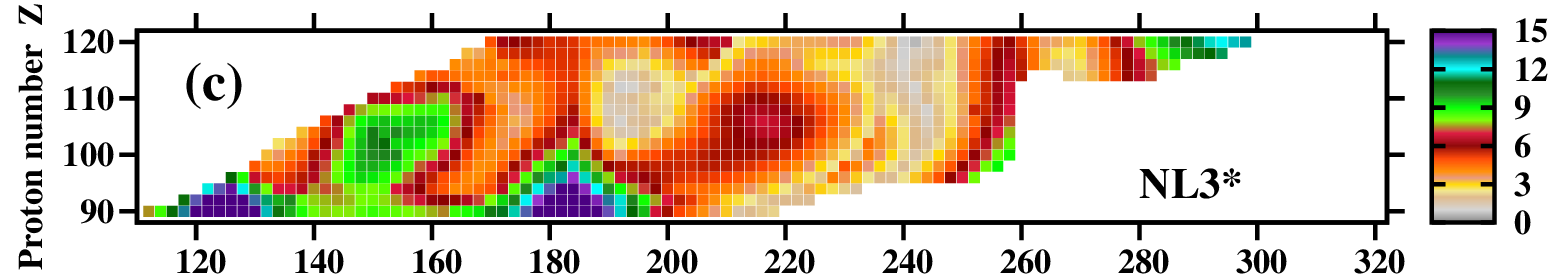}\\
\includegraphics[angle=-90,width=18.0cm]{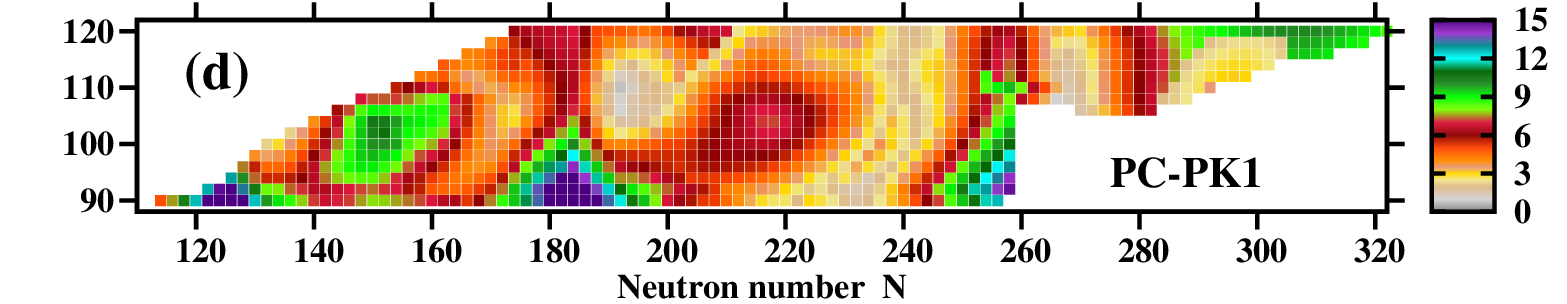}\\
\caption{The heights of primary fission barriers (in MeV) obtained in axial RS-RHB and RA-RHB calculations 
as a function of proton and neutron numbers for nuclei located between two-proton and two-neutron drip lines.
}
\label{FB-height}
\end{figure*}

  Note that general trends in the development of the $Q_{\alpha}$ values as a function  of proton 
and neutron number are similar in all functionals. The major differences are  related to the location
of the two-neutron drip line, the impact of the $N=184$ and $N=258$ shell closures and the 
location of the boundaries between prolate and oblate nuclear shapes. These differences 
between the functionals are summarized in Fig.\ \ref{Q_alpha_spread}  which shows the $Q_{\alpha}$ 
spreads $\Delta Q_{\alpha}(Z,N)$ as a function of proton and neutron number.  The largest spread in
the predictions exists in the island centered around $Z\sim 110, N \sim 198$
in which $\Delta Q_{\alpha}(Z,N)> 3$ MeV. This spread emerges from different predictions of
the boundaries in the $(Z,N)$ plane between prolate and oblate shapes (see Fig.\ \ref{deformations}) 
and coincides with the largest spread in calculated ground state deformations (see Fig.\
 \ref{deformation-spreads}). The next region with the largest differences in the predictions is located  
between neutron numbers N=236 and N=258 (see Fig.\ \ref{Q_alpha_spread}).  However, these 
 differences are not critical because (i) this region is not expected to play a role in the r-process, 
 (b) expected $\alpha$-decay half-lives  exceed 10$^{20}$ s (see Fig.\ 
 \ref{T_alpha_VSS_2005}) and (c) many of the nuclei in this region are not expected to decay 
 by $\alpha$-emission.  High $\Delta Q_{\alpha}$ values 
($\Delta Q_{\alpha}(Z,N) \approx 1.5$ MeV) are observed near shell closure at $N=184$ and 
in very neutron-rich nuclei near two-neutron drip line. This is a consequence of the difference
in the predictions of the ground state properties such as deformations in the nuclei near $N=184$
(see Ref.\ \cite{AANR.15}) and general deterioration of predictive power of nuclear models on
approaching neutron drip line (see Ref.\ \cite{AARR.14}).  In other regions of nuclear chart,
$\Delta Q_{\alpha}(Z,N) \leq 1.0$ MeV with smallest spreads seen in the $N<180$ nuclei.

   Note that the inclusion of  dynamical correlations (for example, by means of 5 dimensional 
collective Hamiltonian) can locally modify the binding energies and $Q_\alpha$ values 
\cite{PNLV.12,ZNLYM.14,SALM.19}  but they have the largest impact on transitional nuclei which 
represent only minor part of the nuclei under study. For well deformed nuclei, the impact of 
dynamical correlations on $Q_{\alpha}$ values is rather modest \cite{PNLV.12}.
Thus, their inclusion is not expected to change drastically global picture for the behavior of 
$Q_{\alpha}$.

   Calculated $\alpha$-decay half-lives $\tau_{\alpha}$ (in logarithmic scale) obtained 
with the VSS-2005 empirical formula for four CEDFs are shown in Fig.\ 
\ref{T_alpha_VSS_2005}.  Other phenomenological formulas such as VSS-1989, MB-2016 
and Royer-2000 give similar results; thus, they are not shown.  For a given isotope chain
the calculated half-lives generally increase with increasing neutron number. This trend is
interrupted only at the $N=184$ and $N=258$ shell closures. The consequence of this feature
is the fact that traditional experimental technique of detecting superheavy nuclei by 
$\alpha$-decay will not  work in neutron-rich nuclei because they can decay faster by 
spontaneous  fission.  Note that $\alpha$-decay is energetically forbidden for a 
large group of very neutron-rich nuclei located in the vicinity of neutron-drip line. In
such nuclei as well as in those which have very large $\tau_{\alpha}$ values,
the competition of spontaneous fission,  neutron induced fission, $\beta$-decay,  and 
neutron emission will define the leading  channel of decay in the r-process calculations.

\begin{figure*}[htb]
\centering
\includegraphics[angle=-90,width=18.0cm]{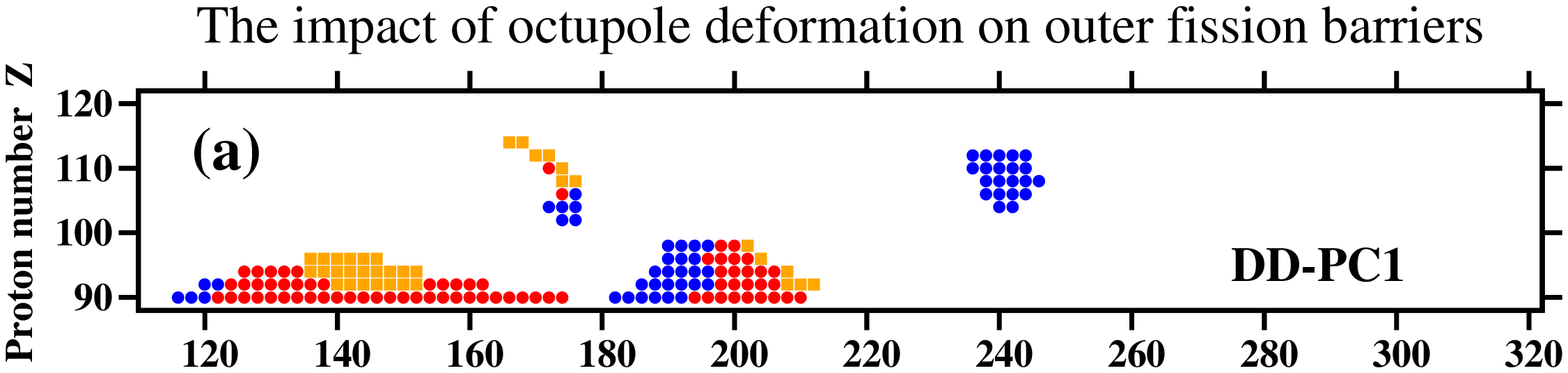}\\
\includegraphics[angle=-90,width=18.0cm]{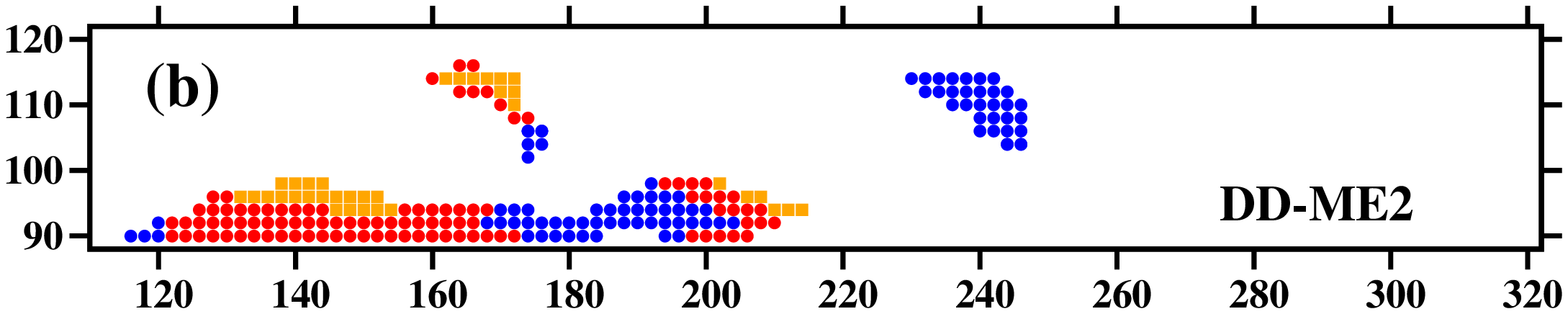}\\
\includegraphics[angle=-90,width=18.0cm]{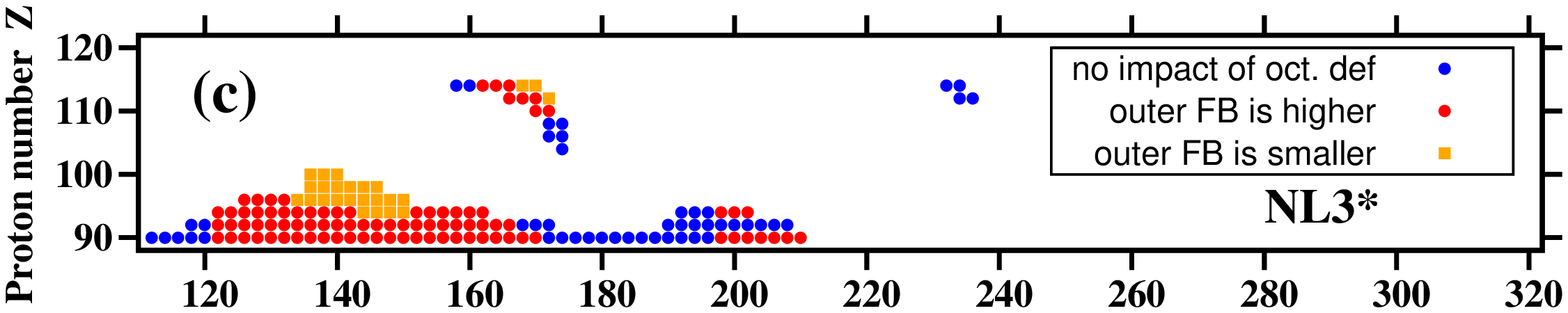}\\
\includegraphics[angle=-90,width=18.0cm]{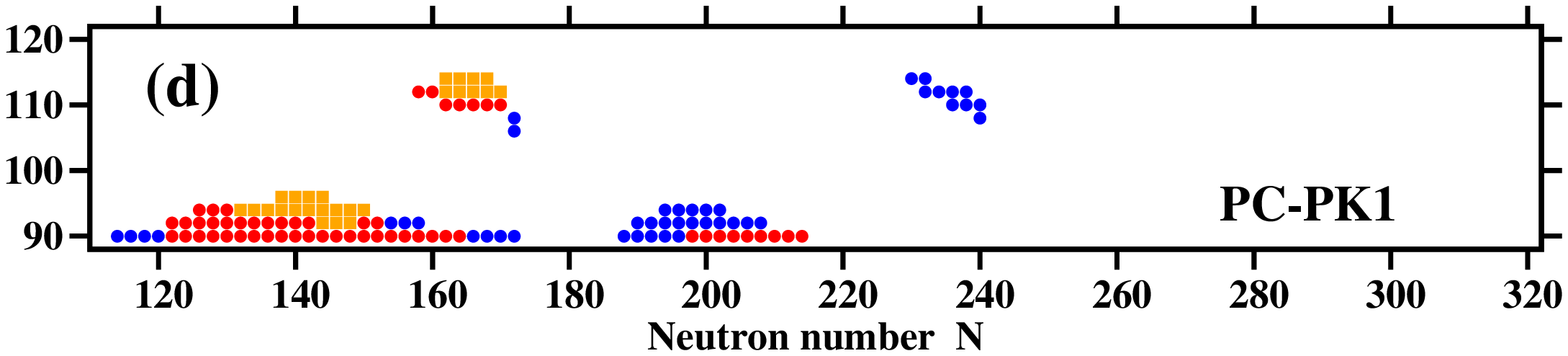}\\
\caption{The distributions of the nuclei,  in which the outer fission barrier is higher than 
inner one in the RS-RHB calculations, in the $(Z,N)$ plane for four employed CEDFs.
Different types of symbols are used to indicate the impact of octupole deformation on 
outer fission barriers of these nuclei.  Blue solid circles are used for the nuclei the heights of outer
fission barriers of which are not affected by the inclusion of octupole deformation. 
Solid red circles/orange squares are used for 
the nuclei in which outer fission barrier is affected by octupole deformation in the
RA-RHB calculations and is higher/lower than inner fission barrier.
}
\label{out-vs-in-bar}
\end{figure*}

     It is important to remember that the typical  timescale of the r-process is in 
the seconds range \cite{KMBQR.17,LMHCF.18,r-process-review-19}. 
Fig.\ \ref{T_alpha_VSS_2005} clearly illustrates 
that with few exceptions the nuclei located to the right of white lines have $\alpha$-decay 
half-lives exceeding 10 s. Thus, $\alpha$-decay half-lives of these nuclei are longer than
the typical timescale of the  r-process and, as a consequence, their alpha-decays will 
not affect the r-process simulations.  These white lines in Fig.\ \ref{T_alpha_VSS_2005}
also  outline the region of nuclear chart in which traditional experimental measurements 
of superheavy nuclei based on the $\alpha$-decays are possible: these are the regions 
located near and to the left of these white lines.

    Theoretical uncertainties in the predictions of $\alpha$-decay half-lives given via 
the $\Delta (log_{10}(\tau_{\alpha}))$ spreads are summarized in Fig.\  \ref{T_alpha_spreads}. 
The comparison of the panels (a) and (b) clearly shows that these uncertainties mostly 
emerge from the differences in the predictions of the $Q_{\alpha}$ values by different functionals. 
These uncertainties exceed 50 orders of magnitude in the nuclei located in the vicinity of two-neutron 
drip line  and  in some nuclei around $Z\approx 108, N\approx 198$ (see Fig.\ \ref{T_alpha_spreads}).
The uncertainties in $\tau_{\alpha}$ originating from different empirical formalas [see Eqs.\ 
(\ref{Viola-Seaborg-eq}), (\ref{Brown-eq}) and (\ref{Royer-eq})] are significantly  smaller [see
Fig.\  \ref{T_alpha_spreads}(b)]. For absolute  majority of the nuclei they are smaller than 5 orders 
of magnitude and for proton-rich nuclei and the nuclei located close to the beta-stability they are 
very small. They are larger than 10 orders of magnitude only in neutron-rich nuclei located in close 
vicinity of two-neutron drip line.  Combined theoretical uncertainties in $\tau_{\alpha}$ emerging 
from  the use of four empirical formulas and four CEDFs are summarized in Fig.\ 
\ref{T_alpha_spreads}(c). One can see that for almost half of nuclei they exceed 10 orders 
of magnitude; these nuclei are located on neutron-rich side of nuclear chart.
However, these uncertainties are not very critical since the $\alpha$-decay
lifetimes become extremely large in such nuclei (see Fig.\ \ref{T_alpha_VSS_2005}) so 
$\alpha$-decay can compete neither with fission nor with $\beta$-decay. 
Note also that the removing of the PC-PK1 functional from consideration does not
change appreciably theoretical uncertainties in the predictions of $\alpha$-decay 
half-lives (compare Figs.\ \ref{T_alpha_spreads} (c) and (d)).

\section{Fission properties}
\label{sect-fission}

\subsection{Primary fission barriers}

   The distributions of primary fission barriers\footnote{The highest 
in energy fission  barrier (among inner and outer ones) is called primary and it 
plays an important role in the r-process modeling (see Ref.\ \cite{AG.20}).} (PFB) 
heights in the $(Z,N)$ plane obtained with  employed functionals are shown in 
Fig.\ \ref{FB-height}.  Fig.\ \ref{out-vs-in-bar} presents the maps of the nuclei in 
the region under study in which outer fission barriers are higher than inner 
ones in the RS-RHB calculations (see Figs.\ \ref{deformation-curve-Ds} and
\ref{deformation-curve-Th} for more details).  It also illustrates that the importance of outer fission barriers
in stabilization of nuclei in general decreases on going from light actinides to superheavy nuclei because
of increased importance of Coulomb interaction (compare Figs.\ \ref{deformation-curve-Th}
and \ref{deformation-curve-Ds}).
Fig.\ \ref{out-vs-in-bar} also demonstrates the impact of octupole deformation (as obtained
in RA-RHB calculations) on the outer fission barriers and on their heights with respect of
inner ones.  The lowering of outer fission barrier due to octupole deformation indicates that
asymmetric fission becomes dominant, while the absence of the impact of octupole deformation
on outer fission barrier height tells that fission will be symmetric.

  Fig.\ \ref{out-vs-in-bar} shows that similar regions in the $(Z,N)$ plane, in which
the outer fission barriers are higher in energy than inner ones in the RS-RHB 
calculations, appear in the calculations with all employed functionals.   However,
these regions are substantially larger in the density dependent functionals  (such 
as DD-PC1 and DD-ME2) as compared  with CEDFs NL3* and
PC-PK1.  Octupole deformation does not affect outer fission barriers in the nuclei
located in the $Z\sim 110, N\sim 240$ region, in the $N\leq 120$ nuclei as well as in the
nuclei located not so far away from $N\approx 180$. On going away from the latter two 
regions,  octupole deformation first starts to reduce the heights of outer fission
barriers but they still remain higher in energy than inner ones.  Further transition away
from these regions leads to the reduction of the heights of outer fission barriers below 
the inner ones due to the impact of octupole deformation.

\begin{figure*}[htb]
\centering
\includegraphics[angle=-90,width=18.0cm]{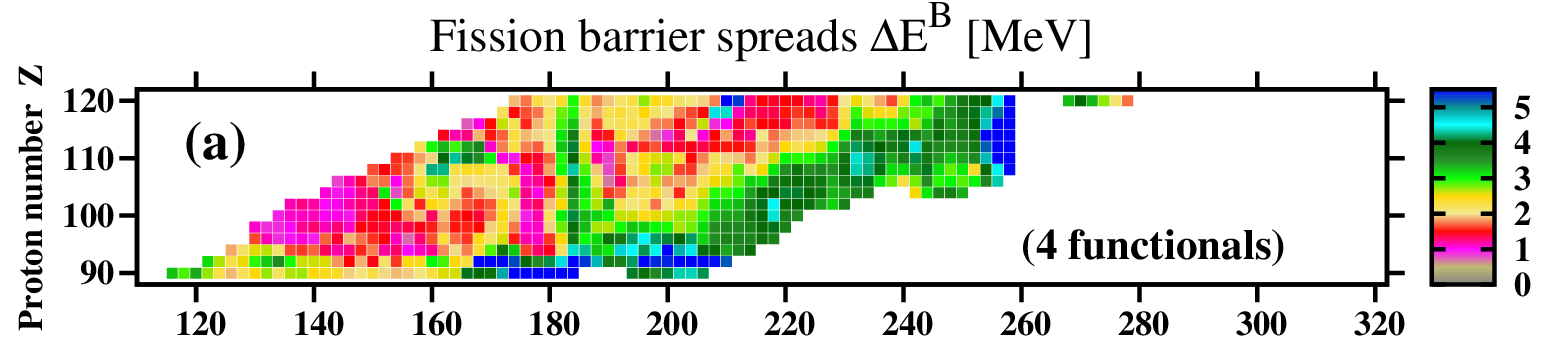}\\
\includegraphics[angle=-90,width=18.0cm]{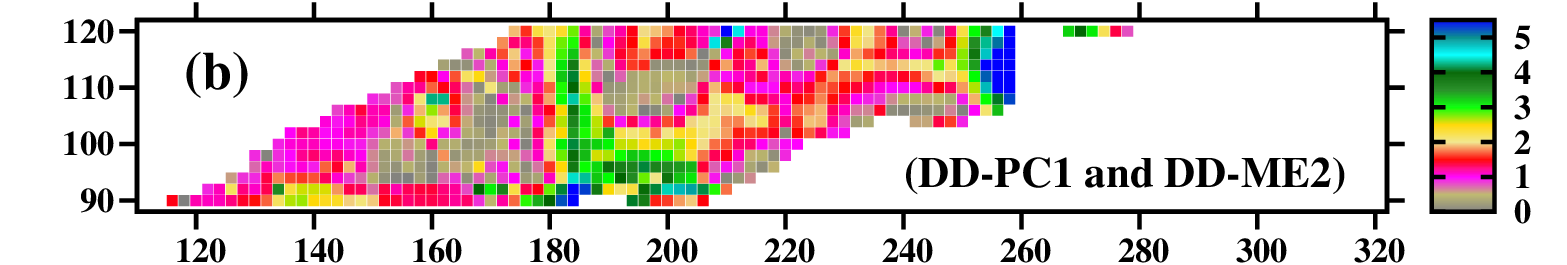}\\
\includegraphics[angle=-90,width=18.0cm]{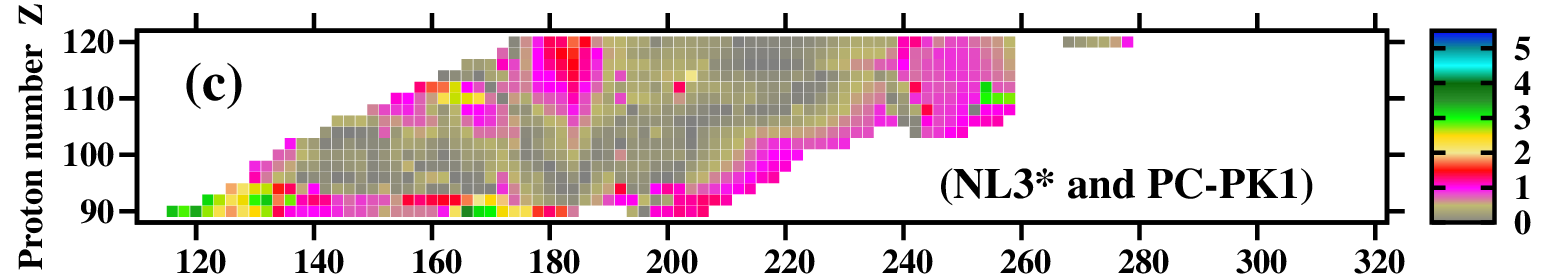}\\
\includegraphics[angle=-90,width=18.0cm]{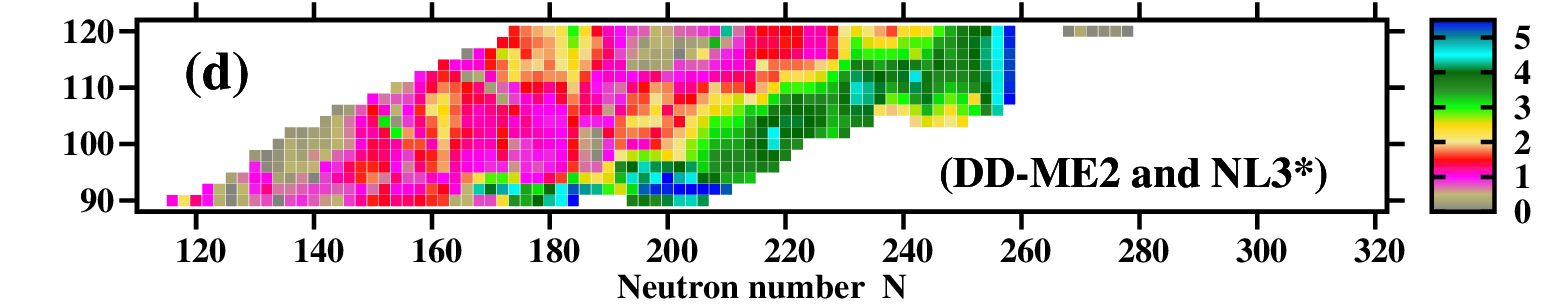}\\
\caption{(a) The spreads $\Delta E^B$ of the heights of primary fission barriers as a function 
of proton and neutron numbers. $\Delta E^B(Z,N) = |E^B_{max}(Z,N) - E^{B}_{min}(Z,N)|$,
where, for given $Z$ and $N$ values, $E^B_{max}(Z,N)$ and  $E^{B}_{min}(Z,N)$ are
the largest and smallest heights of inner fission barriers obtained with employed set 
of four functionals. (b-d) The spreads $\Delta E^B$ obtained for indicated pairs of
the functionals.
}
\label{FB-spreads}
\end{figure*}

Fig. \ref{FB-height} reveals a lot of similarities in the predictions of the global structure of 
the maps of fission barrier heights obtained with 
employed functionals. The highest PFBs
are predicted in the islands of low-$Z$ nuclei centered around spherical shell closures with 
$N=126$ and $N=184$ (and $N=258$ in the case of the PC-PK1 functional). Fission 
barriers reach 15 MeV in the centers of these islands.  Next island with high fission barriers
exists around $Z\approx 100, N\approx 150$. Left bottom part of this island coincides with
the region of actinides (see, for example, Fig.\ 7 in Ref.\ \cite{AAR.10}) in which the heights of fission 
barriers have been experimentally measured. Relativistic mean field calculations
with the NL3*, PC-PK1 and DD-PC1 functionals performed by different groups rather well 
describe inner and outer fission barriers in actinides \cite{AAR.10,PNLV.12,LZZ.12,LZZ.14}.
Note also that the spreads 
 of the heights of inner  fission barriers obtained
with these three functionals  in the $Z\approx 100, N\approx 150$ island are relatively small
(approximately 1 MeV or less) for the majority of nuclei in this island  (see Fig.\ 3b in Ref.\ \cite{AARR.17}). 
On the  contrary, the DD-ME2 functional predicts somewhat higher fission barriers in this island
(see Fig.\ \ref{FB-height}b) which leads to somewhat higher spreads $\Delta E^B$ in 
the heights of primary fission barriers (see Fig.\  \ref{FB-spreads}a).  
The island of low fission 
barriers is seen near $Z\approx 108, N\approx 192$ in all functionals. Then another island of high 
fission barriers centered around $Z\approx 104, N\approx 216$ is formed. The highest fission 
barriers reaching $10-11$ MeV in the center of this island are predicted by the DD-ME2 functional 
(see Fig.\ \ref{FB-height}b). Somewhat lower fission barriers (with approximately 9 MeV height in 
the center of the island) are predicted by the  DD-PC1 functional (see Fig.\ \ref{FB-height}a).
Fission barriers with height of approximately 6 MeV appear in broad region of this island in the 
calculations with the NL3* and PC-PK1 CEDFs (see Figs.\ \ref{FB-height}c and d).

    With increasing neutron number the predictions start to differ substantially. The NL3*
and PC-PK1 functionals predict extremely low fission barriers with heights of around
2 MeV or less for the band of nuclei around $N\approx 240$ (see Figs.\ \ref{FB-height}c and d).
No such band is formed in the calculations with DD-ME2 and DD-PC1 functionals (see Figs.\ 
\ref{FB-height}a and b).  This could have a drastic impact on the creation of superheavy elements 
in the r-process  because the nuclear flow during most of neutron irradiation step of the 
r-process follows the neutron drip line and produces in tens of ms the heaviest drip line nuclei 
(see the discussion  in Sec. 4 of Ref.\ \cite{StefG.15}). However, this nuclear flow will most likely 
be terminated at $N\approx 240$ nuclei in the calculations with NL3* and PC-PK1 since fission 
will be  much faster than neutron capture.  Thus, no superheavy nuclei are expected to be 
formed  in the r-process calculations based on fission barriers obtained with these two functionals. 
This is similar to the results of the r-process simulations based on non-relativistic models (such as 
Skyrme DFT with HFB-14 functional, Thomas-Fermi (TF) model and Finite Range Droplet Model 
(FRDM)) with low fission barriers in the vicinity of neutron drip line \cite{PLMPRT.12,StefG.15}.  
On the contrary, the formation of superheavy elements in the r-process is more likely in the
calculations based on the DD-ME2 and DD-PC1 functionals since the ($Z,N$) region near 
neutron drip line is characterized by relatively high fission barriers (see Figs.\ \ref{FB-height}a 
and b). As a consequence, neutron capture is expected to proceed faster than fission and
nuclear flow during neutron irradiation step of the r-process could extend to higher proton
numbers \cite{PLMPRT.12,StefG.15}.  To a degree this is similar to the r-process simulations 
based on the models (such as Extended Thomas Fermi (ETFSI) model) with high fission 
barriers near neutron drip line (see Ref.\ \cite{PLMPRT.12}). 

Further increase of neutron number leads to a rise of the heights of primary fission 
barriers (PFBs) and the formation of the band of nuclei near shell closure at $N=258$ with 
the heights of  PFBs exceeding 6 MeV (see Fig.\ \ref{FB-height}b, c and d). In some nuclei 
they even reach 12-15 MeV in the calculations with PC-PK1 and NL3* CEDFs (see Fig.\ 
\ref{FB-height}c and d). Note that this process is suppressed in the calculations with 
CEDF DD-PC1 (see Fig.\ \ref{FB-height}a) because of reduced impact of the $N=258$
spherical shell gap on the ground state deformations (see Fig.\ \ref{deformations}a).
Nuclear landscape extends substantially beyond  $N=258$ in the calculations with NL3* 
and PC-PK1  (see Fig.\ \ref{FB-height}c and d). In this region we again see the alteration 
of the regions of low (near $N\approx 268$) and high (near $N\approx280$ and above) PFBs.  Note 
that in the $N>258$ region toroidal shapes \cite{Wong.73,AAG.18} could become the lowest in energy solutions. 
This has been verified in the calculations with DD-PC1 functional  in Ref.\ \cite{AATG.19} 
but has not been checked for other functionals.

\subsection{Theoretical uncertainties in primary fission barriers
and their sources}
\label{Theor-uncer-FB}

 The spreads $\Delta E^B$ of the heights of primary fission barriers as a function of proton 
and neutron numbers for four employed functionals are shown in Fig.\ \ref{FB-spreads}(a).
One can see that on the average they are moderate (on the level of 1.0-1.5 MeV) in the 
neutron poor region of actinides centered around $(Z\sim 100, N\sim 140$).
Then the spreads start to increase with a small island of high $\Delta E^B \approx 4.0$ 
MeV values seen in superheavy nuclei around $Z\approx 110, N\approx 164$.
Further increase of neutron number leads to the  band of 
high $\Delta E^B \approx  4.0$ MeV values for the nuclei with $N\approx 184$.
The sources of these uncertainties are directly related to the differences in the predictions of the ground 
state properties of these nuclei: spherical ground states are predicted in these nuclei in the 
calculations with PCPK1, DD-ME2 and NL3* functionals [see Figs.\ \ref{deformations}(b),(c) and (d)]  
while  the DD-PC1 functional prefers oblate ground states in superheavy $N\approx 184$
nuclei (see Fig.\ \ref{deformations}(a)).   After crossing  this band, theoretical uncertainties 
in the heights of primary fission barriers substantially decrease and for the majority of the 
nuclei located inside the triangle with the sides $N=188$ [for $Z=96-120$], $Z=120$ [for $N=188-240$]
and [$Z=96,N=188$ to  $Z=120, N=240$] 
they are in general  better than 2.5 MeV and in many nuclei they are 
even better than 1.5 MeV.  However, the $\Delta E^B$ spreads start to increase again on approaching
two-neutron drip line. Here they form quite wide band of the nuclei parallel to two-neutron drip 
line in which   $\Delta E^B$ is close to 4.0 MeV. Even higher spreads reaching 5.5 MeV are 
seen near shell closure at $N=258$. 

  The analysis of the spreads $\Delta E^B$ allows to identify major sources of theoretical
uncertainties in the predictions of the heights of PFBs. These could be reduces to two major 
contributors, namely, underlying single-particle structure mostly affecting the ground state 
properties and nuclear matter properties of employed CEDFs. To facilitate the discussion we 
will consider the $\Delta E^B$ spreads for the pairs of selected functionals.

  The lowest spreads exist for the pair of the NL3* and PC-PK1 functionals (see Fig.\ 
\ref{FB-spreads}c):   $\Delta E^B\leq 0.5$ MeV for absolute majority of the nuclei  and 
only in specific regions of nuclear chart it is higher.  Even in those regions it is higher than 
1 MeV only for limited set of the nuclei.  These regions are:  (i) the actinides around 
$N=126$,  (ii) the $Z=90$ and 92 actinides with $N\approx 170$, (iii) superheavy nuclei in the vicinity of the $Z=120$ and
$N=184$ lines, (iv) very neutron-rich nuclei in the vicinity of two-neutron drip line and
(v) the band of the nuclei around $N\approx 246$.  
The differences in the predictions of the heights of outer fission barriers are responsible for 
the spreads in the region (i). At present, their source is not clear. The spreads  $\Delta E^B$ seen 
in the region (ii) are due both to different proton and neutron dependencies of the impact of octupole 
deformation on outer fission barriers in the NL3* and PC-PK1 functionals
(compare panels (c) and (d) of Fig.\ \ref{out-vs-in-bar}) and 
the fact that in some nuclei we compare the heights of outer and inner fission barriers.
In the region (iii), the large $\Delta E^B$  values are due to the differences in the 
predictions of the spherical shell closures at $Z=120$ and $N=184$ and the densities of 
the single-particle states in their vicinities (see Fig. 1 in Ref.\ \cite{AANR.15} and the 
discussion in this reference). Slightly different isovector properties of the NL3* and
PC-PK1 functionals (see Table \ref{Table-mass-SNMP}) may be responsible for the 
divergence of their predictions in the region (iv). The large  $\Delta E^B$  values 
in the region (v) are due to prolate-oblate-spherical shape coexistence which takes place 
in  slightly different regions of the $(Z,N)$ chart in these two functionals [compare Figs.\ 
\ref{deformations} (c) and (d)]. 

 The comparison of the predictions of the NL3* and PC-PK1 functionals for the fission
barriers is quite illuminating since it shows in a global way a number of important features.
First, apart of the regions (i)-(v)  the comparable (typically within 0.5 MeV)  predictions for the 
heights of PFBs are obtained on a global scale by these two functionals despite the fact
that they differ substantially in the predictions of the ground states energies in neutron-rich 
nuclei (see the discussion of Fig.\ \ref{energy-spreads}). Thus, the description of the ground 
state energies is to a degree decoupled from the description of fission barriers; the latter
depends on the relative energies of the saddle and the ground state. As a consequence, good
description of the ground state energies does not guarantee good description of the fission
barriers and vice versa. Second, the differences in the predictions of the PFB heights seen in 
the regions (iii) and (v) are related to the differences in the predictions of the ground states properties,
which in turn are defined by the differences in the underlying single-particle structure. Third,
if to exclude the regions (i)-(v) from consideration it becomes clear that some differences in 
nuclear matter properties such as the symmetry energy $J$ and its slope $L_0$ (see 
Table \ref{Table-mass-SNMP}) do not lead to important differences in the predictions for
PFBs. Fourth, comparable global predictions for the PFBs are obtained despite underlying 
differences in the basic structure of the functionals and their fitting protocols. The NL3* functional
includes meson exchange of finite range, while PC-PK1 CEDF does not have mesons and
thus it has zero range interactions (see Ref.\ \cite{TAAR.20}). The fitting protocol of the CEDF NL3* is based on 
12 spherical nuclei and includes empirical data on nuclear matter properties (see Ref.\ 
\cite{NL3*}),  while the one for PC-PK1 includes only data on binding energies (60 spherical nuclei) and 
charge radii (17 spherical nuclei) \cite{PC-PK1}.  Note that the NL3* and PC-PK1 functionals
have 6 and 9 parameters, respectively. However, the analysis of parametric correlations
shows that in reality there are only 5 and 6 independent parameters in these two 
functionals \cite{AAT.19,TAAR.20}.

   Next we consider the spreads $\Delta E^B$ obtained with the NL3*/DD-ME2 pair of the
functionals (see Fig.\ \ref{FB-spreads}d). These two functionals have almost identical fitting
protocols (see Refs.\ \cite{NL3*,DD-ME2}). The only difference is the fact that DD-ME2  fitting 
protocol uses 3 experimental data points on neutron skins as compared with 4 in NL3* but the 
impact of this difference is expected  to be small. Thus, larger values of the $\Delta E^B$ 
spreads in the NL3*/DD-ME2 pair as compared with the ones in the NL3*/PC-PK1 pair are related to 
the  basic difference of these two functionals, namely, to the implementation of density
dependence. The DD-ME2 functional has explicit dependence of the meson-nucleon 
coupling on the nucleonic density, while NL3* employs cubic and quartic powers of the 
$\sigma$ meson for density dependence (see Sect. II of Ref.\ \cite{AARR.14} for details). 
In addition, the nuclear matter properties (in particular, the symmetry energy $J$ and its slope 
$L_0$) of these two functionals differ substantially (see Table \ref{Table-mass-SNMP})
and this difference is expected to contribute into the increase of the spreads $\Delta E^B$ 
obtained for the NL3*/DD-ME2 pair as compared with those for the NL3*/PC-PK1
pair.

   Finally, the $\Delta E^B$ spreads for the DD-PC1/DD-ME2 pair of the functionals are
presented in Fig.\ \ref{FB-spreads}b. Nuclear matter properties of these two functionals 
are close to each other and they are located within the limits of the SET2b constraint set 
on the experimental/empirical ranges for the quantities of interest derived in Ref.\ \cite{RMF-nm}
(see Table \ref{Table-mass-SNMP}). However, fitting protocols of these two functionals are 
drastically different: CEDF DD-ME2
is fitted to the properties of 12 spherical nuclei (see Ref.\ \cite{NL3*}) while DD-PC1 is 
defined by the properties of 64 deformed rare-earth nuclei and actinides (see Ref.\ \cite{DD-PC1}). 
As a result of this difference in fitting protocols, the largest $\Delta E^B$ spreads appear in the 
vicinity of spherical shell closures at $N=184$ (with $\Delta E^B$ reaching 4.0 MeV) and 
$N=258$ (with $\Delta E^B$ reaching 5.5 MeV) [see Fig.\ \ref{FB-spreads}b].  Indeed, the 
impact of these shell closures on the equilibrium deformation differs substantially  in these 
two functionals (compare Figs.\ \ref{deformations}(a) and (b) and see Fig.\ \ref{deformation-spreads}(b)) 
and this is a reason for increased $\Delta E^B$ spreads.

\subsection{The comparison with the results obtained in non-relativistic calculations}

  It is interesting to compare the global trends of the heights of PFB in the $(Z,N)$ plane
obtained in the RHB calculations (see Fig.\ \ref{FB-height}) with those obtained in earlier
non-relativistic calculations for which the maps similar to those presented in Fig.\ 
\ref{FB-height} are available. Note that similar to our calculations all these non-relativistic 
calculations have been performed only for axial nuclear shapes.

  Fission barriers obtained in Gogny DFT calculations with D1M* functional are presented 
in Fig. 12 of Ref.\ \cite{RHR.20}. These calculations cover the region from two-proton drip 
line up to the nuclei with two-neutron separation energies of $S_{2n}=4.0$ MeV. In these
calculations, the fission barriers of the $N\leq186$ nuclei typically exceed 6 MeV 
and in a number of these nuclei their heights are close to 12 MeV. Then fission barriers 
in the $N\approx 190-210$ nuclei are lower than 4 MeV but they increase to 
approximately 8 MeV on approaching $S_{2n}=4.0$ MeV line.  The differences in 
the predictions of the heights of PFB obtained in the CDFT and Gogny DFT calculations
(compare Fig.\ \ref{FB-height} with Fig. 12 of Ref.\ \cite{RHR.20}) are in part related 
to the differences in the predictions of ground state properties (compare Fig.\ 
\ref{deformations} in the present paper with Fig.\ 5 in Ref.\ \cite{RHR.20}).

  Our results for fission barriers (Fig.\ \ref{FB-height}) could also be compared with those 
obtained in non-relativistic DFTs with the BCPM and HFB14 functionals and FRLDM 
(see Fig.\ 7 in Ref.\ \cite{GPR.18}). The calculations with HFB14
predict very low fission barriers
(with $E^B <4$ MeV)  for the  $Z\geq 110$ nuclei with exceptionally low fission barriers
($E^B <2$ MeV) in many nuclei located in the $N\approx 184-210$ region (see middle 
panel  of Fig.\ 7 in Ref.\ \cite{GPR.18} and Fig.\ 13 in Ref.\ \cite{AG.20}). The RHB 
calculations predict in general higher fission
barriers (as compared with HFB14 ones), but similar island of low fission barriers is seen 
near $Z\approx 108, N\approx 192$ in all functionals (see Fig.\ \ref{FB-height}). However, 
this island is narrower as compared with the HFB14 one. Fission barriers obtained with the BCPM 
functional and FRLDM are somewhat higher than those obtained with HFB14 
(compare top and bottom panel with middle panel of Fig. 7 in Ref.\ \cite{GPR.18}). 
However, they share the same general structure in the $(Z,N)$ plane. 

  Fission barriers calculated in the DFT framework with Skyrme SLy6, SkI3, SV-min and SV-bas
functionals are presented in Fig.\ 5 of Ref.\ \cite{ELLMR.12}. Unfortunately, the colormap used 
in this figure does not allow to extract the details in the most interesting energy range of $6-10$ 
MeV\footnote{Better colormap for the fission barrier height distribution in the $(Z,N)$ plane
obtained with the Skyrme SV-min functional is used in Fig. 4 of Ref.\ \cite{Reinh.18}.}. However, the 
region of low fission barriers (with $E^B < 4$ MeV) similar  to that discussed above appear in all 
functionals for $N\approx 190-210, Z\approx 94-120$. Fission barriers obtained in the TF and
ETFSI models for the $(Z=84-120, N=140-236)$ and $(Z=84-115, N=140-216)$ regions of nuclear
chart are 
presented in Fig.\ 2 of Ref.\ \cite{PLMPRT.12}. Both of these models show the island of low fission
barriers centered around $Z\approx 110, N\approx 192$.  In general, the ETFSI results are close
to above mentioned results obtained with Skyrme EDFs.

\section{Conclusions}
\label{sect-concl}

  The systematic investigation of the ground state and fission properties of 
even-even actinides and superheavy nuclei with $Z=90-120$ from the 
two-proton up to two-neutron drip lines has been performed for the first time 
in the framework of covariant density functional theory.  Four state-of-the-art 
CEDFs such as DD-PC1, DD-ME2, NL3* and PC-PK1  are used in this study. 
They represent the major classes of the CDFT models which differ by basic 
assumptions and fitting protocols. This allows a proper assessment 
of systematic  theoretical uncertainties for physical observables of interest. 
Obtained results provide a necessary  theoretical input for the r-process 
modeling in heavy nuclei and, in particular, for the study of fission cycling. 
The main results can be summarized  as follows:

\begin{itemize}

\item 
 Quadrupole deformations of calculated ground states and related
theoretical uncertainties have been investigated. It turns out that four 
employed functionals predict very similar  deformations for the majority
of the nuclei.  However, large theoretical uncertainties in quadrupole 
deformation exist for some nuclei but  they are  well localized in the 
$(Z,N)$ plane.  These uncertainties are mostly due to the
uncertainties in the predictions of the underlying single-particle structure.
They are dominated by the uncertainties in the predictions of both spherical
shell closures at $N=184$, $N=258$  and $Z=120$ (for the 
$N<190 $ nuclei) and deformed single-particle structures  leading to somewhat 
different boundaries in the $(Z,N)$ plane between the regions with
oblate and prolate shapes. The differences in nuclear matter properties
of employed functionals have only marginal impact on theoretical 
uncertainties related to calculated quadrupole deformations.

\item 
   Theoretical uncertainties $\Delta E(Z,N)$ in binding energies,
emerging from poorly defined isovector properties of CEDFs,
increase drastically when approaching the neutron drip line and in 
some nuclei they reach 50 MeV. However, they reduce substantially 
(down to  maximum value of $\Delta E(Z,N) \approx 21$ MeV) when
PC-PK1 functional is removed from consideration.  Two-neutron 
drip line of this functional is also located at substantially higher neutron
numbers as compared with the ones obtained with other functionals.
In addition, this functional is the major source of theoretical uncertainties 
in two-neutron separation energies.  Parametric correlations
leading to an over-parametrization of the isoscalar channel is 
a possible reason for such a unique behaviour of the CEDF PC-PK1.
Theoretical uncertainties in two-neutron separation energies 
reveal clear importance of the uncertainties in the $N=184$
and $N=258$ spherical shell closures and in the location of
the boundaries between the regions of prolate and oblate 
shapes.

\item
  $\alpha$-decay properties, such as the $Q_{\alpha}$ values and the lifetimes 
$\tau_{\alpha}$, and related theoretical uncertainties have been investigated employing 
four empirical formulas and four CEDFs.  While the predictive power of the models is 
relatively high on proton-rich side of nuclear chart, it starts to deteriorate  on approaching 
$N=184$.  It is especially low in the nuclei around $Z\approx 108, N\approx 198$ [the 
region of the transition from prolate to oblate ground states] and in very neutron-rich nuclei 
located  in the vicinity of two-proton drip line. However, the uncertainties in the latter
region are not very important since the $\alpha$-decay in these nuclei is not expected
to play any role in the r-process because of extremely large lifetimes.
  
\item
The distributions of the primary fission barriers in the $(Z,N)$ plane have been investigated with
four employed CEDFs. Globally, the highest fission barriers are produced by the CEDF 
DD-ME2, and the lowest ones by the NL3* and PC-PK1 functionals.  The results obtained 
with DD-PC1 are located between these two extremes but closer to the DD-ME2 ones.
The presence of the band of nuclei with $N\approx 240$ in the $(Z,N)$ plane with low fission 
barriers,  obtained in the calculations with the NL3* and PC-PK1 functionals,  could have a 
drastic impact on the creation of superheavy elements in the r-process.  The nuclear flow 
during most of neutron irradiation step of the r-process  follows the neutron drip line and 
produces in tens of ms the heaviest drip line nuclei. However,  this nuclear flow will most
likely be terminated at $N\approx 240$ nuclei since fission will be much faster than 
neutron capture.  On the contrary, the formation of superheavy elements in the r-process is 
more likely in the calculations based on the DD-ME2 and DD-PC1 functionals since the 
$(Z, N)$ region near neutron drip line is characterized by relatively high fission barriers 
and the band of nuclei with low fission barriers (similar to the one at $N\approx 240$ in the
NL3* and PC-PK1 functionals) is absent.

\item
   There are two major sources of theoretical uncertainties in the predictions of the heights 
of  PFBs, namely, underlying single-particle structure mostly affecting the ground state 
properties and nuclear matter properties of employed CEDFs.  For example, the 
increase  of  theoretical uncertainties for the ground states of the nuclei in the vicinity 
of the $N=184$  and  $N=258$ spherical shell closures leads to an increase of theoretical uncertainties 
for their fission barriers.  The functionals with nuclear matter properties located in the
vicinity of empirical SET2b estimates \cite{RMF-nm}
tend to produce higher fission barriers as compared 
with the predictions of the functionals the nuclear matter properties of which are located
outside the limits of the SET2b constraint set.  The problem of finding
the best functional for the description of fission barriers is further complicated   by the 
fact that the description of the ground state energies  is to a degree decoupled from 
the description of fission barriers; the latter depends on the relative energies of the 
saddle and the ground state.  As a consequence, good description of the ground state 
energies does not guarantee good description of the fission barriers and vice versa.

\end{itemize}

This is first ever systematic attempt within the covariant density functional theory 
to provide both the input for the r-process calculations which includes the ground 
state and fission properties of actinides and superheavy nuclei and the
assessment of systematic theoretical uncertainties in the physical quantities
of interest.  As such it follows the ideology of all previous non-relativistic calculations 
of relevance for the r-process of heavy and superheavy nuclei, which depend
also on the fission processes, and assumes the axial symmetry of nuclei. This is a 
reasonable approximation for the ground state properties of the majority of 
nuclei; the only exception is transitional nuclei which are soft in 
$\gamma$-deformation. However, the restriction to axial symmetry leads to the 
fact that the calculated inner and outer  fission barriers represent the upper limits and can 
be potentially lowered when the triaxiality is taken into account.  The 
r-process simulations with the data obtained in this study will allow to limit the region of 
the $(Z,N)$ plane which has an impact on this process.  The hope is that for this limited 
set of nuclei, systematic refined  calculations taking into account the dynamical correlations 
and the triaxiality in the calculations of the part of the $(\beta_2, \gamma)$-plane covering
ground state, inner fission barrier and second minimum as well as triaixiality and octupole
deformation in the calculations of the part of the $(\beta_2, \beta_3, \gamma)$ plane
covering  second minimum, outer fission barrier and the region beyond that will be 
possible in the era of  exascale computing.
 
  Underlying single-particle structure and nuclear matter properties of CEDFs
emerge  as the major sources of theoretical uncertainties. However, they affect different physical 
observables in a different way. For example,  theoretical uncertainties in the ground state quadrupole
deformations are defined mostly by the uncertainties in the underlying single-particle    
structure. On the contrary, both factors contribute into theoretical uncertainties for
fission barriers.  The existence of appreciable  theoretical uncertainties in the 
ground state and fission properties calls for a better covariant energy density 
functionals.  The reduction of parametric correlations between the parameters
of CEDFs is one possible way in that direction \cite{TAAR.20,NIV.17}.
In addition, experimental studies of superheavy elements in the 
vicinity of the $Z=120$ and $N=184$ lines, planned at new facilities such as 
SHE factory \cite{OD.16}, will hopefully provide critical data which will allow to discriminate 
the predictions of different models.  Such information could be used for a better
constraint of the CEDFs and thus to the reduction of substantial theoretical 
uncertainties in this region of nuclear chart which affect all physical 
observables of interest and have a direct impact on the modeling of the r-process.

\section{Acknowledgement}

This material is based upon work supported by the US Department of Energy, 
National Nuclear Security Administration under Award No. DE-NA0002925, by 
the US Department of Energy, Office of Science, Office of Nuclear Physics under 
Award No. DE-SC0013037 and by Ghana Atomic Energy Commission, National
Nuclear Research Institute, Ghana.

\appendix

\section{Neglect of dynamical correlations in the fission barrier calculations}


   Some of non-relativistic calculations, mentioned in the last paragraph of Sec.\ \ref{sect-theory}, 
take into  account dynamical correlations, but others not. Dynamical correlations are not taken 
into account  in our calculations due to following reasons.

  First, the analysis performed in Ref.\ \cite{SALM.19} in the CDFT-based approach indicates that 
in most of the nuclei dynamical correlations
modify fission barrier heights by less than 1 MeV
but increase substantially computational time.  The only exceptions are the nuclei with soft potential 
energy surfaces the ground state energy minimum of which is located at spherical shape.  Note that 
the absolute majority of the nuclei under consideration are deformed in the ground states (see 
Sect. \ref{sect-ground}). Thus, the errors introduced into the fission barrier heights due to 
neglect of dynamical correlations are expected to be  smaller than the ones which are coming from 
the selection of CEDFs (see Sect.\ \ref{Theor-uncer-FB} below).  In addition, theoretical uncertainties 
in fission barrier heights defined by their spreads (see Eq.\  (\ref{spread})) are not expected to be modified 
much by the neglect of dynamical correlations.  This is 
because in the majority of the cases the topology of potential energy surface  of a given nucleus weakly 
depends on the underlying functional (see, for example, Figs. 7 and 8 in Ref. \cite {AARR.17}, Fig. 8  
in Ref.\ \cite{AAR.12} and supplemental material to Ref.\ \cite{SALM.19}).  As a consequence, dynamical 
correlations are expected to be comparable for different functionals and they will at least partially cancel 
each other in Eq.\ (\ref{spread}). 

  Second, the inner fission barriers are lowered when the triaxiality is taken 
into account (see Ref.\ \cite{AAR.10} and references quoted therein) and the potential energy surfaces of 
the ground states in many superheavy nuclei are soft in $\gamma$-deformation (see Ref.\ \cite{CHN.05}
and Appendix \ref{Triax-gs-fb}.)  In the CDFT  calculations, the outer fission barriers can also be affected by 
triaxial deformation\footnote{The important role of triaxiality in the description of outer fission barriers of 
actinides has also been discussed in the framework of microscopic+macroscopic approach in Ref.\ 
\cite{DPB.07}.} via  two mechanisms. In the first one, the saddle of reflection-symmetric 
triaxial fission path becomes lower in energy than the saddle of  reflection-asymmetric (octupole-deformed) 
axial fission path due to underlying shell structure \cite{AAR.12}. In the second mechanism, 
reflection-symmetric fission path and its saddle loose their axial symmetry and attain some degree of 
triaxiality  \cite{LZZ.12,LZZ.14}. The investigations of the impact of dynamical correlations on fission 
barriers in triaxial calculations are very rare and  quite limited in coverage. In non-relativistic frameworks, 
only limited set of actinides  \cite{SDNSB.14,BMIQ.17,ZLNVZ.16,DGGL.06}
and  superheavy \cite{GSPS.99}
nuclei have been studied so far.   The impact of dynamical correlations on  fission barriers of restricted 
set of superheavy nuclei along the $Z=120$ isotopic and $N=174,184$ isotonic chains has been studied 
in the CDFT-based framework in Ref.\ \cite{SALM.19}.

   Third, dynamical calculations do not provide a unique answer because of underlying
assumptions and approximations 
\cite{SDNSB.14,ZLNVZ.16,SR.16,GR-plb.18,GPR.18,GMNORSSSS.19,RHR.20}. 
For example, there exist substantial differences 
between the predictions based on Adiabatic Time Dependent HFB (ATDHFB) and 
the Generator Coordinate method (GCM) [based on Gaussian  Overlap  Approximation 
(GOA)] schemes \cite{SR.16,GR-plb.18,GPR.18,RHR.20}.  The differences between spontaneous fission half-lives 
$\tau_{SF}$ obtained in these two schemes could reach many orders of magnitude and
increase with the decrease of the fissility-related parameter $Z^2/A$   [which is equivalent to 
the increase of neutron number within a given isotopic chain] (see Fig.\ 2 in Ref.\ \cite{RHR.20}). 
Large differences between  experimental and calculated  $\tau_{SF}$ also exist; for example, in 
the U isotopes these differences reach almost 20 orders of magnitudes when ATDHFB values 
for  $\tau_{SF}$  are used \cite{RHR.20}. As illustrated in Refs.\ \cite{SDNSB.14,ZLNVZ.16} on the
example of the $^{250,264}$Fm and $^{240}$Pu nuclei, the inclusion of pairing fluctuations within a least
action approach improves the agreement between the predicted $\tau_{SF}$ values 
and experiment. However, it remains to be seen whether that is a general conclusion
applicable to all nuclei. In addition, such calculations are prohibitively expensive (in part, 
because of breaking of axial symmetry) and thus are not scalable to global calculations.

  In addition, the treatment of the ground state  energy $E_0$ (which is also tunneling energy for 
fission) relies on simplified approximations  in the majority of the publications  (see discussion in 
Ref.\ \cite{SALM.19}).  In microscopic calculations, tunneling energy is associated with the energy 
of collective ground state defined either in GCM \cite{SEKRH.09} or in five-dimensional collective 
Hamiltonian  (5DCH) \cite{SALM.19}. The energy of collective ground state depends on softness 
(both in quadrupole deformation $\beta_2$ and in triaxial deformation $\gamma$) of collective 
energy surface  in the  vicinity of ground  state minimum. It differs from approximate values 
substantially \cite{SALM.19}; this could modify calculated $\tau_{SF}$ by several orders of magnitude
\cite{GSPS.99,SALM.19}.
These extremely large theoretical uncertainties in $\tau_{SF}$, coming from the selection of the 
method (ATDHFB versus GCM+GOA) and the treatment of the ground state energy, are the reasons 
why we have not attempted to calculate  spontaneous fission half-lives in the present  paper.

 Fourth, there are some indications that the role of triaxiality can be reduced in dynamical calculations 
for some nuclei. For example, it was shown in Ref.\ \cite{SDNSB.14} based on least-action calculations 
with Skyrme EDF SkM* that pairing fluctuations act in the direction of restoration of axial symmetry along the 
fission path in the $^{240}$Pu nucleus. This nucleus is characterized by relatively modest decrease 
(approximately 2 MeV) of inner fission barrier height by triaxiality in static calculations. Similar results
have also been obtained for $^{250,264}$Fm with similar formalism based on DD-PC1 CEDF in 
Ref.\ \cite{ZLNVZ.16}.  The  calculations of Refs.\ \cite{DGGL.06,GSPS.99} based on least-action 
principle also indicate that the axial symmetry of fission pathway is restored in many nuclei; they are 
based on the DFT approach with Gogny D1S functional \cite{DGGL.06} and on macroscopic+microscopic 
method \cite{GSPS.99}. 

   However, not in all nuclei the effect of triaxiality is eliminated by least-action principle. For 
example, the least-action fission pathway in $^{264}$Fm is still  characterized by triaxiality 
(although it is somewhat reduced by enhanced pairing as compared with static fission path) 
in the calculations based on Skyrme SkM* functional \cite{SDNSB.14}. 
Note that in this 
nucleus the triaxiality has a large impact (slightly more than 4 MeV) on the height 
of inner fission barrier in static calculations. The calculations of Ref.\ \cite{DGGL.06} 
performed with Gogny D1S functionals also indicate that in some nuclei the 
least-action fission pathway goes through triaxial saddles.
More systematic 
calculations\footnote{These calculations are simplified as compared with quoted DFT 
calculations since they use fixed single-particle spectrum for all nuclei and neglect
the deformations of higher order such as $\beta_6$ and $\beta_8$.} based
on macroscopic+microscopic method show that the impact of the triaxiality on
the least action fission pathway (and thus on spontaneous fission half-live)
shows up in some nuclei with $Z=114$ and becomes much more pronounced 
in the $Z\geq 120$ nuclei \cite{GSPS.99}.  Note that the tendency towards restoration 
of axial symmetry of the fission pathway in the least-action calculations
may somewhat be underestimated in Refs.\ \cite{GSPS.99,DGGL.06} because of the 
neglect of pairing fluctuations.

   The analysis of these publications suggests two possible situations in which the 
   least-action fission path will most likely
be characterized by triaxiality. In the first one, the decrease of the fission barrier by 
triaxiality in static calculations is substantial being on the order of $3-4$ MeV 
\cite{GSPS.99,SDNSB.14}.  In the second one, the ground state is oblate (or 
possibly soft in oblate-prolate direction \cite{SALM.19}) so that the fission path 
across the $\gamma$-plane is shorter than the one along the 
$\gamma = 0^{\circ}$ line \cite{GSPS.99}.  As discussed in Sec. \ref{sect-ground} 
and Appendix \ref{Triax-gs-fb} only limited number of nuclei satisfy such conditions.  Thus, 
the restriction to axial symmetry should be considered as a reasonable first 
approximation. However, one should keep in mind that the values obtained for
fission barriers represent upper limits since their possible lowering due to
triaxiality is neglected.

\section{Possible impact of triaxiality on inner fission barriers}
\label{Triax-gs-fb}

\begin{figure*}[htb]
\centering
\includegraphics[width=4.3cm]{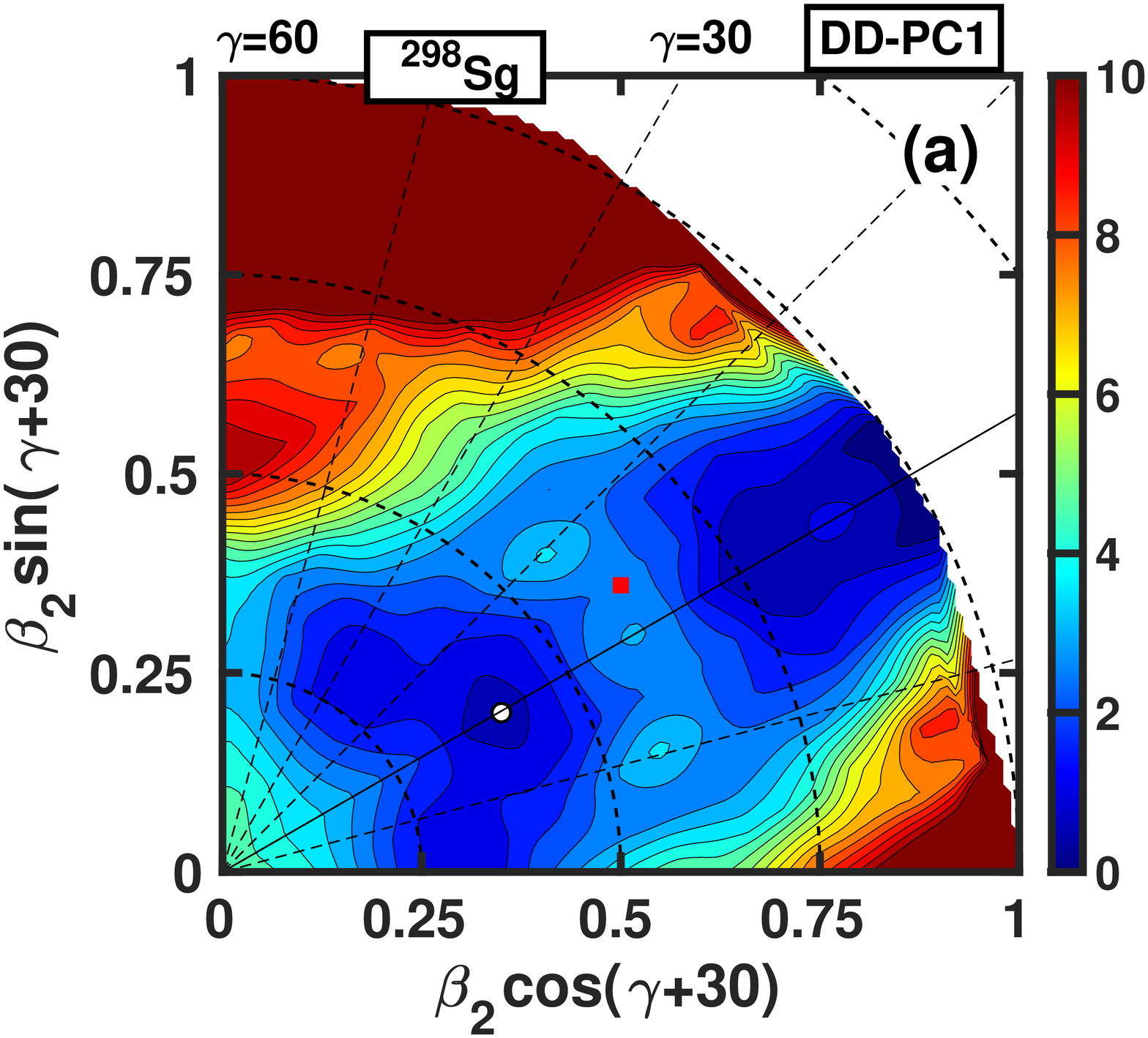}
\includegraphics[width=4.3cm]{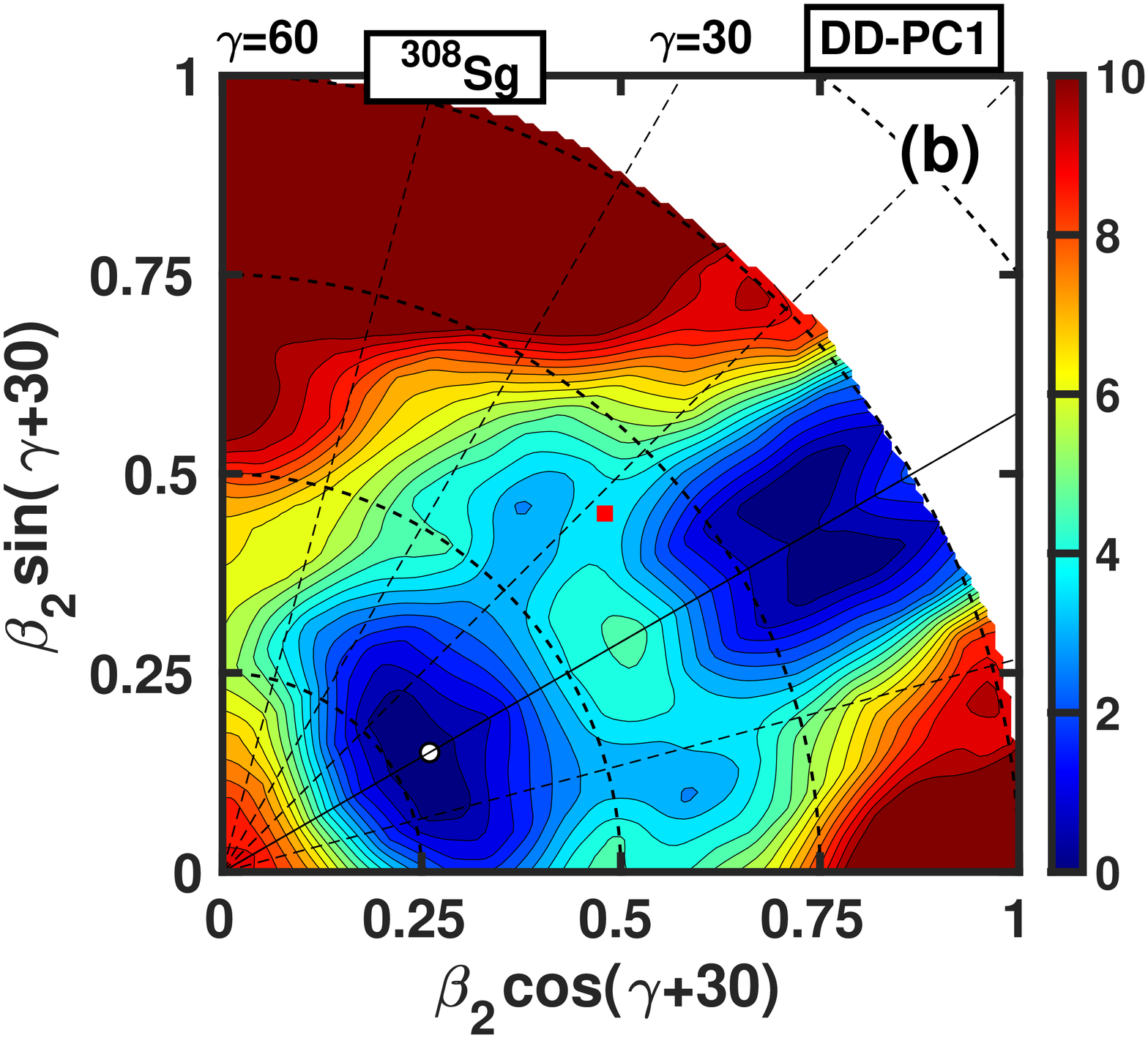}
\includegraphics[width=4.3cm]{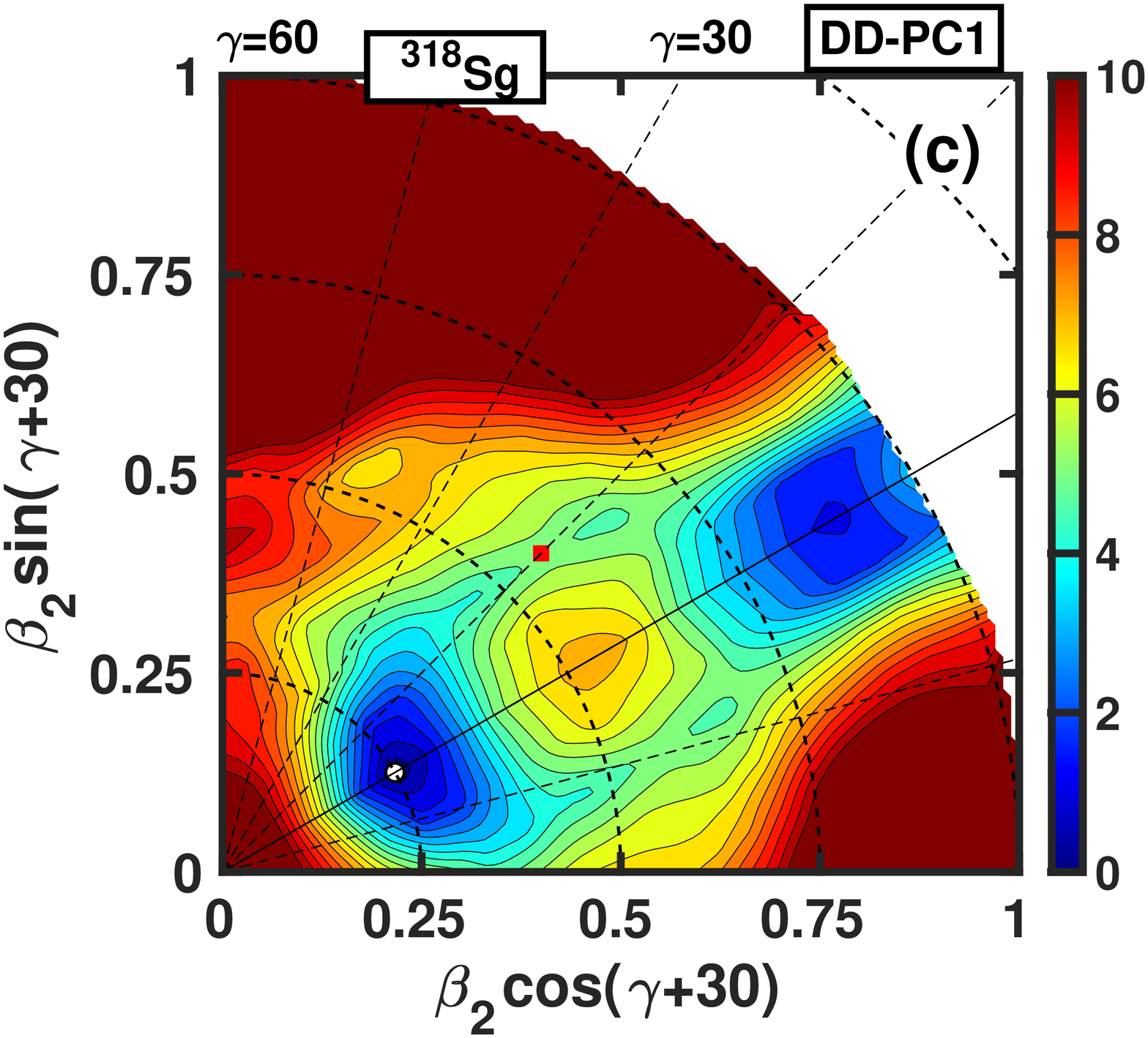}
\includegraphics[width=4.3cm]{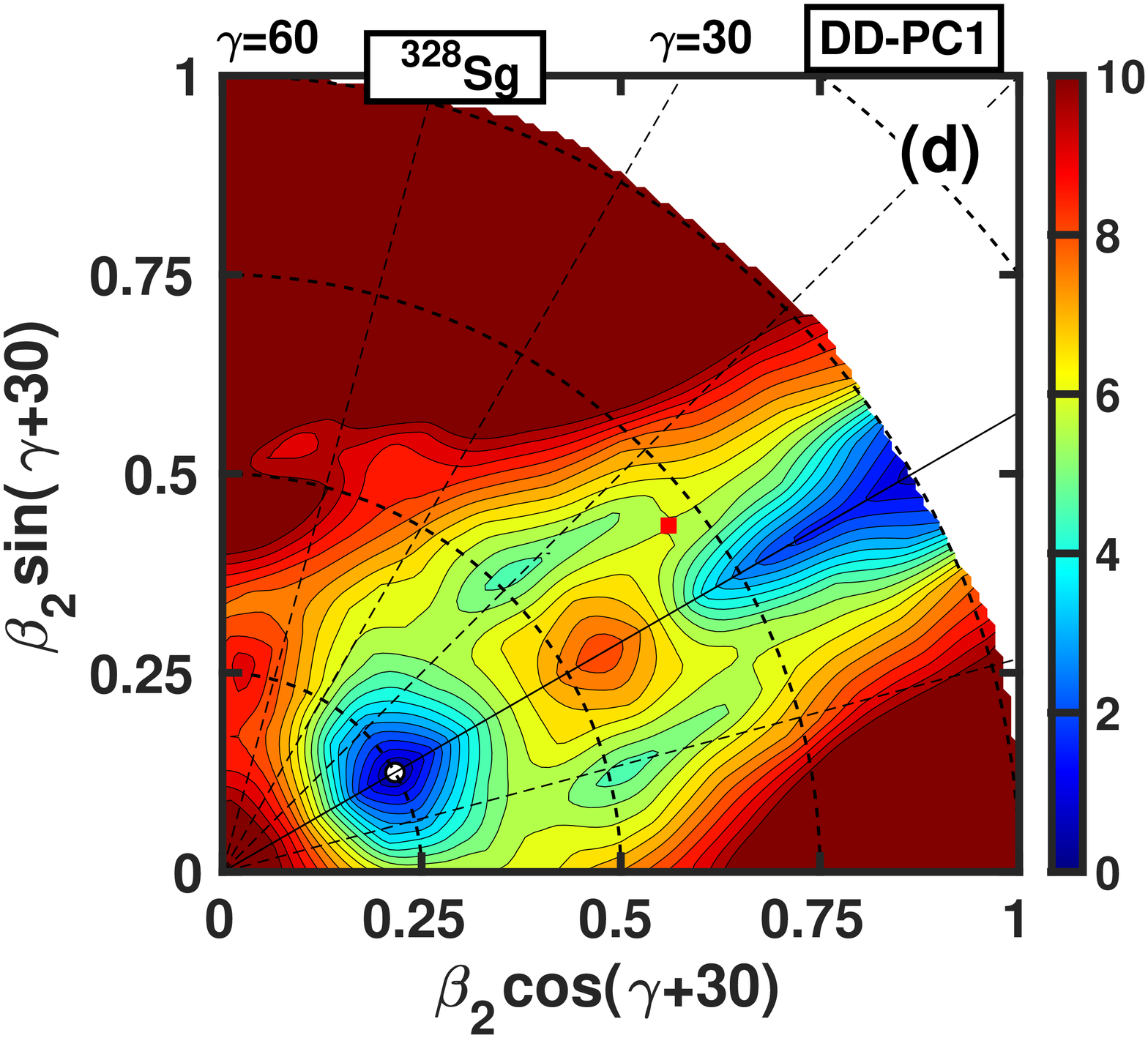}
\includegraphics[width=4.3cm]{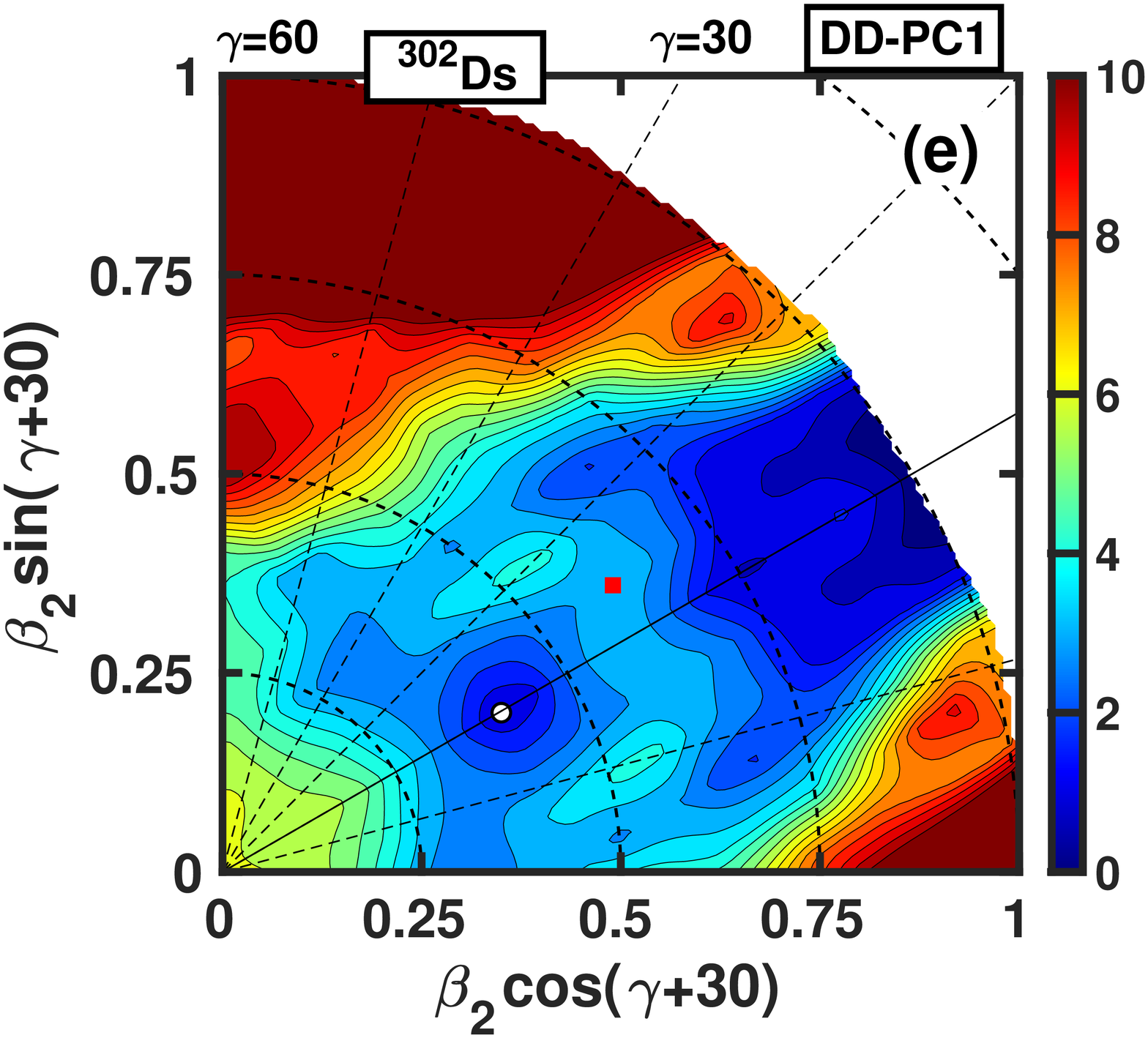}
\includegraphics[width=4.3cm]{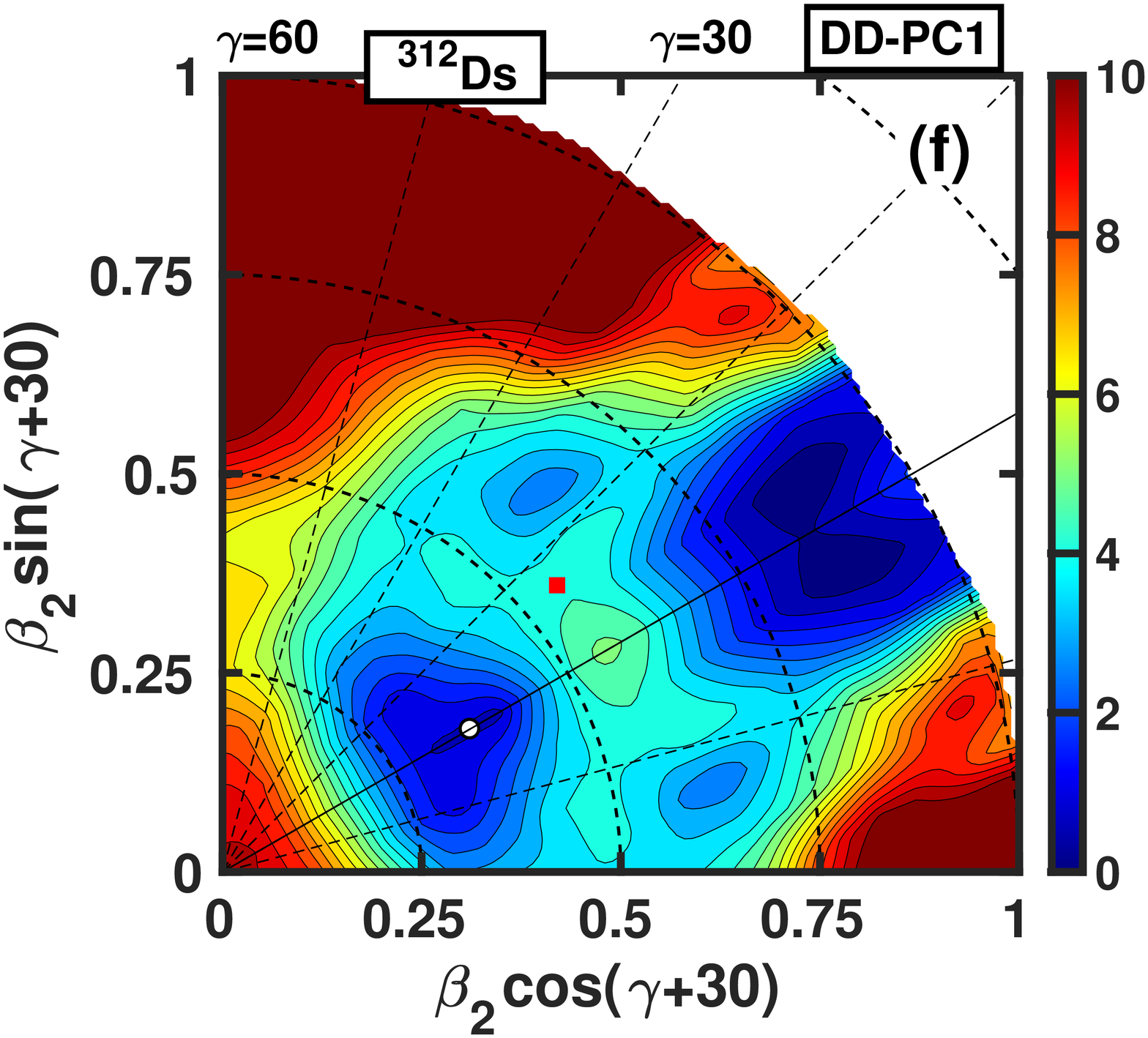}
\includegraphics[width=4.3cm]{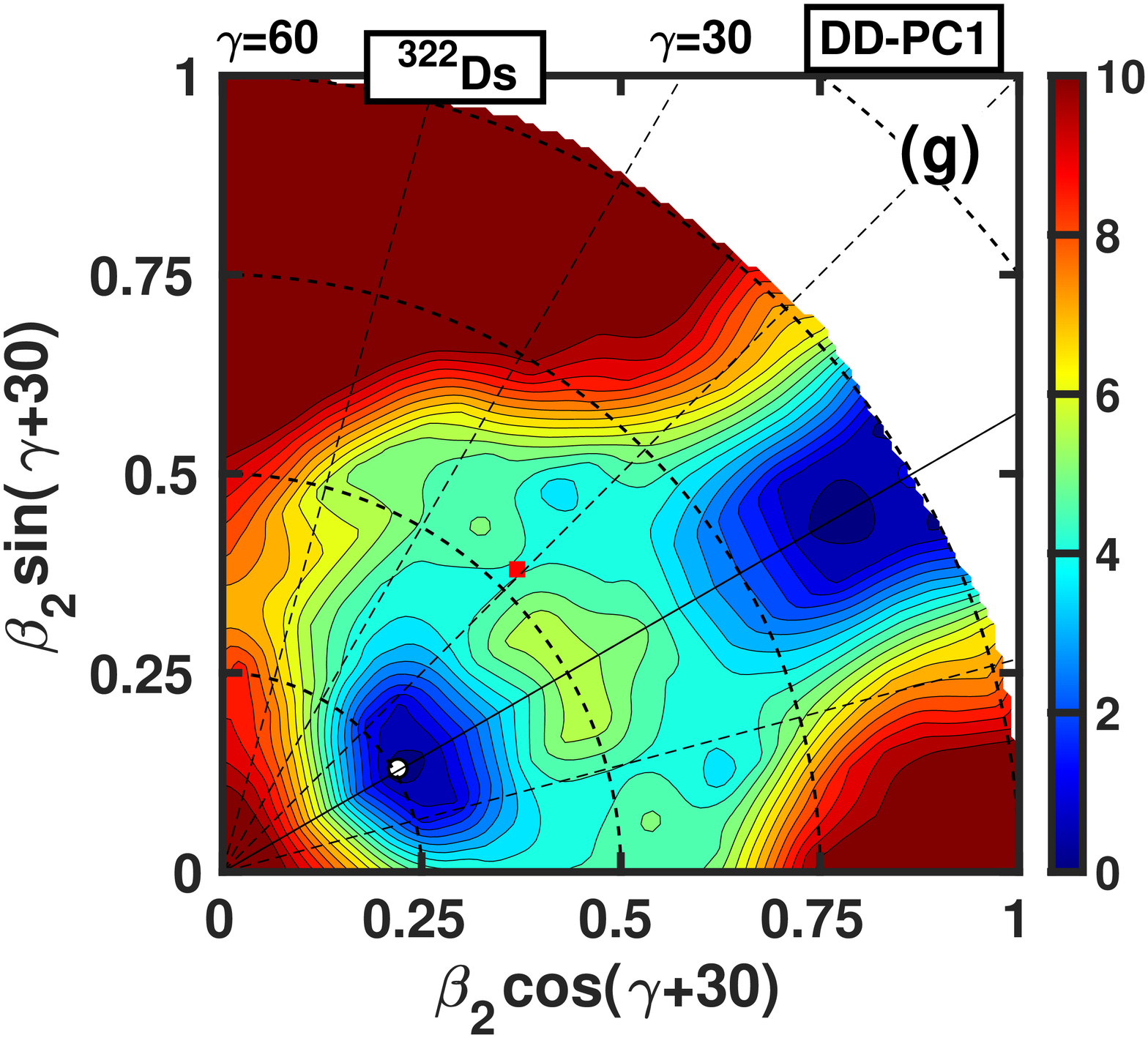}
\includegraphics[width=4.3cm]{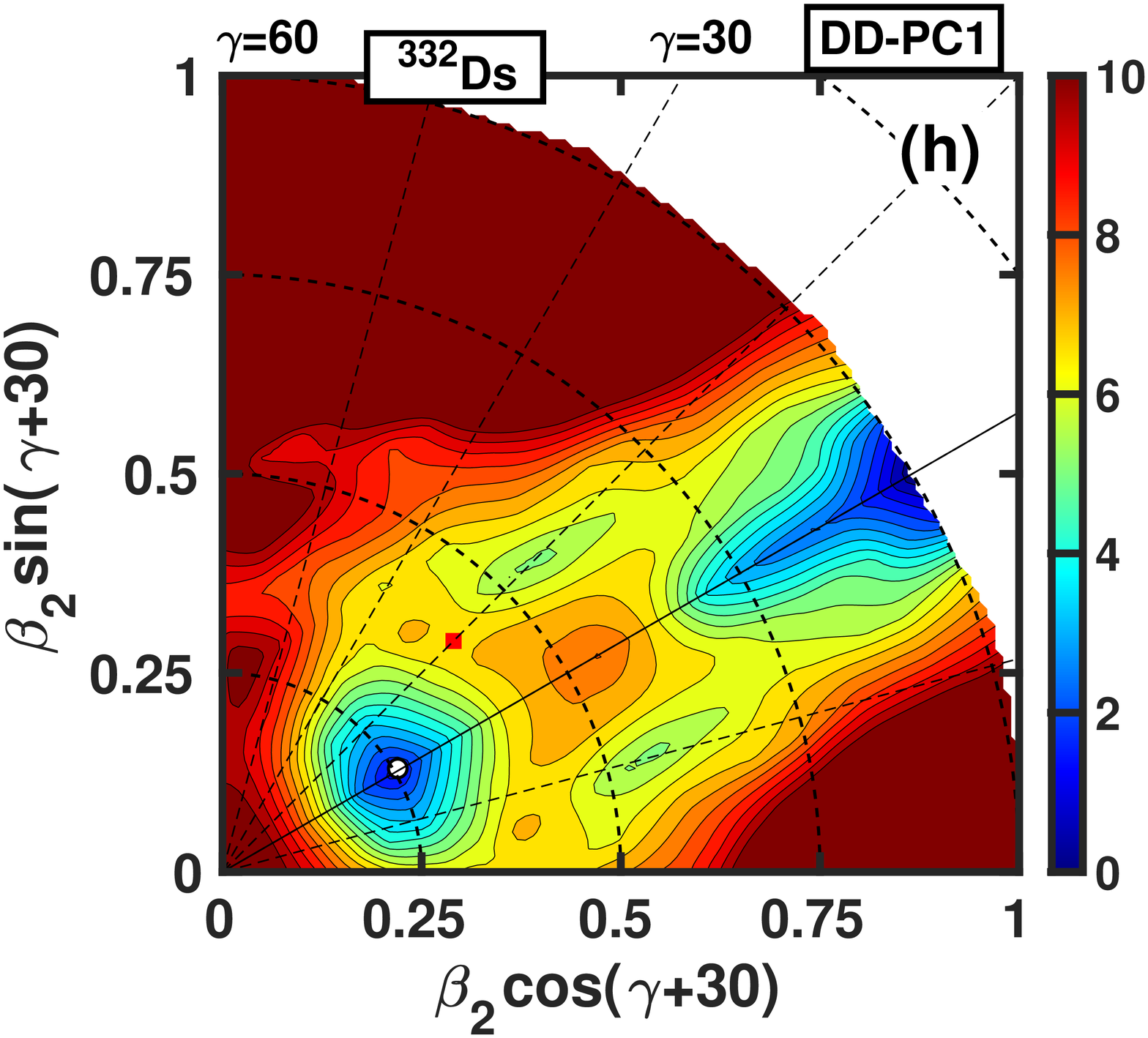}
\includegraphics[width=4.3cm]{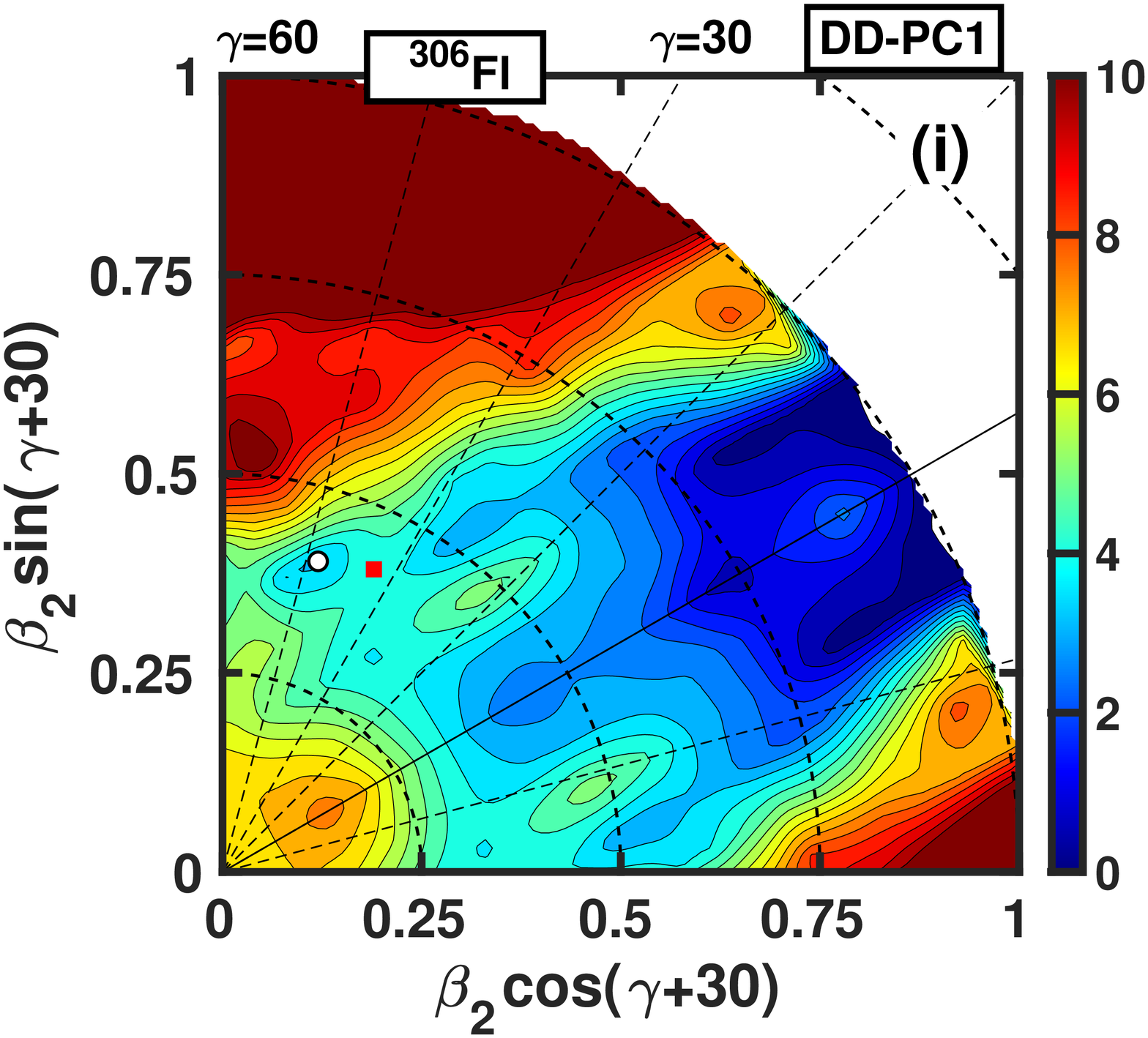}
\includegraphics[width=4.3cm]{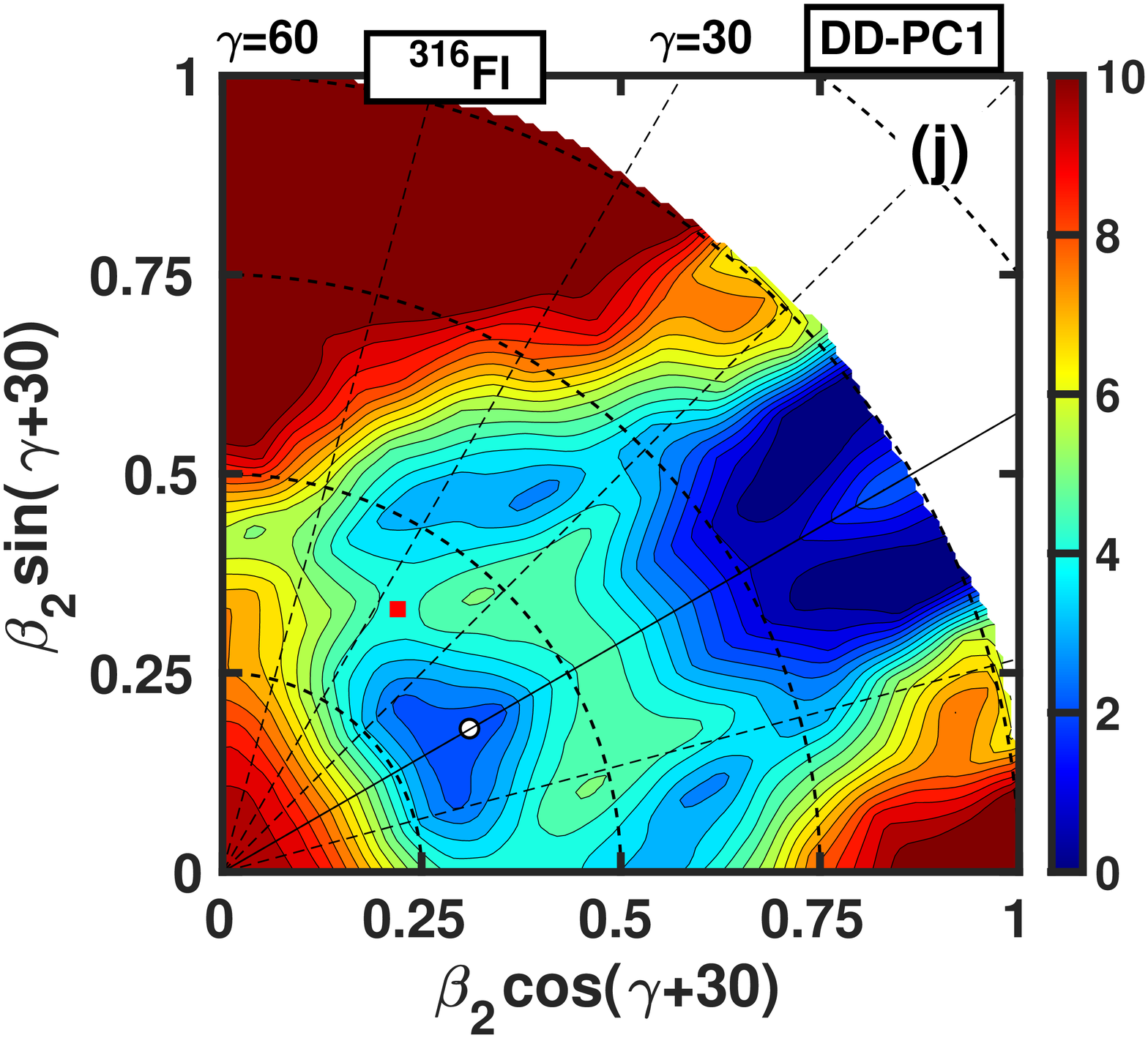}
\includegraphics[width=4.3cm]{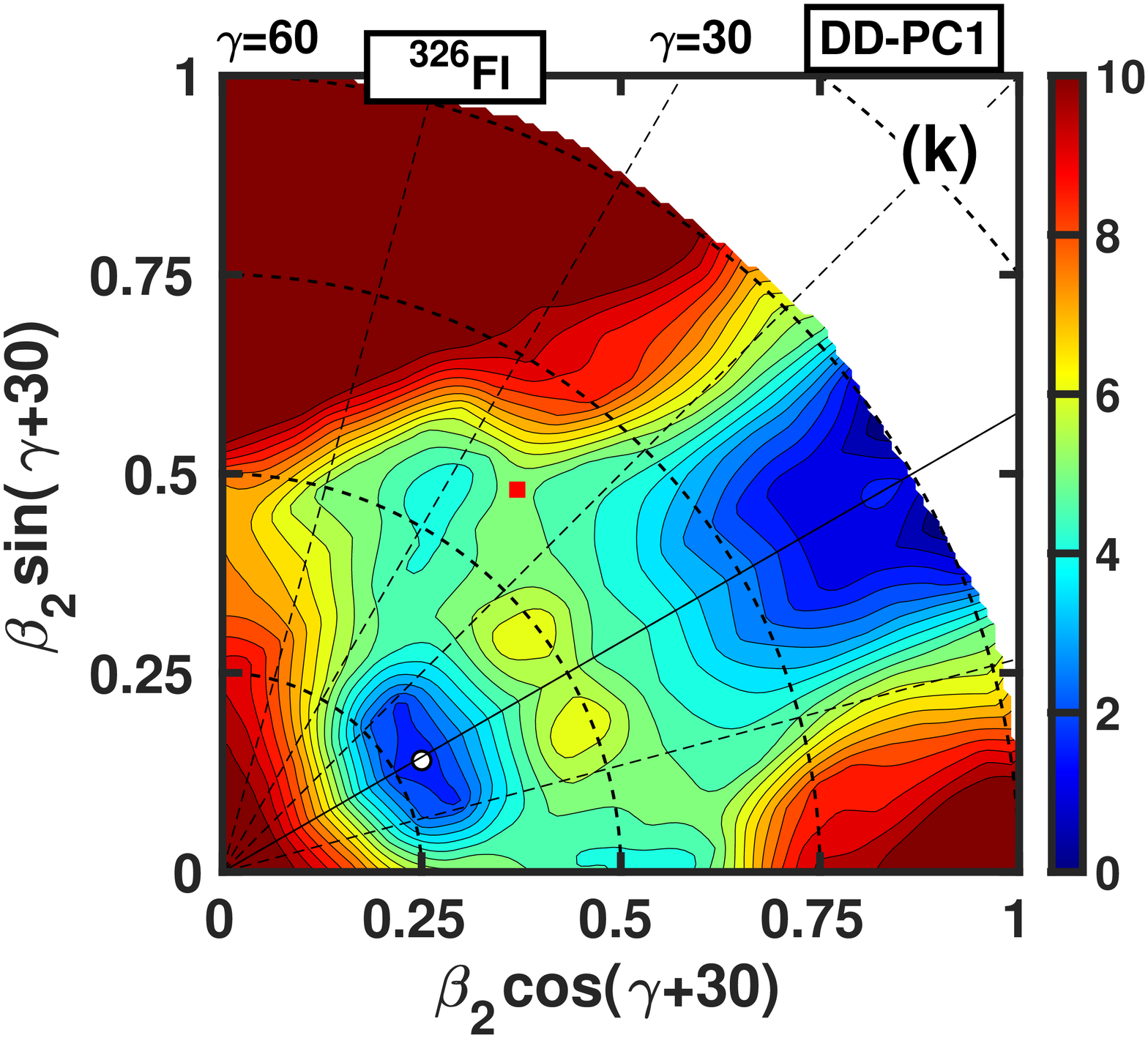}
\includegraphics[width=4.3cm]{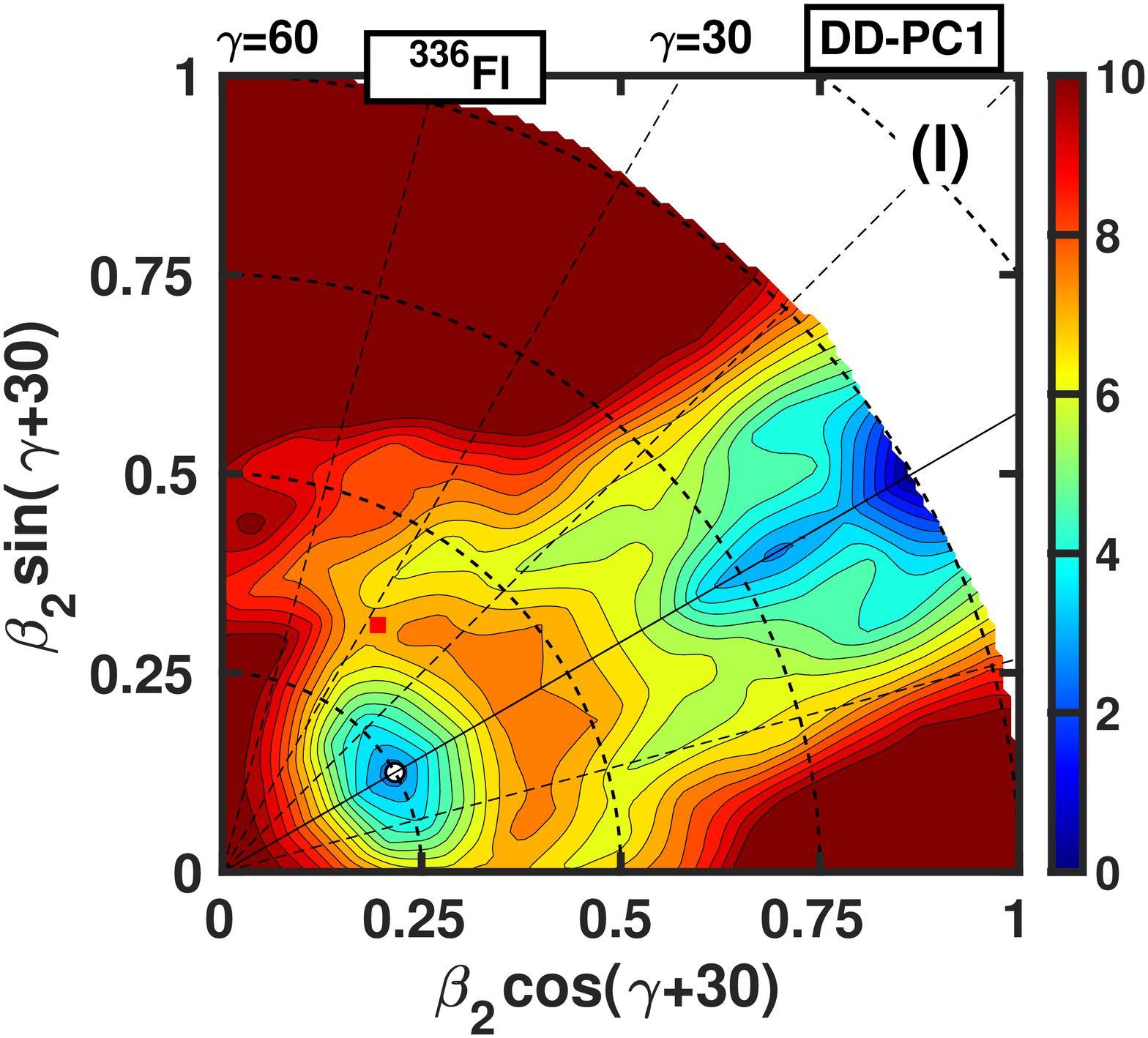}
\includegraphics[width=4.3cm]{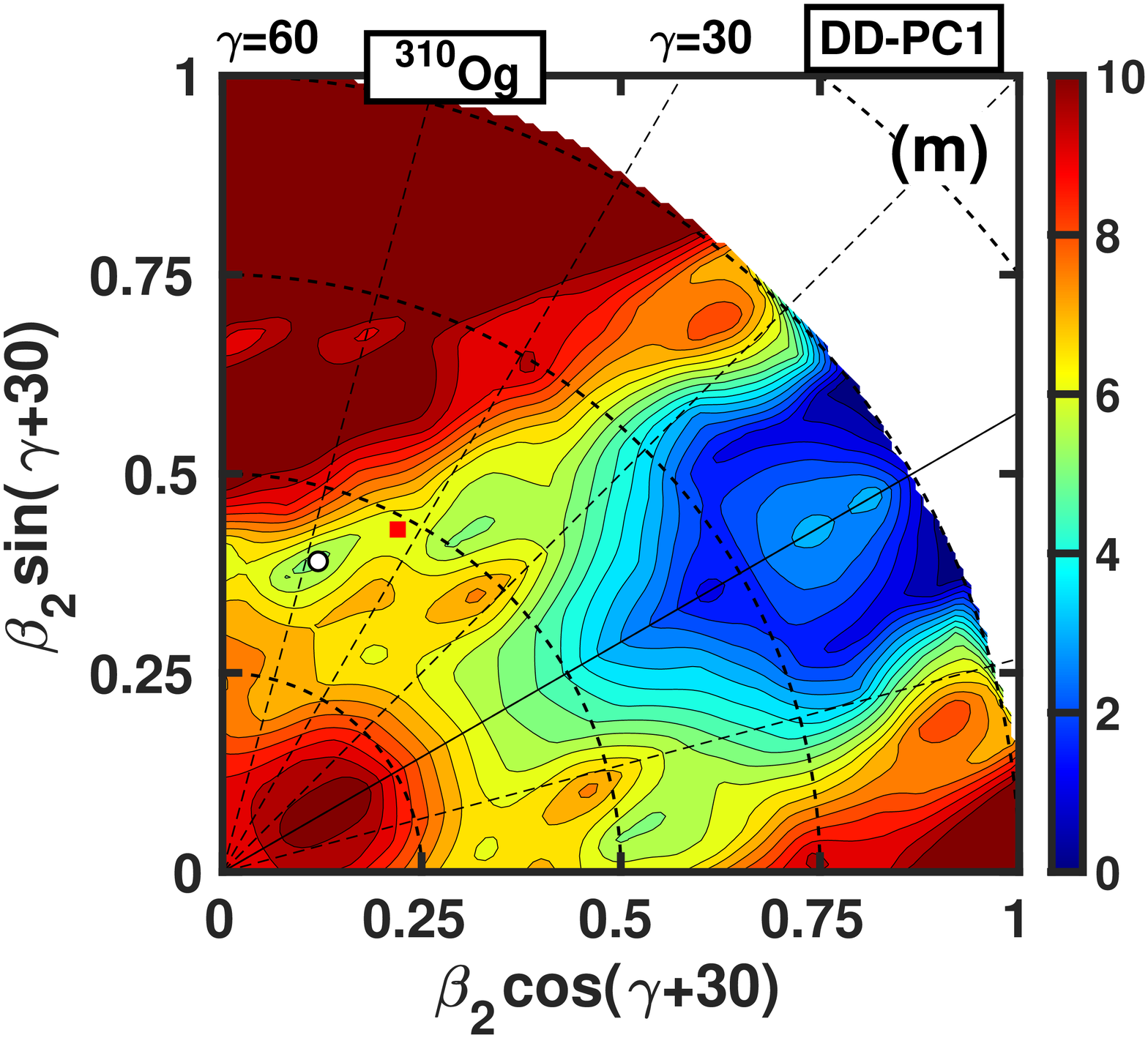}
\includegraphics[width=4.3cm]{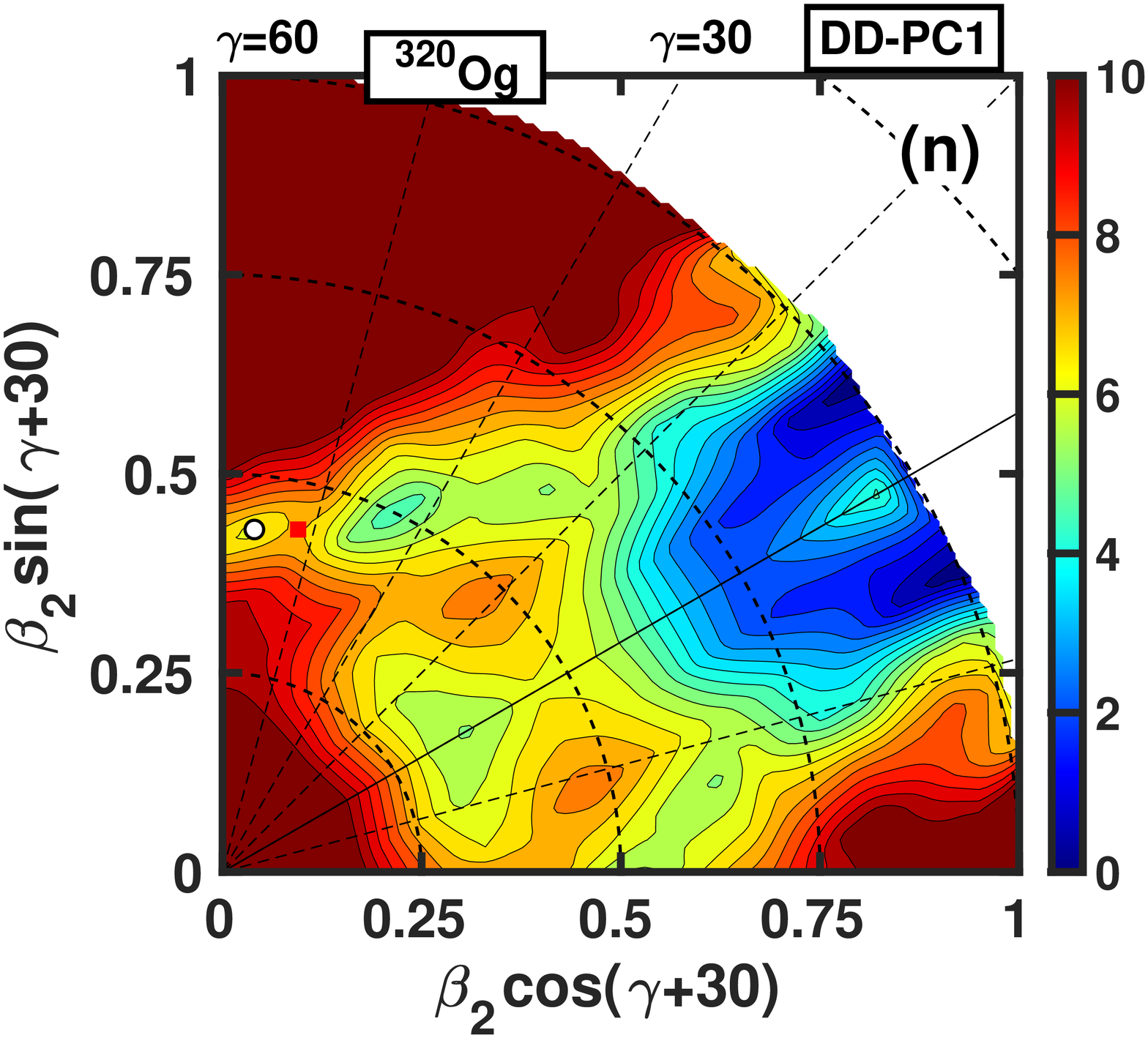}
\includegraphics[width=4.3cm]{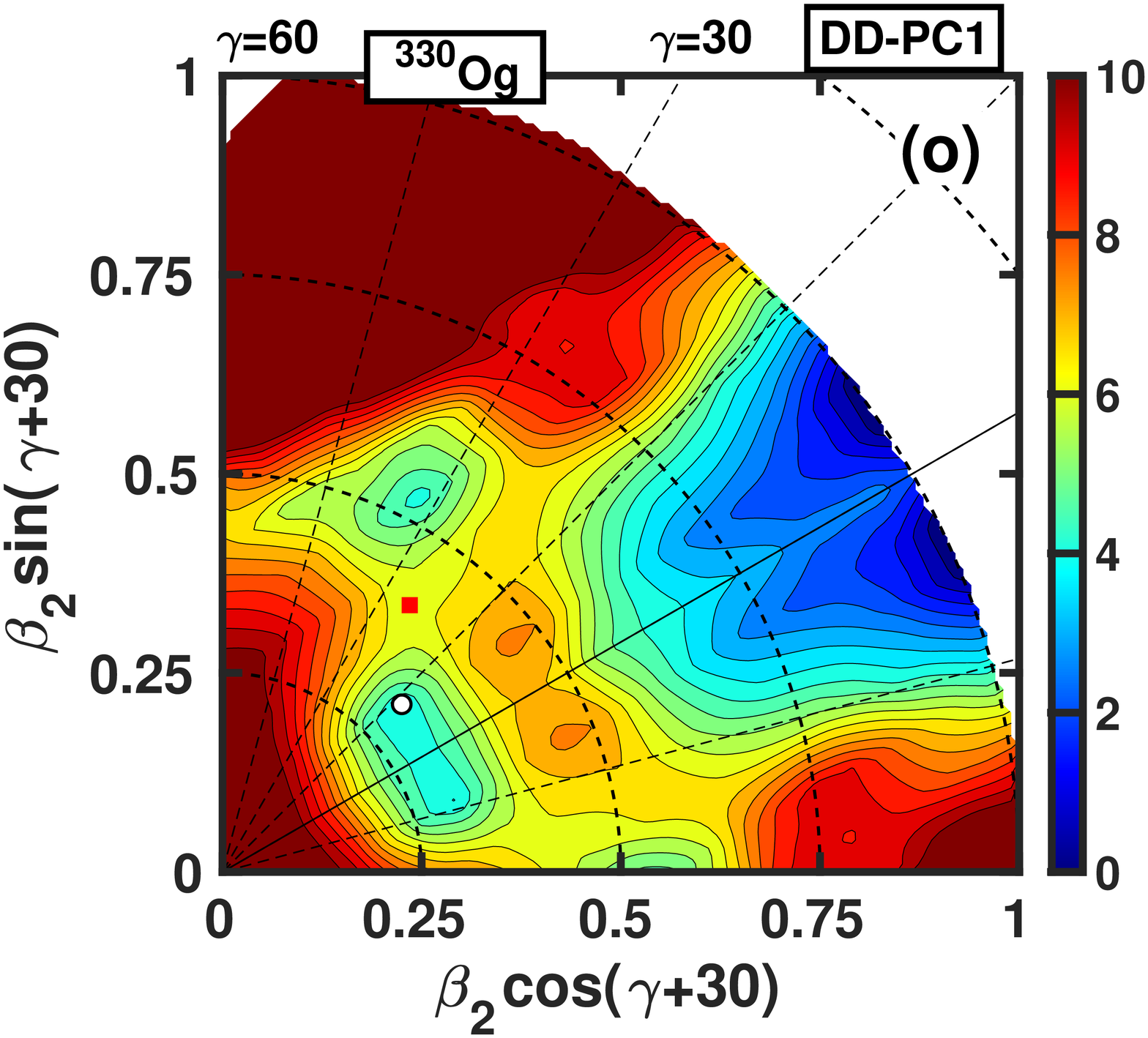}
\includegraphics[width=4.3cm]{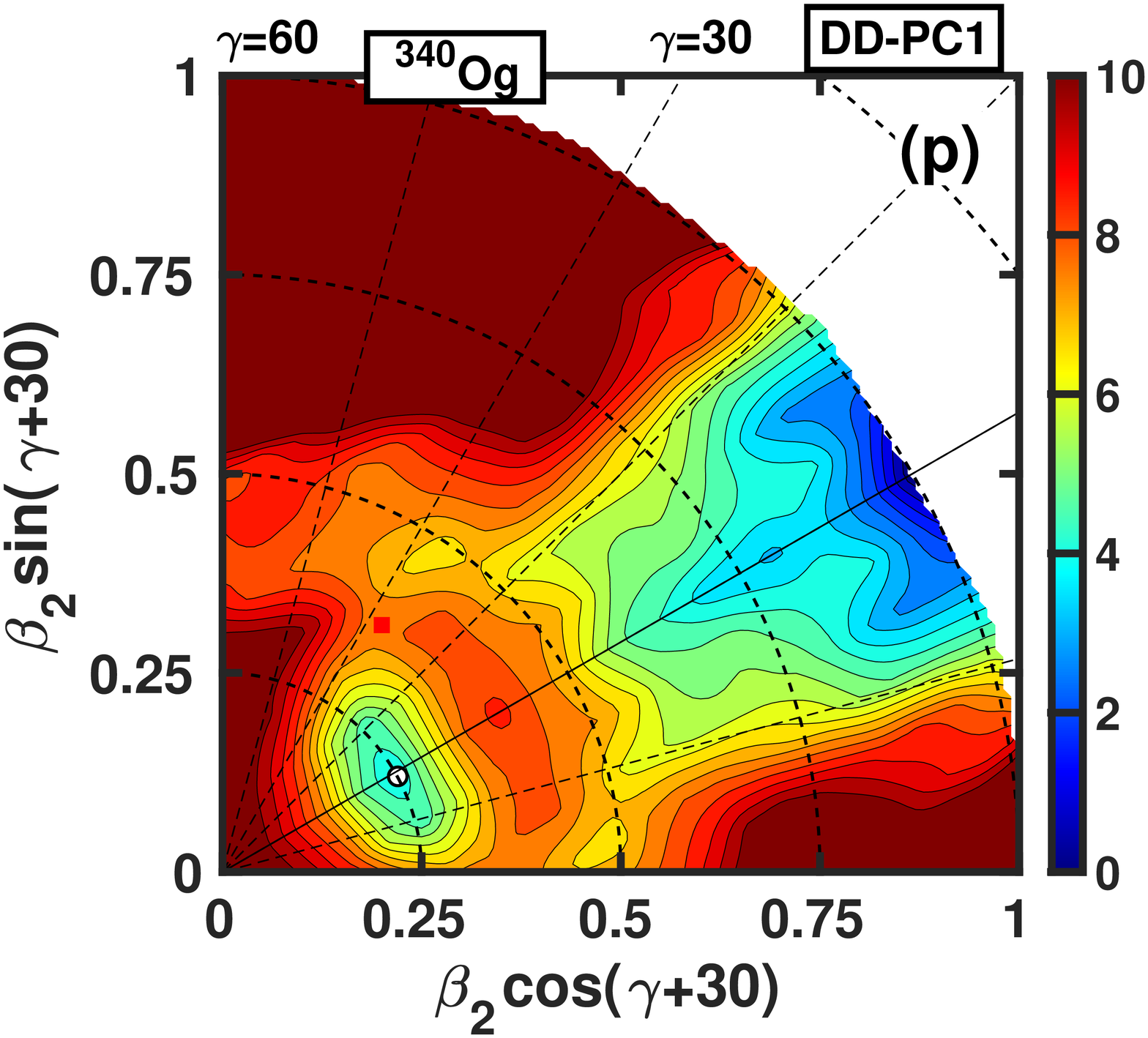}
\caption{Potential energy surfaces of the Sg ($Z=106$), Ds ($Z=110$), Fl ($Z=114$)
and Og ($Z=118$) isotopes with neutron numbers $N=192,202,212$ and 222 obtained
in the triaxial RHB calculations with the DD-PC1 functional. Neutron number is increasing on
going from left to right.  The energy difference between two neighboring equipotential 
lines is equal to 0.5 MeV. The ground state minima and saddle points are shown by 
white circles and red solid squares, respectively. 
\label{PES-triax-RHB}
}
\end{figure*}

    The restriction to axial symmetry is one of the approximations used in the present
study which is a consequence of the global character of the study (see detailed discussion
presented in the end of Sect.\ \ref{sect-theory} and in Appendix A). In order to better 
understand for which nuclei this approximation may be violated (even in least-action 
calculations with pairing fluctuations included such as those presented in Refs.\ \cite{SDNSB.14,ZLNVZ.16})
we consider the examples of potential energy surfaces obtained in triaxial RHB 
calculations with the DD-PC1 functional.  These PES calculated for the Sg ($Z=106$), 
Ds ($Z=110$), Fl ($Z=114$) and Og ($Z=118$) isotopes with neutron numbers 
$N=192,202,212$ and 222 are presented in Fig.\ \ref{PES-triax-RHB}. They represent
the extension of the calculations, executed in a more limited deformation space, the
results of which are discussed in Sect. XI of Ref.\ \cite{AATG.19}.  The summary of the heights 
$E_{triax}^{B}$ of triaxial inner fission barriers and the decreases of the fission barrier 
heights due to triaxiality $\Delta E^{gain}$ are presented in Fig.\ 21 and Table II of 
this reference. Note that these superheavy nuclei are selected in such a way that they cover the 
part of nuclear chart characterized by both oblate and prolate ground states (see Fig.\ 
\ref{deformations}a).

   The review of existing literature presented in Appendix A
suggests two possible scenarios in which the least-action fission path will most likely
be characterized by triaxiality.  In the first one, the decrease of the fission barrier by 
triaxiality in static calculations is substantial being on the order of $3-4$ MeV 
\cite{GSPS.99,SDNSB.14}.  Such decreases are observed in $^{328}$Sg
($\Delta E^{gain}=4.04$ MeV),  $^{310}$Og ($\Delta E^{gain}=3.42$ MeV), 
and $^{320}$Og ($\Delta E^{gain}=4.93$ MeV) (see Figs.\ \ref{PES-triax-RHB}(d), (m) and (n)
and Table II in Ref.\ \cite{AATG.19}).  In the second scenario, the ground state is oblate 
(or possibly soft in oblate-prolate direction \cite{SALM.19}) so that the fission path 
across the $\gamma$-plane is shorter then the one along the  $\gamma = 0^{\circ}$ 
line \cite{GSPS.99}.  This condition is satisfied only in the $^{306}$Fl (see Fig.\ 
\ref{PES-triax-RHB}(i)) and $^{310,320}$Og (see Fig.\ \ref{PES-triax-RHB}(m) 
and (n)) nuclei. Based on general features discussed in Refs.\ 
\cite{SDNSB.14,ZLNVZ.16},  the analysis of PES of remaining nuclei (see 
Figs.\ \ref{PES-triax-RHB} a, b, c, e, f, g, h, j, k, l, o and p) suggests that least-action 
fission pathway will be axial in these nuclei when pairing fluctuations are taken into 
account.

    Whether one or another scenario takes place depends on the underlying shell structure 
(both at the ground state and saddle)     
defining the topology of potential energy surfaces in the ($\beta_2, \gamma$)  plane. Fig.
\ref{PES-triax-RHB}  shows that the $N=192$ isotones are extremely soft in the $\gamma$-plane with clear
tendency for the formation of near-oblate triaxial ground state  minimum in the Fl and Og
nuclei. However, static fission pathes from these minima are characterized by low fission
barriers so these nuclei are expected to be unstable. Similar (but slightly less pronounced)  situation 
is also seen for the $N=202$ isotones.  The increase of neutron number to $N=212$ 
and 222 leads to a better localization of the ground state minimum at prolate shape and
to an increase of fission barrier heights.

\bibliography{references-27-PRC-fis-NAstr}
\end{document}